\documentclass[a4paper,onecolumn,12pt,accepted=2021-06-01]{quantumarticle}
\pdfoutput=1
\usepackage[utf8]{inputenc}
\usepackage[english]{babel}
\usepackage[T1]{fontenc}
\usepackage{amsmath}
\usepackage{hyperref}
\usepackage[numbers,sort&compress]{natbib}
\usepackage{amssymb,amsfonts,mathrsfs,dsfont,yfonts,bbm}
\usepackage{xcolor}
\usepackage{braket}
\usepackage[normalem]{ulem}
\usepackage{tabularx}
\usepackage{tikz}
\usepackage{lipsum}
\usepackage{enumitem}

\newcommand{\be}{\begin{equation}}
\newcommand{\ee}{\end{equation}}
\newcommand{\bea}{\begin{eqnarray}}
\newcommand{\eea}{\end{eqnarray}}
\newcommand{\Rep}{{\rm Rep}}
\newcommand{\Ker}{{\rm Ker}}
\newcommand{\DZ}{\mathbb{Z}}
\newcommand{\CL}{\mathcal{L}}
\newcommand{\CA}{\mathcal{A}}

\newcommand{\CW}{\mathcal{W}}
\newcommand{\CD}{\mathcal{D}}

\newcommand{\CC}{\mathcal{C}}
\newcommand{\CZ}{\mathcal{Z}}
\newcommand{\CO}{\mathcal{O}}
\newcommand{\CT}{\mathcal{T}}

\newcommand{\CN}{\mathcal{N}}
\newcommand{\CS}{\mathcal{S}}
\newcommand{\CM}{\mathcal{M}}

\newcommand*{\boxcoloro}{orange}
\makeatletter
\newcommand{\boxedo}[1]{\textcolor{\boxcoloro}{%
\tikz[baseline={([yshift=-1ex]current bounding box.center)}] \node [rectangle, minimum width=1ex,rounded corners,draw] {\normalcolor\m@th$\displaystyle#1$};}}
 \makeatother
\newcommand*{\boxcolorr}{red}
\makeatletter
\newcommand{\boxedr}[1]{\textcolor{\boxcolorr}{%
\tikz[baseline={([yshift=-1ex]current bounding box.center)}] \node [rectangle, minimum width=1ex,rounded corners,draw] {\normalcolor\m@th$\displaystyle#1$};}}
 \makeatother
\newcommand*{\boxcolorb}{blue}
\makeatletter
\newcommand{\boxedb}[1]{\textcolor{\boxcolorb}{%
\tikz[baseline={([yshift=-1ex]current bounding box.center)}] \node [rectangle, minimum width=1ex,rounded corners,draw] {\normalcolor\m@th$\displaystyle#1$};}}
 \makeatother
\newcommand*{\boxcolorg}{green}
\makeatletter
\newcommand{\boxedg}[1]{\textcolor{\boxcolorg}{%
\tikz[baseline={([yshift=-1ex]current bounding box.center)}] \node [rectangle, minimum width=1ex,rounded corners,draw] {\normalcolor\m@th$\displaystyle#1$};}}
 \makeatother
 \newcommand*{\boxcolorp}{purple}
\makeatletter
\newcommand{\boxedp}[1]{\textcolor{\boxcolorp}{%
\tikz[baseline={([yshift=-1ex]current bounding box.center)}] \node [rectangle, minimum width=1ex,rounded corners,draw] {\normalcolor\m@th$\displaystyle#1$};}}
 \makeatother
  \newcommand*{\boxcolorc}{cyan}
\makeatletter
\newcommand{\boxedc}[1]{\textcolor{\boxcolorc}{%
\tikz[baseline={([yshift=-1ex]current bounding box.center)}] \node [rectangle, minimum width=1ex,rounded corners,draw] {\normalcolor\m@th$\displaystyle#1$};}}
 \makeatother
  \newcommand*{\boxcolory}{yellow}
\makeatletter
\newcommand{\boxedy}[1]{\textcolor{\boxcolory}{%
\tikz[baseline={([yshift=-1ex]current bounding box.center)}] \node [rectangle, minimum width=1ex,rounded corners,draw] {\normalcolor\m@th$\displaystyle#1$};}}
 \makeatother
 
\begin{document}

\title{$a\times b=c$ in 2+1D TQFT}

\author{Matthew Buican}
\email{m.buican@qmul.ac.uk}
\author{Linfeng Li}
\email{linfeng.li@qmul.ac.uk}
\author{Rajath Radhakrishnan}
\email{r.k.radhakrishnan@qmul.ac.uk}
\affiliation{CRST and School of Physics and Astronomy, 
Queen Mary University of London, London E1 4NS, UK}

\maketitle

\begin{abstract}
 We study the implications of the anyon fusion equation $a\times b=c$ on global properties of $2+1$D topological quantum field theories (TQFTs). Here $a$ and $b$ are anyons that fuse together to give a unique anyon, $c$. As is well known, when at least one of $a$ and $b$ is abelian, such equations describe aspects of the one-form symmetry of the theory. When $a$ and $b$ are non-abelian, the most obvious way such fusions arise is when a TQFT can be resolved into a product of TQFTs with trivial mutual braiding, and $a$ and $b$ lie in separate factors. More generally, we argue that the appearance of such fusions for non-abelian $a$ and $b$ can also be an indication of zero-form symmetries in a TQFT, of what we term ``quasi-zero-form symmetries" (as in the case of discrete gauge theories based on the largest Mathieu group, $M_{24}$), or of the existence of non-modular fusion subcategories. We study these ideas in a variety of TQFT settings from (twisted and untwisted) discrete gauge theories to Chern-Simons theories based on continuous gauge groups and related cosets. Along the way, we prove various useful theorems.
\end{abstract}

\tableofcontents

\section{Introduction}
Topological quantum field theories (TQFTs) in $2+1$ dimensions and their anyonic excitations lie at the heart of important physical \cite{Moore:1991ks}, mathematical \cite{Witten:1988hf}, and computational \cite{wang2010topological} systems and constructions. In principle, these TQFTs can be fully characterized by solving a set of polynomial consistency conditions \cite{Moore:1989vd,bakalov2001lectures,kitaev2006anyons}, although proceeding in this way is often quite difficult as a practical matter (however, see \cite{rowell2009classification,bruillard2019modular} for examples of some results; see also \cite{Cho:2020ljj} for a potentially very different approach). More generally, it is interesting to understand aspects of the global structure of a TQFT and its symmetries without the need to fully solve the theory (e.g., see \cite{muger2003structure}).

Proceeding in this way, we will study anyonic fusions $a\times b$ that have a unique product anyon, $c$
\begin{equation}\label{abcfusion0}
a\times b=c~, \ \ \ a,\ b,\ c\in\CT~,
\end{equation}
in a general $2+1$ dimensional TQFT, $\CT$.\footnote{Throughout what follows, we only consider non-spin TQFTs. These are theories that do not require a spin structure in order to be well-defined.} Our main questions is: what does \eqref{abcfusion0} tell us about the global structure of $\CT$ and its symmetries?

For invertible $a$ and $b$ (i.e., $a$ and $b$ are abelian anyons), fusion rules of the form \eqref{abcfusion0} describe the abelian 1-form symmetry group of the theory \cite{Gaiotto:2014kfa} (the closely related modular $S$ matrix characterizes its 't Hooft anomalies \cite{Hsin:2018vcg}). In the case in which, say, $a$ is abelian and $b$ is non-abelian,\footnote{In this case, $b$ is non-invertible, and the fusion $b\times \bar b=1+\cdots$, where $\bar b$ is the anyon conjugate to $b$, necessarily contains at least one more anyon in the ellipses.} the equation \eqref{abcfusion0} gives the fixed points of the fusion of anyons in the theory with the one-form generator, $a$. Such equations have important consequences for anyon condensation / one-form symmetry gauging in TQFT \cite{bais2009condensate,Hsin:2018vcg} as well as for orbifolding and coset constructions in closely related 2D rational conformal field theories (RCFTs) (e.g., see \cite{Intriligator:1989zw,Schellekens:1990xy}).

Although these cases will play a role below, we will be more interested in the situation in which both $a$ and $b$ are non-abelian
\begin{equation}\label{abcfusion}
a\times b=c~, \ \ \ d_a~,\ d_b>1~.
\end{equation}
Here $d_{a,b}$ denote the quantum dimensions of $a$ and $b$ (given they are larger than one, neither $a$ nor $b$ are invertible). Since both $a$ and $b$ are non-abelian, one typically finds that the right-hand side of \eqref{abcfusion} has multiple fusion products. For example, fusions as in \eqref{abcfusion} do not occur in $SU(2)_k$ Chern-Simons (CS) theory for any value of $k\in\mathbb{N}$.\footnote{In section \ref{cosets}, we will discuss the situation for more general $G_k$ CS theories.} As we will see, when fusions of non-abelian $a$ and $b$ do have a unique outcome, there are consequences for the global structure of $\CT$.

The most trivial case in which a fusion of the type \eqref{abcfusion} occurs is when $\CT$ factorizes (not necessarily uniquely) as
\begin{equation}\label{factorization}
\CT=\CT_1\boxtimes\CT_2~,
\end{equation}
with $\CT_1$ and $\CT_2$ two separate TQFTs that have trivial mutual braiding, $a\in\CT_1$, and $b\in\CT_2$.\footnote{Note that $\CT_{1,2}$ may factorize further. Moreover, $a$ may contain an abelian component in $\CT_2$, and $b$ may contain an abelian component in $\CT_1$.} Here \lq\lq$\boxtimes$" denotes a categorical generalization of the direct product called a \lq\lq Deligne product" that respects some of the additional structure present in TQFT.

As we will discuss in section \ref{cosets}, precisely such a situation arises in the modular tensor categories (MTCs) related to unitary $A$-type Virasoro minimal models with $c>1/2$.\footnote{Note that in the case of the Ising model ($c=1/2$), at least one of the anyons in the fusion $a\times b=c$ is abelian (and the corresponding MTC does not factorize). We thank I.~Runkel for drawing our attention to the $a\times b=c$ fusion rules for non-abelian fields in Virasoro minimal models.} MTCs are mathematical descriptions of TQFTs, and, for the theories in question, they encapsulate the topological properties of the Virasoro primary fields. One may think of the, say, left-movers in these RCFTs as arising at a 1+1 dimensional interface between 2+1 dimensional CS theories with gauge groups $SU(2)_1\times SU(2)_k$ and $SU(2)_{k+1}$. In the minimal models, we have
\begin{equation}\label{virabc}
\varphi_{(r,1)}\times\varphi_{(1,s)}=\varphi_{(r,s)}~,
\end{equation}
where $2\le r<p-2$ and $2\le s<p-1$ are Kac labels that give Virasoro primaries with non-abelian fusion rules (here we have $(r,s)\sim(p-1-r,p-s)$, and $p>4$ is an integer labeling the unitary minimal model).\footnote{The abelian field $\varphi_{(p-2,1)}\sim\varphi_{(1,p-1)}$ satisfies the fusion rule $\varphi_{(1,p-1)}\times\varphi_{(1,p-1)}=\varphi_{(1,1)}=1$.}  Thinking in terms of cosets, we will see that \eqref{virabc} arises because the Virasoro MTC factorizes as in \eqref{factorization}.\footnote{Note that this factorization does not extend to one of the RCFT.}

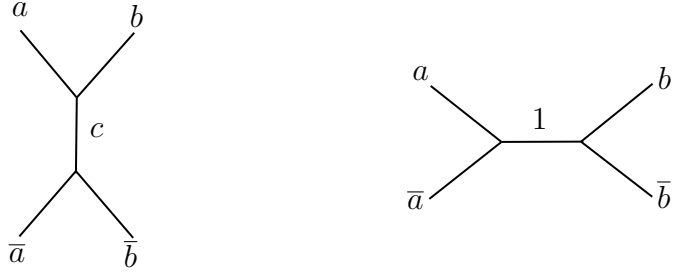
\begin{figure}[h!]

\tikzset{every picture/.style={line width=0.75pt}} 
\tikzset{every picture/.style={line width=0.75pt}} 
\centering
\begin{tikzpicture}[x=0.75pt,y=0.75pt,yscale=-1,xscale=1]

\draw [color={rgb, 255:red, 0; green, 0; blue, 0 }  ,draw opacity=1 ]   (357.49,157.7) -- (393.49,129.05) ;
\draw [color={rgb, 255:red, 0; green, 0; blue, 0 }  ,draw opacity=1 ]   (358.17,100.89) -- (393.49,129.05) ;
\draw [color={rgb, 255:red, 0; green, 0; blue, 0 }  ,draw opacity=1 ]   (393.49,129.05) -- (434.36,128.9) ;
\draw [color={rgb, 255:red, 0; green, 0; blue, 0 }  ,draw opacity=1 ]   (468.92,99.82) -- (433.18,128.92) ;
\draw [color={rgb, 255:red, 0; green, 0; blue, 0 }  ,draw opacity=1 ]   (468.75,156.64) -- (433.18,128.92) ;
\draw [color={rgb, 255:red, 0; green, 0; blue, 0 }  ,draw opacity=1 ]   (153.38,72.93) -- (181.5,106.75) ;
\draw [color={rgb, 255:red, 0; green, 0; blue, 0 }  ,draw opacity=1 ]   (210.18,74.13) -- (181.5,106.75) ;
\draw [color={rgb, 255:red, 0; green, 0; blue, 0 }  ,draw opacity=1 ]   (181.5,106.75) -- (181.07,144.83) ;
\draw [color={rgb, 255:red, 0; green, 0; blue, 0 }  ,draw opacity=1 ]   (209.64,177.31) -- (181.06,143.73) ;
\draw [color={rgb, 255:red, 0; green, 0; blue, 0 }  ,draw opacity=1 ]   (152.83,176.59) -- (181.06,143.73) ;

\draw (347.5,89.4) node [anchor=north west][inner sep=0.75pt]    {$a$};
\draw (344.93,151.9) node [anchor=north west][inner sep=0.75pt]    {$\overline{a}$};
\draw (470.11,89.9) node [anchor=north west][inner sep=0.75pt]    {$b$};
\draw (469.69,146.9) node [anchor=north west][inner sep=0.75pt]    {$\overline{b}$};
\draw (407.09,110.9) node [anchor=north west][inner sep=0.75pt]    {$1$};
\draw (147.5,57.29) node [anchor=north west][inner sep=0.75pt]    {$a$};
\draw (206.5,58.49) node [anchor=north west][inner sep=0.75pt]    {$b$};
\draw (146,178.51) node [anchor=north west][inner sep=0.75pt]    {$\overline{a}$};
\draw (203.5,176.91) node [anchor=north west][inner sep=0.75pt]    {$\overline{b}$};
\draw (186.5,118.4) node [anchor=north west][inner sep=0.75pt]    {$c$};
\end{tikzpicture}
\caption{The fusions $a\times b$ and $\bar a\times\bar b$ have unique outcomes $c$ and $\bar c$ respectively. In the left diagram, we connect the corresponding fusion vertices. To get to the diagram on the right, we perform an $F^{\bar a a b}_{\bar b}$ transformation. Just as the left diagram has a unique internal line, so too does the diagram on the right (in this latter case, the internal line must be the identity).}\label{ftrans1}
\end{figure}

\begin{figure}[h!]
\centering

\tikzset{every picture/.style={line width=0.75pt}} 

\begin{tikzpicture}[x=0.75pt,y=0.75pt,yscale=-1,xscale=1]

\draw [color={rgb, 255:red, 0; green, 0; blue, 0 }  ,draw opacity=1 ]   (352.13,174.7) -- (387.09,146.05) ;
\draw [color={rgb, 255:red, 0; green, 0; blue, 0 }  ,draw opacity=1 ]   (352.79,117.89) -- (387.09,146.05) ;
\draw [color={rgb, 255:red, 0; green, 0; blue, 0 }  ,draw opacity=1 ]   (387.09,146.05) -- (426.77,145.9) ;
\draw [color={rgb, 255:red, 0; green, 0; blue, 0 }  ,draw opacity=1 ]   (460.33,116.82) -- (425.63,145.92) ;
\draw [color={rgb, 255:red, 0; green, 0; blue, 0 }  ,draw opacity=1 ]   (460.17,173.64) -- (425.63,145.92) ;
\draw [color={rgb, 255:red, 0; green, 0; blue, 0 }  ,draw opacity=1 ]   (150.38,86.98) -- (178.5,120.91) ;
\draw [color={rgb, 255:red, 0; green, 0; blue, 0 }  ,draw opacity=1 ]   (207.18,88.18) -- (178.5,120.91) ;
\draw [color={rgb, 255:red, 0; green, 0; blue, 0 }  ,draw opacity=1 ]   (178.5,120.91) -- (178.07,159.1) ;
\draw [color={rgb, 255:red, 0; green, 0; blue, 0 }  ,draw opacity=1 ]   (206.64,191.69) -- (178.06,158) ;
\draw [color={rgb, 255:red, 0; green, 0; blue, 0 }  ,draw opacity=1 ]   (149.83,190.97) -- (178.06,158) ;

\draw (342.24,106.4) node [anchor=north west][inner sep=0.75pt]    {$a$};
\draw (339.74,168.9) node [anchor=north west][inner sep=0.75pt]    {$\overline{b}$};
\draw (461.3,106.9) node [anchor=north west][inner sep=0.75pt]    {$b$};
\draw (460.88,165.9) node [anchor=north west][inner sep=0.75pt]    {$\overline{a}$};
\draw (400.11,127.9) node [anchor=north west][inner sep=0.75pt]    {$d$};
\draw (144.5,71.31) node [anchor=north west][inner sep=0.75pt]    {$a$};
\draw (203.5,72.52) node [anchor=north west][inner sep=0.75pt]    {$b$};
\draw (143,191.52) node [anchor=north west][inner sep=0.75pt]    {$\overline{b}$};
\draw (200.5,193.71) node [anchor=north west][inner sep=0.75pt]    {$\overline{a}$};
\draw (183.5,132.62) node [anchor=north west][inner sep=0.75pt]    {$c$};

\end{tikzpicture}
\caption{By rotating the bottom vertex in the left diagram of figure \ref{ftrans1}, we arrive at the above diagram on the left. Again, we have a single internal line labeled by $c$. We get to the diagram on the right by performing an $F^{\bar b a b}_{\bar a}$ transformation. Just as the left diagram has a unique internal line, so too does the diagram on the right.}\label{ftrans2}
\end{figure}
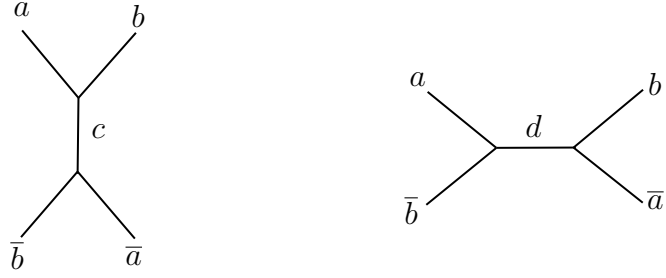

To gain further insight into more general situations in which \eqref{abcfusion} occurs, it is useful to imagine connecting a fusion vertex involving the $a$, $b$, $c$ ayons with a fusion vertex involving the $\bar a$, $\bar b$, and $\bar c$ anyons via a $c$ internal line as in the left diagram of figure \ref{ftrans1}. Using associativity of fusion (via a so-called $F^{\bar a a b}_{\bar b}$ symbol) we arrive at the right diagram of figure \ref{ftrans1}. The relation between these two diagrams can be thought of as a change of basis on the space of internal states. Since, by construction, the left diagram in figure \ref{ftrans1} can only involve a $c$ internal line, the right diagram in figure \ref{ftrans1} can also only involve a single internal line. On general grounds, this line must be the identity.\footnote{By rotating the $\bar a$, $\bar b$, and $\bar c$ vertex, we see that $a\times b=c$ is equivalent to requiring $a\times\bar b=d$ and $\bar a\times b=\bar d$ (see figure \ref{ftrans2}). This logic also explains why, for non-abelian $a$, it is impossible to have $a\times a=c$ even if $a\ne\bar a$.}  This result can also be derived by looking at decomposition of fusion spaces. Consider the fusion space $V_{ba\overline{a}}^b$. It can be decomposed in the following different ways
\be
\label{fusdecomp}
V_{ba\overline{a}}^b \simeq \sum_c V_{ba}^c \otimes V_{c\overline{a}}^b \simeq \sum_x V_{bx}^b \otimes V_{a\overline{a}}^x   \simeq  \sum_x V_{b\overline{b}}^{\overline{x}} \otimes V_{a\overline{a}}^x ,
\ee
where, in the last equality above, we have used the fusion space isomorphism, $V_{bx}^b\simeq V_{b\overline{b}}^{\overline{x}}$. If we have the fusion rule $a \times b=c$, then \eqref{fusdecomp} simplifies to
\be
V_{ba\overline{a}}^b \simeq V_{ba}^c \otimes V_{c\overline{a}}^b \simeq  \sum_x V_{b\overline{b}}^{\overline{x}} \otimes V_{a\overline{a}}^x
\ee
Moreover, we know that $V_{ba}^c$  and $V_{c\overline{a}}^b$ are 1-dimensional. Hence, the dimension of direct sum of fusion spaces $\sum_x V_{b\overline{b}}^{\overline{x}} \otimes V_{a\overline{a}}^x$ should be 1-dimensional. It follows that the sum should be over a single element and that the fusion spaces $V_{a\overline{a}}^x$ and $V_{b\overline{b}}^{\overline{x}}$ should be 1-dimensional. Since the trivial anyon $1$ is always an element in the fusions $a \times \overline{a}$ and $b \times \overline{b}$, we have
 \be
V_{ba\overline{a}}^b \simeq V_{ba}^c \otimes V_{c\overline{a}}^b \simeq V_{b\overline{b}}^{1} \otimes V_{a\overline{a}}^1 
\ee
Therefore, we learn that a fusion rule of the form \eqref{abcfusion} is equivalent to the following
\begin{eqnarray}\label{aabar}
&&a\times\bar a=1+\sum_{a_i\ne1}N_{a\bar a}^{a_i}\ a_i~, \ \ \ b\times\bar b=1+\sum_{b_j\ne1}N_{b\bar b}^{b_j}\ b_j~, \ \ \ b_j\in b\times \bar b\ \Rightarrow\ b_j\not\in a\times\bar a~,\cr &&a_i\in a\times\bar a\ \Rightarrow\ a_i\not\in b\times\bar b\ \ \ \forall\ i, j~.
\end{eqnarray}
In other words, the fusion of $a\times b$ has a unique outcome if and only if the only fusion product that $a\times \bar a$ and $b\times \bar b$ have in common is the identity.

Reformulating the problem as in \eqref{aabar} immediately suggests scenarios in which fusions of the form \eqref{abcfusion} occur beyond cases in which $\CT$ factorizes into prime TQFTs. For example, if $a\in\CC_1\subset\CT$ and $b\in\CC_2\subset\CT$ lie in non-modular fusion subcategories of $\CT$, $\CC_{1,2}$, with trivial intersection (i.e., $\CC_1\cap\CC_2=1$ only contains the trivial anyon), then we have \eqref{abcfusion} and $\CT$ need not factorize.\footnote{In other words, fusion of anyons in $\CC_i$ is closed. Moreover, the $\CC_i$ inherit associativity and braiding from $\CT$, but the Hopf link evaluated on anyons in these subcategories is degenerate (as a matrix). By modularity, the $\CC_i$ will have non-trivial braiding with some anyons $x_A\not\in\CC_{1,2}$ (where $A$ is an index running over such anyons). On the other hand, if the Hopf links for the $\CC_i$ are non-degenerate, M\"uger's theorem \cite{muger2003structure} guarantees that they will in fact be separate TQFTs and so we are back in the situation of \eqref{factorization}.} More generally, when $a\in\CC\subset\CT$ is a member of a non-modular subcategory that does not include $b$ (i.e., $b\not\in\CC$), we expect it to be more likely to find fusions of the form \eqref{aabar} and \eqref{abcfusion} since $a\times\bar a\in\CC$, but $b\times\bar b$ will generally include elements outside $\CC$. In fact, we will see that we can often say more when the fusion of a non-abelian Wilson line carrying charge in an unfaithful representation of a discrete gauge group is involved.

Another scenario in which we can imagine \eqref{aabar}---and therefore \eqref{abcfusion}---arising is one in which zero-form symmetries act non-trivially on $a$ (i.e., $g(a)\ne a$ for some zero-form generator $g\in G$, where $G$ is the zero-form group) and the $a_i\ne1$ but not on $b$.\footnote{By definition, the symmetry also acts non-trivially on $\bar a$ so that $\overline{g(a)}=g(\bar a)\ne \bar a$. On the other hand, note that one-form symmetry will act trivially on the product $a\times\bar a$.} In this case, combinations of $a_i$ that do not form full orbits under $G$ are forbidden from appearing in $b\times\bar b$. Given a particular $G$, this argument may suffice to show that, for all $i$, $a_i\not\in b\times\bar b$. More generally, symmetries constrain what can appear as fusion products of $a\times\bar a$ and $b\times \bar b$. The more powerful these symmetries, the more likely to find fusion rules of the type \eqref{aabar}.

Interestingly, there is a close connection between the existence of symmetries and the existence of subcategories in TQFT. For example, as we will discuss further in section \ref{0formsec}, for TQFTs corresponding to discrete gauge theories \cite{Dijkgraaf:1989pz,Roche:1990hs}, certain \lq\lq quantum symmetries" or electric-magnetic self-dualities arise when we have particular non-modular subcategories $\CC_i\subset\CT$ (see \cite{nikshych2014categorical} for a general theory of such symmetries and \cite{beigi2011quantum} for the case of $S_3$ discrete gauge theory).

We will also find various other, more subtle, connections between symmetries and fusion rules of the form \eqref{aabar} and \eqref{abcfusion}. Moreover, we will see that symmetry is ubiquitous: in all the theories with fusion rules of the form \eqref{aabar} and \eqref{abcfusion} we analyze, either there is a zero-form symmetry present or else there is, at the very least, a symmetry of the modular data that exchanges anyons (in cases where this action does not lift to the full TQFT, we call these symmetries \lq\lq quasi zero-form symmetries").

We will study fusions of the above type in two typically very different classes of $2+1$D TQFTs:\footnote{Note that there are sometimes dualities between theories in these two classes.} discrete gauge theories and cosets built out of CS theories with continuous gauge groups (we will refer to these latter theories simply as \lq\lq cosets"). Discrete gauge theories are always non-chiral, whereas Chern-Simons theories and their associated cosets are typically chiral.\footnote{By a chiral TQFT, we mean one in which the topological central charge satisfies $c_{\rm top}\ne0\ ({\rm mod}\ 8)$.}

In the context of discrete gauge theories, whenever we have a (full) zero-form symmetry present, we will see that fusion rules of the type \eqref{aabar} and \eqref{abcfusion} have simple interpretations in certain parent theories gotten by gauging the zero-form symmetry, $G_0$. We go from the parent theories back to the original theories by gauging a \lq\lq dual" one-form symmetry, $G_1$, that is isomorphic (as a group) to $G_0$ (see \cite{Barkeshli:2014cna} for a more general review of this procedure). In this reverse process, we produce the $a\times b=c$ fusion rules of the corresponding discrete gauge theories via certain fusion fixed points of the one-form symmetry generators in the parent theories.

Similarly, in the context of our coset theories, we will see that fusion rules of the form $a\times b=c$ arise due to certain fixed points in the coset construction (though these fixed points do not generally involve $a$, $b$, and $c$). Cosets corresponding to the Virasoro minimal models lack such fixed points and so, as discussed above, they factorize. On the other hand, more complicated cosets do sometimes have such fixed points, and we will construct an explicit example of such a prime TQFT that has fusion rules of the form \eqref{aabar} and \eqref{abcfusion}.

To summarize, this discussion leads us to the following questions we will answer in subsequent sections:
\noindent
\begin{enumerate}
\item{Does \eqref{abcfusion} imply a factorization of TQFTs
\begin{equation}
\CT=\CT_1\boxtimes\CT_2~,
\end{equation}
with $a\in\CT_1$ and $b\in\CT_2$? As has been hinted at above, we will see in sections \ref{discrete} and \ref{cosets} that the answer is generally no.}
\item{Does \eqref{abcfusion} imply that $a$ belongs to one fusion subcategory and $b$ to another and that the intersection of these subcategories is trivial? In other words, do we have
\begin{equation}\label{fusfact}
a\in\CC_1\subset\CT~, \ \ \ b\in\CC_2\subset\CT~, \ \ \ \CC_1\cap\CC_2=1~?
\end{equation}
As we will see in section \ref{discrete}, the answer is generally no, even if we relax the requirement of trivial intersection. However, we will explicitly construct such examples (with non-modular $\CC_{1,2}\subset\CT$, where $\CT$ is prime) in the case of discrete gauge theories.}
\item{Does \eqref{abcfusion} imply that $a$ is in some subcategory $\CC\subset\CT$ that $b$ is not a member of? In other words, do we have
\begin{equation}\label{fuspartialfact}
a\in\CC\subset\CT~, \ \ \ b\notin\CC~?
\end{equation}
As we will see in section \ref{discrete}, the answer is generally no. However, we will argue that such constructions are quite easy to engineer in the context of discrete gauge theories, and we will explain when they arise. We will see that these constructions often have interesting interactions with symmetries.}
\item{Given $a$ and $b$ as in \eqref{abcfusion}, do they have trivial mutual braiding? In other words, do we have
\begin{equation}
{S_{ab}\over S_{0b}}={d_a}~,
\end{equation}
where $S$ is the modular $S$-matrix? This is true in the context of discrete gauge theories with a simple gauge group \cite{Buican:2020oyy}. However, non-trivial braiding does arise naturally in the context of the fusion of non-abelian electrically charged lines with non-abelian magnetically charged lines.
}
\item{Given $a$ and $b$ as in \eqref{abcfusion}, does $\CT$ have a non-trivial zero-form symmetry acting on either $a$ or $b$? Does the TQFT have a zero-form symmetry that acts more generally? We will see in section \ref{discrete} the answer to both these questions is no. However, in cases in which this is true, it seems to always be related to the existence of a certain fusion fixed point of one-form symmetry generators in a parent TQFT. Of the infinitely many examples of untwisted discrete gauge theories we study, only gauge theories based on the Mathieu groups $M_{23}$ and $M_{24}$ fail to have zero-form symmetries.}
\item{Given $a$ and $b$ as in \eqref{abcfusion}, does $\CT$ have a non-trivial symmetry of the modular data? As we will see in sections \ref{discrete} and \ref{cosets}, the answer seems to be yes. Clearly, it would be interesting to see if it is possible to define parents of such theories that generalize the relationship in (5). Note that the Mathieu gauge theories discussed in the previous point do have symmetries of their modular data (however, these symmetries do not lift to symmetries of the full TQFTs).}
\end{enumerate}

As we will see, many of these questions have simpler answers when studying discrete gauge theories. The reason is that powerful statements in these TQFTs can often be deduced from simple reasoning in the underlying theory of discrete groups. On the other hand, intuition one gains from taking products of representations in various continuous groups, like $SU(N)$, turns out to be somewhat misleading for our questions above.

The plan of this paper is as follows. In the next section, we start with discrete gauge theories and explain how intuition in the theory of finite groups leads us to various answers to the above questions. Along the way, we prove various theorems about discrete gauge theories and fusion rules of the form \eqref{abcfusion} and \eqref{aabar} generalizing our work in \cite{Buican:2020oyy}. Moreover, we discuss the role that subcategories and symmetries of discrete gauge theories play in such fusion rules. In the final part of the paper, we go to continuous groups and discuss coset theories. We tie the existence of fusion rules of type \eqref{abcfusion} and \eqref{aabar} to certain fixed points in the coset construction. We then finish with some conclusions and future directions.

\section{Discrete gauge theories and $a\times b=c$: from groups to TQFT}\label{discrete}
In this section we test the ideas presented in the introduction on discrete gauge theories \cite{Dijkgraaf:1989pz,Roche:1990hs}. These TQFTs are characterized by a choice of discrete gauge group, $G$, and a Dijkgraaf-Witten twist, $\omega\in H^3(G,U(1))$.\footnote{There are redundancies / dualities in this description: see \cite{naidu2007categorical}.} The basic degrees of freedom are anyonic line operators of the following three types
\begin{enumerate}
\item{{\it Wilson lines}, $\CW_{\pi}$, carrying electric charge labeled by a linear representation, $\pi$, of $G$ and trivial magnetic charge. This set of operators exists no matter the value of $\omega$.}
\item{{\it Magnetic flux lines}, $\mu_{[g]}$, carrying magnetic charge labeled by a conjugacy class, $[g]$, of a representative element, $g\in G$, but having trivial electric charge. In general, their existence depends on the choice of $\omega$.}
\item{{\it Dyonic lines}, $\CL_{([g],\pi^{\omega}_g)}$, carrying both magnetic flux and electric charge. In general, they carry a projective representation of $G$.}
\end{enumerate}
These theories have the advantage that we can prove many theorems about them. At the same time, they are very broad and so we can gain some insight into the physical and mathematical questions we are asking.\footnote{Discrete gauge theories are, however, necessarily non-chiral. We will consider chiral coset theories in section \ref{cosets}.}

As we will see in the subsequent subsections, the physics of the various operators listed above is qualitatively different. In order to take the shortest route to answering some of the questions posed in the introduction and in order to establish the existence of fusion rules of the form \eqref{abcfusion} in prime TQFTs, we will start with an analysis of Wilson lines. These objects form a closed fusion subcategory that is particular easy to analyze.\footnote{For other degrees of freedom, the story is more complicated. For example, in section \ref{subcat}, we will see that in non-abelian discrete gauge theories, full sets of magnetic fluxes do not form fusion subcategories.} As we explain, these are the most \lq\lq group theoretical" and least anyonic objects in a discrete gauge theory (in addition, as we see from the discussion of the above list of operators, they are the most robust). As a result, we can borrow various useful results from the study of finite groups.

In order to study the physics of other sectors of discrete gauge theories, we will find it convenient to introduce some additional machinery for discussing subcategories (in section \ref{subcat}) and symmetries (in section \ref{0formsec}). We also discuss quasi zero-form symmetries and their appearance in various discrete gauge theories of interest based on large Mathieu groups. Finally, we move beyond Wilson lines in section \ref{nonW} and discuss fusions of the form \eqref{abcfusion} involving non-abelian fluxes, magnetic fluxes, and dyons.

\subsection{Fusion rules and modular data}\label{fusionrules}
In this section, we briefly review how to construct a discrete gauge theory given a finite gauge group, $G$ (for a more succinct version of the review below, see \cite{Buican:2020oyy}). Although much of what we say is a consequence of \cite{Dijkgraaf:1989pz}, we will follow the perspective in \cite{Barkeshli:2014cna}. The main reason for this choice is that this latter perspective lends itself to generalizations to cases in which we want to gauge global symmetries of theories that already posses topological order (i.e., theories that already have non-trivial anyons).

To construct the $G$ discrete gauge theory we take a set of surface defects labeled by elements $g\in G$ (these are elements of a zero-form symmetry and therefore may be non-abelian). These objects comprise a $G$-symmetry protected topological phase ($G$-SPT). The fusion of the defects satisfies the corresponding group multiplication law, $g\times h=gh$. To complete the definition of the fusion category, we need to choose how to implement associativity. This boils down to choosing an element $\omega(g,h,k)\in H^3(G, U(1))$.

We are now ready to gauge the symmetry. To do this we pair conjugacy classes, $[g_i]$, with irreducible representations, $\pi_{g_i}^{\omega}$, of the corresponding centralizers, $N_{g_i}$ (for different choices of the conjugacy class representatives, the centralizers are isomorphic). These two objects combine to give anyons in the gauged theory
\begin{equation}
([g_i],\pi^{\omega}_{g_i})\in\CZ({\rm Vec}_G^{\omega})~,
\end{equation}
where $[g_i]$ and $\pi_{g_i}^{\omega}$ label the electric and magnetic charges of the discrete gauge theory, which we denote as $\CZ({\rm Vec}_G^{\omega})$.\footnote{Our choice of notation $\CZ({\rm Vec}_G^{\omega})$ for the discrete gauge theory reflects an alternative way to think about the theory: as the Drinfeld center of the fusion category of $G$ graded vector spaces with associator $\omega\in H^3(G,\mathbb{C}^{\times})$. The Drinfeld center of a spherical fusion category is an MTC.}

If $\omega$ is non-trivial in cohomology, we have a \lq\lq twisted" discrete gauge theory with Dijkgraaf Witten 3-cocycle $\omega$. This twisting typically leads to the $\pi^{\omega}_{g_i}$ being projective (as opposed to linear) representations of the centralizers. More precisely, to determine the projectivity of these representations, we should compute
\begin{equation}\label{H2f3}
\eta_g(h,k):={\omega(g,h,k) \omega(h,k,g)\over \omega(h,g,k)}\in H^2(N_g,U(1))~,\ \ \ h,k\in N_g~,
\end{equation}
as this is the phase that appears in $\pi_g^{\omega}(h)\pi_g^{\omega}(k)=\eta_g(h,k)\pi_g^{\omega}(hk)$. If $\eta_g$ is non-trivial in cohomology, then the representation $\pi_g^{\omega}$ is projective. Note that whenever $g=1$, the representations are linear. This means that, regardless of the twisting, the sector of Wilson lines is unchanged.\footnote{As a result, all of the statements that we arrive at for Wilson lines apply for both twisted and untwisted discrete gauge theories.} More generally, even if $g\ne 1$ and $\omega$ is non-trivial, we may still have linear representations.\footnote{More precisely, if $\eta_g$ is a non-trivial 2-coboundary, we will obtain projective representations that are in one-to-one correspondence with linear representations. We can remove these projective factors via a symmetry gauge transformation of the type described in \cite{Barkeshli:2014cna}. Note that while linear representations can be one-dimensional (e.g., if the centralizer is an abelian group), projective representations resulting from $\eta_g$ cohomologically non-trivial are necessarily higher dimensional.} As an example, we may consider $G=PSL(2,4)$ and the $\mathbb{Z}_3$ centralizer of the length twenty conjugacy class. In this case, we have $H^2(\mathbb{Z}_3,U(1))=\mathbb{Z}_1$, so the resulting $\eta_g$ (with $g$ in the length twenty conjugacy class) is cohomologically trivial no matter the choice of $\omega\in H^3(PSL(2,4),U(1))=\mathbb{Z}_6\times\mathbb{Z}_{10}$.

The most important things for us to focus on in what follows are the fusion coefficients appearing in
\begin{equation}
([g],\pi_g^{\omega})\times([h],\pi_h^{\omega})=\sum_{k,\pi_k^{\omega}}N_{([g],\pi^{\omega}_g),([h], \pi^{\omega}_h)}^{([k],\pi^{\omega}_k)}([k],\pi_k^{\omega})~.
\end{equation}
To arrive a description of such a process we must combine conjugacy classes and representations. In particular, we need to multiply elements in $[g]$ and $[h]$ and determine the corresponding conjugacy classes. At the same time, we must decompose the product of irreducible representations of the corresponding centralizers into irreducible representations of centralizers of $G$. A simple prescription for doing this is given in \cite{Barkeshli:2014cna}
\begin{eqnarray}\label{genfusion}
N_{([g],\pi^{\omega}_g),([h], \pi^{\omega}_h)}^{([k],\pi^{\omega}_k)}&=&\sum_{(t,s)\in N_g \backslash G /N_h} m(\pi^{\omega}_k |_{N_{{}^tg} \cap N_{{}^sh} \cap N_k}, {}^t\pi^{\omega}_{g}|_{N_{{}^tg} \cap N_{{}^sh} \cap N_k} \otimes {}^s\pi^{\omega}_{h}|_{N_{{}^tg} \cap N_{{}^sh} \cap N_k} \cr&&\hspace{8.7cm} \otimes ~ \pi^{\omega}_{({}^tg,{}^sh,k)})~,
\end{eqnarray}
where the sum is over the double coset, we define ${}^tg:=t^{-1}gt$, and ${}^t\pi^{\omega}_{g}|_{N_{{}^tg} \cap N_{{}^sh} \cap N_k} \otimes {}^s\pi^{\omega}_{h}|_{N_{{}^tg} \cap N_{{}^sh} \cap N_k} \otimes \pi^{\omega}_{({}^tg,{}^sh,k)}$ and $\pi^{\omega}_k |_{N_{{}^tg} \cap N_{{}^sh} \cap N_k}$ are restrictions of irreducible representations of $N_{{}^tg}$, $N_{{}^sh}$, and $N_k$  to the triple intersections of these normalizers. These restrictions are generally (though crucially for us below not always) reducible representations of $N_{{}^tg} \cap N_{{}^sh} \cap N_k$. The $m(a,b)$ function computes inner products of the representations $a$ and $b$ (we will fill in further details of this function as needed later in this section). Crucially, $a$ and $b$ must be the same type of representation (i.e., they should both be linear or else transform with the same set of projective weights) in order to be meaningfully compared.

We can determine the projectivity of the ${}^t\pi^{\omega}_{g}$, ${}^s\pi^{\omega}_{h}$, and $\pi^{\omega}_k$  representations by a computation in the relevant cohomology as in \eqref{H2f3}. The representation $\pi^{\omega}_{({}^tg,{}^sh,k)}$ is one dimensional (it is a representation of the action of symmetries on the one-dimensional $V_{{}^tg{}^sh}^{k}$ fusion space in the $G$-SPT) and ensures that the arguments entering $m(a,b)$ involve the same type of representations. Therefore, $\pi^{\omega}_{({}^tg,{}^sh,k)}$ satisfies
\begin{eqnarray}
\pi^{\omega}_{({}^tg,{}^sh,k)}(\ell)\pi^{\omega}_{({}^tg,{}^sh,k)}(m)&=&{\eta_{k}(\ell,m)\over\eta_{{}^tg}(\ell,m)\eta_{{}^sh}(\ell,m)}\cdot ~ \pi^{\omega}_{({}^tg,{}^sh,k)}(\ell m)~.
\end{eqnarray}

A more basic quantity of interest to us in what follows is the modular data of the discrete gauge theory. It is given by \cite{Hu:2012wx}
\begin{eqnarray}\label{STdg}
S_{([g],\pi_{g}^{\omega}),([h],\pi_{h}^{\omega})}&=&\frac{1}{|G|}\sum_{\substack{k \in [g],\ \ell \in [h],\\ k \ell=\ell k}} \chi^{k}_{\pi_g^{\omega}}(\ell)^{*} \chi^{\ell}_{\pi_{h}^{\omega}}(k)^{*}~, \cr \theta_{([g], \pi_{g}^{\omega})}&=&\frac{\chi_{\pi^{\omega}_{g}}(g)}{\chi_{\pi^{\omega}_{g}}(e)}~,
\end{eqnarray}
where we define $\chi^{h}_{\pi_g^{\omega}}(\ell)$ as follows 
\begin{equation}
\chi^{xgx^{-1}}_{\pi_g^{\omega}}(xhx^{-1}):= \frac{\eta_g(x^{-1},xhx^{-1})}{\eta_g(h,x^{-1})} \chi_{\pi^{\omega}_g}(h)~.
\end{equation}
 Here, $\theta$ is the topological spin, and $S$ is the modular $S$ matrix. It follows from these definitions that quantum dimensions are given by
\begin{equation}
d_{([g],\pi^{\omega}_g)}={S_{([g],\pi_g^{\omega})([1],1)}\over S_{([1],1)([1],1)}}=|[g]|\cdot|\pi^{\omega}_g|~,
\end{equation}
where $|[g]|$ is the size of the conjugacy class, and $|\pi^{\omega}_g|$ is the dimension of the representation. Non-abelian anyons have $d_{([g],\pi^{\omega})}>1$. As a consequence, they must satisfy
\begin{equation}
([g],\pi^{\omega}_g)\times([g^{-1}],(\pi^{\omega}_g)^*)=([1],1)+\cdots~,
\end{equation}
where the ellipses necessarily contain additional terms (otherwise we would have $d_{([g],\pi^{\omega})}=1$), $1$ is the trivial representation of $G$, and $(([g^{-1}],(\pi^{\omega}_g)^*)$ is the anyon conjugate to $([g],\pi^{\omega}_g)$.

We may write a dictionary between the non-abelian Wilson lines, flux lines, and dyons discussed in the previous sections and the objects discussed in this section as follows
\begin{eqnarray}\label{dictionary}
\CW_{\pi_1}&\leftrightarrow&([1],\pi_1)~, \ \ \ |\pi_1|>1~,\cr
\mu_{[g]}&\leftrightarrow&([g],1_g^{\epsilon})~, \ \ \ |[g]|>1~,\cr
\CL_{([h],\pi_h^{\omega})}&\leftrightarrow&([h],\pi_h^{\omega})~,\ \ \ |[h]|\cdot|\pi_h^{\omega}|>1~.
\end{eqnarray}
We have dropped the $\omega$ superscript from $\pi_1$ in order to emphasize, as discussed above, that Wilson lines always transform under linear representations of $G$. We include an $\epsilon$ superscript on the trivial representation of the flux line because these objects only exist when the relevant $\eta_g$ in \eqref{H2f3} is trivial in cohomology. This triviality means that $\eta_g(h,k)$ can be expressed in terms of a one co-chain as follows: $\eta_g(h,k)= \frac{\epsilon_g(h)\epsilon_g(k)}{\epsilon_g(h\cdot k)}$.\footnote{Note that $1_g^{\epsilon}$ is the irreducible projective representation of $N_g$ whose character is proportional to the trivial representation of $N_g$.}

\subsection{Non-abelian Wilson lines and $a\times b=c$}\label{WilsonSec}
We would like to recast the problem of constructing discrete gauge theories with fusion rules \eqref{abcfusion} and \eqref{aabar} in terms of the closely related problem of finding irreducible products of irreducible finite group representations. To make this connection as direct as possible, it is useful to focus on Wilson lines of the discrete gauge theories we are studying. Indeed, by specializing \eqref{genfusion} to Wilson lines, we find
\begin{eqnarray}\label{Wfusion}
N_{(1,\pi),(1, \pi')}^{(1,\pi'')}&=&m(\pi'' ,\pi\otimes\pi')={1\over|G|}\sum_{g\in G}\chi_{\pi''}(g)\chi_{\pi}^*(g)\chi_{\pi'}^*(g)=\langle \chi_{\pi''},\chi_{\pi}\chi_{\pi'}\rangle~,
\end{eqnarray}
where $\langle\cdot,\cdot\rangle$ is the standard inner product on characters. Therefore, the Wilson lines form a closed fusion subcategory of the discrete gauge theory, $\CC_{\CW}$. Moreover, the fusion rules of the Wilson lines are those of the representation semiring of the gauge group.\footnote{In fact, we have $\CC_{\CW}\simeq\Rep(G)$, where $\Rep(G)$ is the category of finite dimensional representations of $G$ over $\mathbb{C}$.} Note that $\CC_{\CW}$ is, in some sense, the \lq\lq least anyonic" part of the theory: it is easy to check from \eqref{STdg} that the Wilson lines are bosonic, so $\theta_{\CW_i}=1$, and that the braiding of Wilson lines amongst themselves is trivial,\footnote{The Wilson lines braid non-trivially with other anyons in the theory (more formally: the Wilson line subcategory is Lagrangian and so the M\"uger center of $\CC_{\CW}$ is $\CC_{\CW}$ itself).} so $S_{\CW_1\CW_2}=d_{\CW_1}d_{\CW_2}/\CD$ (here $\CD=\sqrt{\sum_{i=1}^Nd_i^2}$, and the sum is over all the anyons).\footnote{In fact, \cite{deligne2002categories} guarantees that any such subcategory is equivalent to $\Rep(H)$ for some group $H$.} To summarize, we see that if we can find representations of some group, $G$, satisfying
\begin{equation}\label{gpabc}
\chi_{\pi}\cdot\chi_{\pi'}=\chi_{\pi''}~, \ \ \ |\pi|,\ |\pi'|,\ |\pi''|>1~,
\end{equation}
where $\pi$, $\pi'$, and $\pi''$ are irreducible, then, in the corresponding $G$ discrete gauge theory, we will have non-abelian Wilson lines satisfying
\begin{equation}\label{abcW}
\CW_{\pi}\times\CW_{\pi'}=\CW_{\pi''}~.
\end{equation}

Since, by Cayley's theorem, every finite group is isomorphic to a subgroup of the symmetric group, $S_N$, (for some $N$) it is natural to start our discussion with $S_N$. In particular, to check whether $\pi''$ is irreducible, we want to perform the group theory analog of the $F$ transformation discussed in the introduction (see figure \ref{ftrans1})
\begin{equation}\label{GroupF}
\langle \chi_{\pi}\cdot\chi_{\pi'},\chi_{\pi}\cdot\chi_{\pi'}\rangle=\langle \chi_{\pi}^2,\chi_{\pi'}^2\rangle~,
\end{equation}
where we have used the fact that $S_N$ is ambivalent ($g$ and $g^{-1}$ are in the same conjugacy class for all $g\in S_N$) so that the characters are real. A theorem of Zisser \cite{zisser1993irreducible} shows that $\chi_{[N-2,2]}\in\chi_{\alpha}^2$, where $[N-2,2]$ is a partition of $N$ labeling the corresponding representation of $S_N$, and $\alpha$ is any irreducible representation of dimension larger than one, $|\alpha|>1$. Moreover, since $S_N$ is ambivalent, this means that $\chi_{[N]}\in\chi_{\alpha}^2$, where $\chi_{[N]}$ is the trivial representation of $S_N$. As a result, we see that the analog of \eqref{aabar} yields
\begin{equation}\label{charaabar}
\chi_{\pi}\cdot\chi_{\pi}=\chi_{[N]}+\chi_{[N-2,2]}+\cdots~,\ \ \chi_{\pi'}\cdot\chi_{\pi'}=\chi_{[N]}+\chi_{[N-2,2]}+\cdots\ \Rightarrow\ \langle \chi_{\pi}\cdot\chi_{\pi'},\chi_{\pi}\cdot\chi_{\pi'}\rangle>1~,
\end{equation} 
and so products of non-abelian representations of $S_N$ are never irreducible. Therefore, we cannot have \eqref{abcW} in $S_N$ discrete gauge theory.

\subsubsection{Discrete gauge theories of finite simple groups}\label{simpleW}
Since we have $A_N\triangleleft S_N$ (i.e., the alternating group, $A_N$, is a normal subgroup of $S_N$), it is natural to consider $A_N$ discrete gauge theories as the next possibility for realizing \eqref{gpabc} \cite{zisser1993irreducible} and hence \eqref{abcW}. Moreover, since $A_N$ is simple, only pure Wilson lines can be involved in fusions of the form \eqref{abcfusion} \cite{Buican:2020oyy}, and the $A_N$ discrete gauge theories are guaranteed to be prime \cite{naidu2009fusion} (we will return to the question of primality in greater generality in section \ref{subcat}). Therefore, finding an example of \eqref{abcW} in $A_N$ discrete gauge theories is sufficient to answer question (1) from the introduction in the negative.

To understand if going to $A_N$ is a fruitful direction, we note that there are two types of characters that arise in going from $S_N$ to $A_N$:
\begin{enumerate}[label=(\Alph*)]
\item{Characters that are restrictions of $S_N$ characters satisfying $\chi_{\lambda}\ne\chi_{[1^N]}\cdot\chi_{\lambda}$, where $\chi_{[1^N]}$ corresponds to the sign representation of $S_N$. Let us call these \lq\lq type A" characters: $\tilde\chi_{\lambda}:=\chi_{\lambda}|_{A_N}$.}
\item{Characters that descend from $S_N$ characters satisfying $\chi_{\rho}=\chi_{[1^N]}\cdot\chi_{\rho}$. As representations of $A_N$, they split into two representations of the same dimension, $\lambda_{\pm}$. Let us call these \lq\lq type B" characters: $\chi_{\rho}^{(B)}=\chi_{\rho_{+}}+\chi_{\rho_{-}}=\chi_{\rho}|_{A_N}$.}
\end{enumerate}

In going from $S_N$ to $A_N$, we perform a group-theoretical version of gauging the \lq\lq one-form symmetry" generated by $\chi_{[1^N]}$: we identify characters related by multiplication with $\chi_{[1^N]}$, and we split characters that are invariant under multiplication with $\chi_{[1^N]}$. Clearly, products of type A characters cannot be irreducible since they will always contain $\chi_{[N]}^{(A)}$ and $\chi_{[N-2,2]}^{(A)}$ after performing the F-transformation and computing \eqref{charaabar}.\footnote{In this discussion, we have implicitly assumed that $N\ne 4$ (although, for $N=3$, we should take $[N-2,2]\to[2,1]$ to conform to usual conventions). For $N=4$, we have $\chi_{[N-2,2]}^{(A)}\to\chi_{[N-2,2]}^{(B)}=\chi_{[N-2,2],+}+\chi_{[N-2,2],-}$.}

A little more work in \cite{zisser1993irreducible} shows that we can obtain \eqref{gpabc} for $A_N$ if and only if $N=k^2\ge9$ by taking the product of the following type A and type B representations
\begin{equation}\label{ABirred}
\tilde\chi_{[N-1,1]}\cdot\chi_{[k^k]_{\pm}}=\tilde\chi_{[k^{k-1},k-1,1]}~.
\end{equation}
Moreover, the $\mathbb{Z}_2$ outer automorphism of $A_N$ acts on the type B characters as 
\begin{equation}\label{outactgp}
g\left(\chi_{[k^k]_{\pm}}\right)=\chi_{[k^k]_{\mp}}~,\ \ 1\ne g\in{\rm Out}(A_N)\simeq\mathbb{Z}_2~. 
\end{equation}
Therefore, at the level of the non-abelian Wilson lines in the corresponding $A_N$ discrete gauge theory, we learn that
\begin{equation}\label{Anabc}
\CW_{[N-1,1]}\times \CW_{[k^k]_{\pm}}=\CW_{[k^{k-1},k-1,1]}~.
\end{equation}
Finally, ${\rm Out}(A_N)$ lifts to a full zero-form symmetry of the discrete gauge theory \cite{nikshych2014categorical}, since, according to corollary 7.8 of \cite{nikshych2014categorical}
\begin{equation}\label{symAn}
{\rm Aut}^{\rm br}(\CZ({\rm Vec}_{A_N}))\simeq H^2(A_N, U(1))\rtimes{\rm Out}(A_N)\simeq \mathbb{Z}_2\times\mathbb{Z}_2~,
\end{equation} 
where the group on the left hand side is the group of braided tensor auto-equivalences of the MTC underlying the discrete gauge theory, $\CZ({\rm Vec}_{A_N})$. As a result, we learn that the symmetries of the discrete gauge theory exchange the $\CW_{[k^k]_{\pm}}$ lines
\begin{equation}\label{outactW}
g\left(\CW_{[k^k]_{\pm}}\right)=\CW_{[k^k]_{\mp}}~,\ \ 1\ne g\in{\rm Out}(A_N)\triangleleft{\rm Aut}^{\rm br}(\CZ({\rm Vec}_{A_N})~. 
\end{equation}

In other words, we have found that, in an infinite number of prime theories, fusion rules of the type \eqref{abcfusion} are generated in pairs related by symmetries of the discrete gauge theory. This discussion shows that TQFTs with fusions of the form \eqref{abcfusion} need not factorize and so the answer to question (1) in the introduction is \lq\lq no."

Let us now drive home the importance of symmetries in arriving at \eqref{Anabc} and, at the same time, gain insight that will be useful later. To that end, let us consider gauging the $\mathbb{Z}_2$ outer automorphism symmetry of the $A_N$ discrete gauge theory. Note that this gauging is allowed since the \lq\lq defectification" obstruction described physically in \cite{Barkeshli:2014cna} is trivial here: $H^4(\mathbb{Z}_2,U(1))=\mathbb{Z}_1$. Moreover, since $A_N$ is simple, the discrete gauge theory has no non-trivial abelian anyons (i.e., $\CA=\CW_{[N]}$) and so $H^3(\mathbb{Z}_2,\CA)=\mathbb{Z}_1$. Therefore, \eqref{symAn} is a genuine zero-form symmetry group (as opposed to being a 2-group).

More abstractly, let us consider a generalization of the fusion rules in \eqref{genfusion} to the case of gauging a zero form group, $H$, of a more general G-crossed braided theory, $\CT_{G^{\times}}$ (as worked out in \cite{Barkeshli:2014cna})
\begin{eqnarray}\label{genfusionA}
N_{([a],\pi_a),([b], \pi_b)}^{([c],\pi_c)}=\sum_{(t,s)\in N_a \backslash H /N_b} m(\pi_c |_{N_{{}^ta} \cap N_{{}^sb} \cap N_c},{}^t\pi_{a}|_{N_{{}^ta} \cap N_{{}^sb} \cap N_c} \otimes {}^s\pi_{b}|_{N_{{}^ta} \cap N_{{}^sb} \cap N_c} \otimes \pi^{\omega}_{({}^ta,{}^sb,c)})~,\ \ \
\end{eqnarray}
where $a,b,c\in\CT_{G^{\times}}$, $[a]:=\left\{h(a),\ \forall h\in H \right\}$, $N_a:=\left\{h\in H|h(a)=a\right\}$, and $\pi_a$ is a representation of $N_a$.

In our case at hand, $\CT_{G^{\times}}=\CZ({\rm Vec}_{A_N})_{H^{\times}}$ is the $A_N$ discrete gauge theory extended by surface defects implementing the $H=\mathbb{Z}_2$ global symmetry. Moreover, $a=\CW_{[N-1,1]}$, $b=\CW_{[k^k]_{\pm}}$, $N_a=\mathbb{Z}_2$, and $N_b=\mathbb{Z}_1$. As a result, $t=s=1$, the summation in \eqref{genfusionA} is trivial, the various representations are all restricted to the trivial subgroup, and $\pi^{\omega}_{a,b,c}=1$ (this latter statement follows from the fact that the action of $H$ on the $V_{ab}^c$ fusion space via $U_{1}(a,b,c)$ is trivial). In particular, we have
\begin{eqnarray}\label{genfusionA2}
N_{([\CW_{[N-1,1]}],\pm),([\CW_{[k^k]_{\pm}}], +)}^{([\CW_{[k^{k-1},k-1,1]}],\pm)}=m(\pm |_{\mathbb{Z}_1},\pm|_{\mathbb{Z}_1} \otimes +|_{\mathbb{Z}_1})=m(1,1)=1~,
\end{eqnarray}
where $\pm$ denote the two representations of $\mathbb{Z}_2$. Therefore, we learn that when we gauge the outer automorphism group of $A_N$, we have
\begin{equation}\label{Wirredlift}
([\CW_{[N-1,1]}],\pm)\times([\CW_{[k^k]_{\pm}}], +)= ([\CW_{[k^{k-1},k-1,1]}],+)+([\CW_{[k^{k-1},k-1,1]}],-)~,
\end{equation}
which is the TQFT version of the lift of \eqref{ABirred} to $S_N$. This is what we expect, since we can always fix our choice of parameters so that gauging $\mathbb{Z}_2$ yields \cite{cui2016gauging}
\begin{equation}
\CZ({\rm Vec}_{A_N})_{\mathbb{Z}_2^{\times}}\ \overset{\rm gauge}{\longrightarrow}\ \CZ({\rm Vec}_{A_N\rtimes\mathbb{Z}_2})=\CZ({\rm Vec}_{S_N})~,
\end{equation}
where we have used the fact that $S_N\simeq A_N\rtimes\mathbb{Z}_2$.

Finally, from the general rules above, it is not hard to check that the trivial Wilson line in the $A_N$ theory lifts to a $\mathbb{Z}_2$ one-form symmetry in the $S_N$ gauge theory. The resulting non-trivial one-form symmetry generator acts as
\begin{eqnarray}\label{oneformAng}
([\CW_{[N]}],-)\times([\CW_{[N-1,1]}],\pm)&=&([\CW_{[N-1,1]}],\mp)~,\cr ([\CW_{[N]}],-)\times ([\CW_{[k^k]_{\pm}}], +)&=&([\CW_{[k^k]_{\pm}}], +)~,\cr ([\CW_{[N]}],-)\times([\CW_{[k^{k-1},k-1,1]}],\pm)&=&([\CW_{[k^{k-1},k-1,1]}],\mp)~,
\end{eqnarray}
where $([\CW_{[N]}],-)=\CW_{[1^N]}$.

To summarize, we learn that, in order to generate the fusion rule \eqref{Anabc}, we can gauge a $\mathbb{Z}_2$ one-form symmetry in the $S_N$ (with $N=k^2\ge9$) discrete gauge theory with fusion rules \eqref{Wirredlift} and \eqref{oneformAng}. Crucially, we need a fixed point of the one-form symmetry (as in the second line in \eqref{oneformAng}) in order to generate the fusion rule of the form \eqref{Anabc} in the $A_N$ discrete gauge theory. We will return to the existence of fixed points of various kinds repeatedly throughout this paper.

One may wonder if zero-form gaugings always resolve fusion rules of the form $a\times b=c$ into fusion rules with multiple outcomes. Taking $G=O(5,3)$, one can see the answer is no.\footnote{This is the group $O(5)$ over the field $\mathbb{F}_3$. It has order 25920 and is the smallest simple group whose discrete gauge theory has a fusion of non-abelian Wilson lines with a unique outcome.} Indeed, in this theory, one can check that we have the following analogs of \eqref{Anabc}
\begin{equation}\label{O53Wabc}
\CW_{5_i}\times\CW_{6}=\CW_{30_i}~,\ \ \ i=1,2~,
\end{equation}
where $5_i$ are the two five-dimensional representations of $O(5,3)$, $6$ is the unique six-dimensional representation, and $30_i$ are the two complex thirty-dimensional representations (there is also a third, real, thirty-dimensional representation that does not appear in \eqref{O53Wabc}). As in the previous case, ${\rm Out}(O(5,3))=\mathbb{Z}_2$ and it acts non-trivially on the Wilson lines involved in the fusion above. In particular, we have 
\be
\label{O53Symaction}
\CW_{5_1} \leftrightarrow \CW_{5_2} ~ 
\text{ and } ~ \CW_{30_1} \leftrightarrow \CW_{30_2}
\ee
 under the action of the non-trivial element in $\text{Out}(O(5,3))$. This symmetry lifts to a symmetry of the discrete gauge theory that we can gauge. Doing so, we can choose parameters such that
\begin{equation}
\CZ({\rm Vec}_{O(5,3)})_{\mathbb{Z}_2^{\times}}\ \overset{\rm gauge}{\longrightarrow}\ \CZ({\rm Vec}_{O(5,3)\rtimes\mathbb{Z}_2})~.
\end{equation}
We may again apply \eqref{genfusionA} to find
\begin{eqnarray}\label{genfusionO532}
N_{([\CW_{5_i}],+),(\CW_{6}, \pm)}^{([\CW_{30_i}],+)}=m(+|_{\mathbb{Z}_1},+|_{\mathbb{Z}_1} \otimes \pm|_{\mathbb{Z}_1})=m(1,1)=1~,
\end{eqnarray}
and conclude
\begin{equation}\label{O53WabcE}
([\CW_{5_i}],+)\times (\CW_{6}, \pm)=([\CW_{30_i}],+)~.
\end{equation}
Such a situation arises whenever $N_c=\mathbb{Z}_1=N_a\cap N_b$. This equality is special since, more generally, we have $N_a\cap N_b\subseteq N_c$.

Before moving on to discuss other phenomena, let us note that the above discrete gauge theories based on simple groups also provide answers to questions (2) and (3) from the introduction. Indeed, as we will see in greater detail in section \ref{subcat}, a discrete gauge theory with a simple gauge group has no non-trivial proper fusion subcategories except the subcategory of Wilson lines. Therefore, our above examples are enough to answer questions (2) and (3) generally in the negative (although we will see interesting examples of some of these ideas below).

\subsubsection{Non-simple groups and unfaithful higher-dimensional representations}\label{unfaithful}
Let us now consider discrete gauge theories with unfaithful higher-dimensional (i.e., non-abelian) representations. The corresponding gauge groups are necessarily non-simple because the kernel of a non-trivial unfaithful representation is a non-trivial proper normal subgroup. As we will explain at a more pedestrian level below (and in a somewhat more sophisticated way in section \ref{subcat}), these examples illustrate the appearance of non-trivial fusion subcategories in the Wilson line sector. As a result, they demonstrate some of the ideas---described in the introduction---behind constraints from subcategory structure leading to fusion rules of the type \eqref{abcfusion}. In particular, these theories provide examples where ideas in questions (2) and (3) of the introduction are realized.

To that end, let us consider some unfaithful higher-dimensional irreducible representation of the gauge group, $\pi\in{\rm Irrep}(G)$. Since $\pi$ is unfaithful, it has a non-trivial kernel, $\Ker(\pi)\triangleleft G$. Let us also define the set of characters whose kernel includes $\Ker(\pi)$ as follows
\begin{equation}\label{Kpi}
K_{\pi}=\left\{\chi_{\rho}:\ \chi_{\rho}|_{{\rm Ker}(\pi)}=\deg\chi_{\rho}\right\}~,
\end{equation}
where $\deg\chi_{\rho}=|\rho|$ is the degree of the character. Now, consider $\chi_{\lambda},\chi_{\lambda'}\in K_{\pi}$. We claim $\chi_{\lambda}\cdot\chi_{\lambda'}\in K_{\pi}$. To see this, let us study
\begin{equation}
\chi_{\lambda}|_{{\rm Ker}(\pi)}\cdot\chi_{\lambda'}|_{{\rm Ker}(\pi)}=\deg\chi_{\lambda}\cdot\deg\chi_{\lambda'}=\sum_{\lambda''}\chi_{\lambda''}|_{{\rm Ker}(\pi)}\le\sum_{\lambda''}\left|\chi_{\lambda''}|_{{\rm Ker}(\pi)}\right|~.
\end{equation}
Evaluating this expression on the identity element shows that $\deg\chi_{\lambda}\cdot\deg\chi_{\lambda'}=\sum_{\lambda''}\deg\chi_{\lambda''}$. Therefore, we have $\chi_{\lambda''}|_{{\rm Ker}(\pi)}=\deg\chi_{\lambda''}$, and $\lambda''\in K_{\pi}$. In particular, we see that 
\begin{equation}
\chi_{\lambda}\cdot\chi_{\lambda'}=\sum_{\lambda''\in K_{\pi}}\chi_{\lambda''}~.
\end{equation}
As a result, the Wilson lines with charges in $K_{\pi}$ form a closed fusion subcategory\footnote{Such Wilson lines recently played an interesting role in \cite{Rudelius:2020orz}. Indeed, when one adds non-topological matter charged under these representations, the corresponding Wilson lines can end on a point. Magnetic flux lines or dyons with flux supported in ${\rm Ker}(\pi)$ remain topological while  lines carrying other fluxes do not.}
\begin{equation}\label{CKpi}
\CW_{{\lambda}}\times\CW_{{\lambda'}}=\sum_{\lambda''\in K_{\pi}}\CW_{{\lambda''}}\in\CC_{K_{\pi}}\simeq{\rm Rep}(G/\Ker(\pi))~.
\end{equation}

If we now consider the fusion of $\CW_{\pi}\in\CC_{K_{\pi}}$ with a non-abelian Wilson line $\CW_{\gamma}\not\in\CC_{K_{\pi}}$, we see that the subcategory structure makes it more likely to find a unique outcome. Indeed, $\CW_{\pi}\times\CW_{\bar\pi}\in\CC_{K_{\pi}}$ whereas $\CW_{\gamma}\times\CW_{\bar\gamma}$ will typically include lines not in $\CC_{K_{\pi}}$. 

In fact, we can go further if we take $\gamma|_{\Ker(\pi)}$ to be an irreducible representation of $\Ker(\pi)$. Since we are assuming that $\gamma$ is a higher-dimensional representation, irreducibility of $\gamma|_{\Ker(\pi)}$ implies that $\Ker(\pi)$ is a non-abelian group. Invoking Gallagher's theorem (e.g., see corollary 6.17 of \cite{isaacs1994character}), we see that, for $\gamma,\pi\in{\rm Irrep}(G)$, $\gamma\otimes\pi$ is an irreducible representation if the restriction $\gamma|_{\Ker(\pi)}$ is irreducible. Then, we are guaranteed to have the following fusion rule of non-abelian Wilson lines
\be\label{ufabc}
\CW_{\pi}\times\CW_{\gamma}=\CW_{\pi\gamma}~.
\ee

To understand this statement, let us first prove that $\gamma\not\in K_{\pi}$. Suppose this were not the case: then we arrive at a contradiction since $|\gamma|>1$ would imply that $\gamma|_{\Ker(\pi)}$ is reducible. As a result, $\CW_{\gamma}\not\in\CC_{K_{\pi}}$. Let us now consider the product
\be
\chi_{\gamma} \cdot \chi_{\overline{\gamma}}= \chi_1 + \sum_i \chi_{\alpha_i}~,
\ee
where $\alpha_i$ are irreps of $G$. Then we have
\be\label{irredprod}
(\chi_{\gamma} \cdot\chi_{\overline{\gamma}})|_{\Ker(\pi)}= \chi_1|_{\Ker(\pi)} + \sum_i \chi_{\alpha_i}|_{\Ker(\pi)}~.
\ee
Here, $\chi_1|_{\text{ker}(\pi)}$ corresponds to the trivial irreducible representation of $\Ker(\pi)$, $\chi_{\alpha_i}|_{\Ker(\pi)}$ corresponds to an, in general, reducible representation of $\Ker(\pi)$. Suppose that $\alpha_i|_{\Ker(\pi)}$ contains the trivial irreducible representation of $\Ker(\pi)$ for some $i$, then we will have at least two copies of the trivial character of $\Ker(\pi)$ on the right hand side of \eqref{irredprod}. However, we know that $(\gamma \otimes \overline{\gamma})|_{\Ker(\pi)}=\gamma|_{\Ker(\pi)} \otimes \overline{\gamma}|_{\Ker(\pi)}$. Therefore, we cannot have more than one copy of the trivial character in the decomposition \eqref{irredprod}. Hence, $\alpha_i|_{\Ker(\pi)}$ cannot contain the trivial representation for any $i$. It follows that $\alpha_i|_{\Ker(\pi)}(h)$ is non-trivial for at least some $h \in \Ker(\pi)$. Therefore, it is clear that $\Ker(\pi)$ cannot be in the kernel of the representations $\alpha_i$ for any $i$. This shows that 
\begin{equation}\label{notCkpi}
\CW_{\alpha_i}\in\CW_{\gamma}\times\CW_{\bar\gamma}\ \Rightarrow\ \CW_{\alpha_i}\not\in\CC_{K_{\pi}}~.
\end{equation}
As a result, the subcategory structure guarantees \eqref{ufabc}.

To better understand the above general discussion (as well as the continuing role of symmetries), let us consider some examples. Note that these results give explicit realizations of the idea in question (3) in the introduction. The simplest discrete gauge theories realizing the above discussion are based on gauge groups of order forty-eight. Interestingly, the existence of subcategory structure in the Wilson line sector, $\CC_{\CW}\simeq{\rm Rep}(G)$, explains the large ratio of orders, $\Delta_{\rm gap}$, between these groups and the smallest simple group, $O(5,3)$, with unique non-abelian fusion outcomes
\begin{equation}\label{groupgap}
\Delta_{\rm gap}={25920\over48}=540\gg1~.
\end{equation}

In this section, we will discuss the examples of the binary octahedral group ($BOG$) and the very closely related general linear group of $2\times 2$ matrices with elements in the finite field $\mathbb{F}_3$, $GL(2,3)$. In appendix \ref{appA} we will consider the remaining cases at order forty-eight.

Let us begin with $BOG$. In this case, we have that $2_1$ is an unfaithful (real) two-dimensional representation and that the restrictions of the other (real and faithful) two-dimensional irreducible representations to $\Ker(2_1)=Q_8\triangleleft BOG$, $2_{2,3}|_{\Ker(2_1)}$, are irreducible. As expected from the general discussion above we have the following Wilson line fusions
\begin{equation}\label{BOGWabc}
\CW_{2_1}\times\CW_{2_2}=\CW_{2_1}\times\CW_{2_3}=\CW_4~.
\end{equation}
Similarly to the simple discrete gauge theories discussed in the previous subsection, $BOG$'s $\mathbb{Z}_2$ outer automorphism again lifts to a non-trivial symmetry of the TQFT, and the non-trivial element $g\ne1$ acts as follows: $g(\CW_{2_2})=\CW_{2_3}$.

Let us note that in this case, the role of symmetries is even more pronounced. Indeed, one can check that
\begin{eqnarray}\label{aabarBOG}
\CW_{2_1}\times\CW_{2_1}&=&\CW_1+\CW_{1_2}+\CW_{2}\in \CC_{K_{2_1}}\simeq\Rep(BOG/Q_8)\simeq \Rep(S_3)~,\cr
\CW_{2_{2}}\times\CW_{2_{2}}&=&\CW_{2_{3}}\times\CW_{2_{3}}=\CW_1+\CW_{3_2}~,
\end{eqnarray}
where $1_2$ is a non-trivial one-dimensional irreducible representation, and $3_2$ is a real three-dimensional irreducible representation.\footnote{Note that since $2_{2,3}$ are faithful representations, a result of Burnside \cite{Burnside} generalized to Wilson lines shows that there exist $n_{1,2}\in\mathbb{N}$ such that $W_{2_{2,3}}^{\times n_1}\supset\CW_{1_2}$ and $W_{2_{2,3}}^{\times n_2}\supset\CW_{2_1}$. Our discussion implies $n_{1,2}>2$.} This latter representation satisfies $\chi_{1_2}\cdot\chi_{3_2}=\chi_{3_1}$ (and similarly $\chi_{1_2}\cdot\chi_{3_1}=\chi_{3_2}$). Therefore, we see that $\CW_{1_2}$ generates a non-trivial one-form symmetry in the $BOG$ discrete gauge theory and that $\CW_{3_{1,2}}$ and $\CW_{2_{2,3}}$ form doublets under fusion with this generator while $\CW_{2_1}$ is fixed
\begin{equation}\label{1formfusion}
\CW_{1_2}\times\CW_{3_2}=\CW_{3_1}~,\ \ \ \CW_{1_2}\times\CW_{2_2}=\CW_{2_3}~, \ \ \ \CW_{1_2}\times\CW_{2_1}=\CW_{2_1}~.
\end{equation}
This non-trivial orbit structure then implies that $\CW_{3_2}\not\in\CW_{2_1}\times\CW_{2_1}$ on symmetry grounds alone. Hence, in this example, both the subcategory structure and the symmetries guarantee the fusion rules \eqref{BOGWabc}.

Before finishing this example, we should check that $\CZ({\rm Vec}_{BOG})$ is indeed prime. After we discuss more formal aspects of subcategory structure in section \ref{subcat}, we will have more tools to use when answering this type of question. For now, let us prove that the Wilson lines must all lie in the same TQFT factor.\footnote{The same pedestrian arguments used below can be extended to the full set of lines in the theory to prove that $\CZ({\rm Vec}_{BOG})$ is prime.} To that end, write down the Wilson lines of the $BOG$ discrete gauge theory
\begin{eqnarray}\label{BOGW}
&&\CW_1~, \ \ \CW_{1_2}~, \ \ \CW_{2_1}~, \ \ \CW_{2_2}~, \ \ \CW_{2_3}=\CW_{2_2}\times\CW_{1_2}~, \ \ \CW_{3_1}~, \ \ \CW_{3_2}=\CW_{3_1}\times\CW_{1_2}~, \cr &&\CW_4=\CW_{2_1}\times\CW_{2_2}=\CW_{2_1}\times\CW_{2_3}~.
\end{eqnarray}
We can consider two cases: (1) $\CW_{3_1}$ is in the same TQFT factor as $\CW_{1_2}$ (call this factor $\CT_0$) or (2) $\CW_{3_1}$ is not in the same TQFT factor as $\CW_{1_2}$. 

Let us consider case (1) first. From the fusion equation involving $\CW_{3_2}$, we immediately see that $\CW_{3_2}$ is also in $\CT_0$. Note that $\CW_{2_1}$ cannot be written as the fusion product of two other Wilson lines. Since there is no Wilson line of quantum dimension six, we also have $\CW_{2_1}\in\CT_0$. Now, we must clearly have that either $\CW_{2_{2,3}}\in\CT_0$ or $\CW_{2_{2,3}}\not\in\CT_0$. However, in the latter case we will again have a Wilson line of quantum dimension six. Therefore, we have that $\CW_{2_{2,3}}\in\CT_0$. Therefore, by the $\CW_4$ fusion rule in \eqref{BOGW}, all Wilson lines are in the same TQFT factor.

Let us now consider case (2). Let $\CW_{3_1}\in\CT_0$ and $\CW_{1_2}\in\CT_1$ with $\CZ({\rm Vec}_{BOG})=\CT_0\boxtimes\CT_1$. As in case (1), $\CW_{2_1}$ cannot be written as the fusion product of two other Wilson lines, and, since there is no Wilson line of quantum dimension six, we have $\CW_2\in\CT_0$. However, this leads to a contradiction because then $\CW_2\times\CW_1'\ne\CW_2$. As a result, we conclude that all Wilson lines must lie in the same factor of $\CZ({\rm Vec}_{BOG})$.

Let us conclude with a brief discussion of the $GL(2,3)$ discrete gauge theory. This gauge group is quite similar to $BOG$. For the purposes of the above discussion, the only difference is that $2_{2,3}$ become complex conjugate two-dimensional irreducible representations (otherwise, the remaining representations and remaining parts of the character tables are the same). Therefore, \eqref{BOGWabc} and \eqref{1formfusion} apply to $\CZ({\rm Vec}_{GL(2,3)})$ as well (by identifying these Wilson lines with their relatives in $\CZ({\rm Vec}_{GL(2,3))}$). The only change is that in the second line of \eqref{aabarBOG}, we should take $\CW_{2_{2,3}}\times\CW_{2_{2,3}}\to\CW_{2_2}\times\CW_{2_3}$. In particular, the roles of subcategory structure (again $\Rep(S_3)\subset \Rep(GL(2,3))$) as well as outer automorphisms and one-form symmetries is the same in both the $BOG$ and the $GL(2,3)$ discrete gauge theories. 

Note that Gallagher's theorem does not exhaust all cases where representations with non-trivial kernel have irreducible products. Another interesting case is given by Gajendragadkar's theorem \cite{gajendragadkar1979characteristic,navarro}. If we have a group $G$ which is both $\pi$-separable as well as $\Sigma$-separable, for two disjoint set of primes $\pi$ and $\Sigma$, then this theorem guarantees that the product of a $\pi$-special character with a $\Sigma$-special character is irreducible. A character $\chi$ is known as $\pi$-special if $\chi(1)$ is a product of powers of primes in $\pi$ (a $\pi$ number) and if, for every subnormal subgroup $N$ of $G$, any  irreducible constituent $\theta$ of $\chi|_{N}$ is such that $o(\theta)$\footnote{$o(\theta)$ is the order of the determinental character $\text{det}(\chi)$ in the group of linear characters.} is a $\pi$-number. Hence, the fusion of Wilson lines corresponding to such characters have a unique outcome. Note that, in this case, one of the characters involved in the fusion is not required to be irreducible in the kernel of the other (unlike in Gallagher's theorem).   

\subsubsection{Some general lessons and theorems}\label{genlessons}
Let us conclude this section with a recapitulation of some of the main points above as well as some general theorems that amplify our discussion:

\begin{itemize}
\item{In all of the infinitely many examples we studied so far, symmetries played an important role. For example, zero-form symmetries had a non-trivial action on Wilson lines involved in the fusion rules of interest in the $A_N$ (with $N=k^2\ge9$) and $O(5,3)$ discrete gauge theories (see \eqref{outactW} and \eqref{O53Symaction}), and similarly in theories based on $BOG$, $GL(2,3)$, and the other order forty-eight groups (e.g., see below \eqref{BOGWabc} and in appendix \ref{appA}). We will revisit some of these discussions after introducing further technical tools for symmetries in section \ref{0formsec}.}

\item{We also saw that we could use $\mathbb{Z}_2$ one-form symmetry gauging in the $S_N$ (with $N=k^2\ge9$) gauge theory to generate fusion rules involving non-abelian Wilson lines with unique outcomes in the $A_N$ discrete gauge theories. We can constrain when such a situation arises with the following theorem:

\smallskip
\noindent
{\bf Theorem 1 (one-form fixed points):} Consider a TQFT, $\CT$, with no fusion rules of the form \eqref{abcfusion}. Suppose we can gauge a non-trivial one-form symmetry of this TQFT, $H$. After performing this gauging, we have fusion rules of the form \eqref{abcfusion} only if there are $a\in\CT$ such that fusion with at least one of the one-form generators, $\alpha\in\Rep(H)$, yields $\alpha\times a =a$.

\smallskip
\noindent
{\bf Proof:} Suppose this were not the case. Then, all anyons are organized into full length orbits under fusion with the one-form symmetry generators. When we gauge the one-form symmetry, we identify these orbits as single elements (if the braiding with one-form symmetry generators is trivial, these orbits become genuine lines of the gauged theory; if the braiding is non-trivial, these orbits become lines bounding symmetry-generating surface operators in the gauged theory). Note that all anyons appearing on the right hand side of fusion rules have the same braiding with the one-form symmetry generators. Therefore, the claim follows. $\square$ 

\smallskip
\noindent
As we will see, this theorem will have echoes in the coset theories we describe in the second half of this paper.
}

\item{In the case of $O(5,3)$ discrete gauge theories, we saw that we could gauge the outer automorphisms and have fusion rules of form \eqref{abcfusion} in this gauged theory as well.  This discussion inspires the following theorem:

\smallskip
\noindent
{\bf Theorem 2 (zero-form fixed points):} Consider a TQFT, $\CT$, and suppose we can gauge a non-trivial zero-form symmetry of this TQFT, $H$. After performing this gauging, we have fusion rules of the form \eqref{abcfusion} only if there are non-trivial $a_i\in\CT$ such that at least one of the non-trivial elements of the zero-form group fixes $a_i$.

\smallskip
\noindent
{\bf Proof:} Suppose that all non-trivial elements of the discrete gauge theory leave all the non-trivial anyons unfixed. Now consider anyons $a,b,c\in\CT$ such that $c\in a\times b$. From the general discussion around \eqref{genfusionA}, we see that $N_{{}^ta}\cap N_{{}^sb}\cap N_c=\mathbb{Z}_1$ and $N_a\backslash H / N_b=H$. Moreover, since the stabilizers are trivial, $\pi_a=\pi_b=\pi_c=1$ are the trivial representations. We then have
\begin{equation}
N_{([a],1),([b],1)}^{([c],1)}=|H|\cdot m(1,1)=|H|>1~.
\end{equation}
Therefore, we cannot produce fusion rules of the desired type. $\square$

\medskip
\noindent
Our discussion of the $O(5,3)$ theory also suggests the following theorem

\smallskip
\noindent
{\bf Theorem 2A:}  Consider a TQFT, $\CT$, with a fusion rule of the form $a\times b=c$ and a zero-form symmetry, $H$. If at least one of $\left\{a,b,c\right\}$ is unfixed by $H$, then the only way for $a\times b=c$ to map to a fusion rule with unique outcome in the gauged theory is for $c$ to be unfixed by $H$.

\smallskip
\noindent
{\bf Proof:} If $c$ is unfixed by $H$, then $N_c=N_a\cap N_b=\mathbb{Z}_1$. If either $a$ or $b$ are unfixed then $N_a\cap N_b=\mathbb{Z}_1$ as well (although we need not have $N_c=\mathbb{Z}_1$). In any case, \eqref{genfusionA} becomes
\begin{eqnarray}\label{genfusionA2}
N_{([a],\pi_a),([b], \pi_b)}^{([c],\pi_c)}=\sum_{(t,s)\in N_a \backslash H /N_b} m(\pi_c |_{\mathbb{Z}_1},{}^t\pi_{a}|_{\mathbb{Z}_1} \otimes {}^s\pi_{b}|_{\mathbb{Z}_1} \otimes \pi^{\omega}_{({}^ta,{}^sb,c)})~.
\end{eqnarray}
We have two cases: {\bf(1)} $N_a\backslash H/N_b\ne\mathbb{Z}_1$ or {\bf(2)} $N_a\backslash H/N_b=\mathbb{Z}_1$. Consider case {\bf(1)} first. In this case, all resulting fusion rules will have multiplicity $|N_a\backslash H/N_b|>1$. Next, consider case {\bf(2)}. If $c$ is fixed by some element of $H$, then we have at least two possible $\pi_c$ (one is the trivial representation). This results in a fusion rules with non-unique outcomes. $\square$
}

\item{In the case of the BOG and $GL(2,3)$ discrete gauge theories we saw that both one-form symmetries and subcategory structure offered an explanation of the existence of the fusion rules \eqref{BOGWabc}. The following theorem further explains and generalizes this connection between symmetries and subcategories of the Wilson line sector:

\smallskip
\noindent
{\bf Theorem 3 (subcategories and symmetries):} Consider a finite group, $G$, with an unfaithful higher-dimensional irreducible representation, $\pi$. Moreover, suppose there are one-dimensional representations, $\pi_i$, with $\Ker(\pi_i)\trianglerighteq\Ker(\pi)$. Then, in the corresponding (twisted or untwisted) discrete gauge theory, Wilson lines charged under representations, $\gamma$, that have $\gamma|_{\Ker(\pi)}$ irreducible transform non-trivially under fusion with the abelian Wilson lines, $\CW_{\pi_i}$.

\smallskip
\noindent
{\bf Proof:}  We have that $\CW_{\pi_i}\in\CC_{K_{\pi}}$, where $\CC_{K_{\pi}}$ was defined around \eqref{CKpi} as the subcategory of Wilson lines charged under representations whose kernels contain $\Ker(\pi)$ (see \eqref{Kpi}). Therefore, we see that the abelian Wilson lines $\CW_{\pi_i}\in\CC_{K_{\pi}}$ .

By the discussion around \eqref{notCkpi}, we also see that all non-identity lines $\CW_{\alpha_i}\in\CW_{\gamma}\times\CW_{\bar\gamma}$ are not elements of $\CC_{K_{\pi}}$. As a result, $\CW_{\pi_i}\not\in\CW_{\gamma}\times\CW_{\bar\gamma}$. On the other hand, the trivial line is clearly in $\CW_{\gamma}\times\CW_{\bar\gamma}$. This logic implies
\begin{equation}
\CW_{\pi_i}\times\CW_{\gamma}\times\CW_{\bar\gamma}\ne\CW_{\gamma}\times\CW_{\bar\gamma}~,
\end{equation}
from which the claim in the theorem trivially follows. $\square$

\medskip
\noindent
This result tells us that the $\CW_{\gamma}$ must transform under fusion with the one-form symmetry generators while $\CW_{\pi}$ need not. In the case of the $BOG$ and $GL(2,3)$ discrete gauge theories, precisely this mechanism gave a symmetry explanation for the $\CW_{\pi}\times\CW_{\gamma}=\CW_{\pi\gamma}$ fusion rule in \eqref{BOGWabc}. Here we see it is somewhat more general.}
\item{Note that the results of this section answer questions (1)-(3) of the introduction negatively in general. Still, we saw that in the BOG and $GL(2,3)$ discrete gauge theories, the ideas in (3) and \eqref{fuspartialfact} do apply in some cases. We will return to a proposal for construct a theory satisfying \eqref{fusfact} in question (2) in section \ref{subcat}.}
\end{itemize}

\subsection{Subgroups, subcategories, and primality}\label{subcat}
In sections \ref{unfaithful} and \ref{genlessons}, we saw the important role subcategories play in generating fusion rules involving non-abelian Wilson lines with unique outcomes (e.g., they explained the hierarchy in \eqref{groupgap}). Moreover, understanding the subcategory structure is crucial to resolving the question of whether a particular discrete gauge theory is prime or not. In the case of theories with simple gauge groups (see section \ref{simpleW}), we used results from \cite{naidu2009fusion}. In the case of the examples of discrete gauge theories with non-simple groups we studied, we used an argument that does not easily generalize. Therefore, in this section, we review some of the more general results of \cite{naidu2009fusion} on subcategories of discrete gauge theories. We then apply these results to generate some useful theorems that will serve us in subsequent sections.

The main power of the results in \cite{naidu2009fusion} is that they rephrase questions about subcategories in discrete gauge theories in terms of data of the underlying gauge group. In particular, we have:

\medskip
\noindent
{\bf Theorem 4 \cite{naidu2009fusion}:} Fusion subcategories of discrete gauge theories with finite group $G$ are in bijective correspondence with triples, $(K,H,B)$. Here $K,H\trianglelefteq G$ are normal subgroups that centralize each other (i.e., they commute element-by-element), and $B:K\times H\to \mathbb{C}^{\times}$ is a $G$-invariant bicharacter. If we have a non-trivial twist, $\omega$, then the same conditions hold except that we demand that $B$ is a $G$-invariant $\omega$-bicharacter.

\medskip
\noindent
{\bf Proof:} See proofs of Theorems 1.1 and 1.2 (though they are phrased using different, but equivalent, terminology) of \cite{naidu2009fusion}. $\square$

\medskip
\noindent
Since $B$ is a bicharacter, it satisfies 
\begin{equation}\label{Bdef}
B(k_1k_2,h)=B(k_1,h)\cdot B(k_2,h)~, \ \ \ B(k,h_1h_2)=B(k,h_1)\cdot B(k,h_2)~.
\end{equation}
Here $G$ invariance means that $B(g^{-1}kg,g^{-1}hg)=B(k,h)$ for all $k\in K$, $h\in H$, and $g\in G$. In fact, \cite{naidu2009fusion} also give a way to construct the subcategory, $\CS(K,H,B)$, in question given the above data:
\begin{equation}\label{subcat1}
\CS(K,H,B):={\rm gen}\left((a,\chi)|\left\{a\in K\cap R~, \ \chi\in{\rm Irr}(N_a)\ {\rm s.t.}\ \chi(h)=B(a,h)\deg\chi~,\ \forall h\in H\right\}\right)~,
\end{equation}
where $R$ is a set of representatives of conjugacy classes, ${\rm Irr}(N_a)$ is the set of characters of irreducible representations of the centralizer $N_a$, and \lq\lq gen$(\cdots)$" means that the category is generated by the simple objects inside the parenthesis. A normal subgroup is a union of conjugacy classes. Hence, $K$ specifies all the conjugacy classes labelling the anyons in the subcategory $\CS(K,H,B)$. Also, all the Wilson lines in $\CS(K,H,B)$ are such that the corresponding representations have kernels which contain $H$. 

If we have non-trivial twist, then \eqref{Bdef} and $G$-invariance become \cite{naidu2009fusion}
\begin{eqnarray}
B(k_1k_2,h)&=&\eta_h(k_1,k_2)\cdot B(k_1,h)\cdot B(k_2,h)~,\ B(k,h_1h_2)=\eta^{-1}_k(h_1,h_2)\cdot B(k,h_1)\cdot B(k,h_2)~,\cr B(g^{-1}kg,h)&=&{\eta_k(g,h)\eta_k(gh,g^{-1})\over\eta_k(g,g^{-1})}B(k,ghg^{-1})~,
\end{eqnarray}
where 
\begin{equation}\label{genH2f3}
\eta_g(h,k):={\omega(g,h,k)\cdot\omega(h,k,k^{-1}h^{-1}ghk)\over\omega(h,h^{-1}gh,k)}~,
\end{equation}
is a generalization of \eqref{H2f3}. For non-trivial twist, we also have that \eqref{subcat1} becomes
\begin{equation}\label{subcat2}
\CS(K,H,B):={\rm gen}\left((a,\chi)|\left\{a\in K\cap R~, \chi\in{\rm Irr}_{\omega}(N_a)\ {\rm s.t.}\ \chi(h)=B(a,h)\deg\chi~,\forall h\in H\right\}\right)~,
\end{equation}
where the $\omega$ in ${\rm Irr}_{\omega}(N_a)$ is a reminder that we should consider characters with projectivity phase given by \eqref{H2f3} or \eqref{genH2f3}. 

We can now immediately see how the subcategories we studied in previous sections arose: $\CS(G,\mathbb{Z}_1,1)\simeq\CZ({\rm Vec}_G^{\omega})$ is the full discrete gauge theory, $\CS(\mathbb{Z}_1,G,1)$ is the trivial subcategory, and $\CS(\mathbb{Z}_1,\mathbb{Z}_1, 1)\simeq\Rep(G)\simeq\CC_{\CW}$ is the full subcategory of Wilson lines. In the case of simple discrete gauge theories, we see that, as claimed in section \ref{simpleW}, these are the {\it only} subcategories. However, in the case of the $\CZ({\rm Vec}_{BOG}^{\omega})$, $\CZ({\rm Vec}_{GL(2,3)}^{\omega})$, and other gauge theories based on gauge groups with unfaithful irreducible representations, $\pi$, we find additional subcategories: $\CS(\mathbb{Z}_1,\Ker(\pi),1)\simeq\Rep(G/\Ker(\pi))$ and $\CS(\Ker(\pi),\mathbb{Z}_1,1)$. Using Lemma 3.11 of \cite{naidu2009fusion}, we have that $\CS(\Ker(\pi),\mathbb{Z}_1,1)$ is the M\"uger center of $\CS(\mathbb{Z}_1,\Ker(\pi),1)$.

Since we will study flux lines and dyons below, it is interesting to ask what the above theorems imply for such operators. One immediate consequence is that magnetic flux lines behave very differently from Wilson lines. For example:

\medskip
\noindent
{\bf Theorem 5:} The set of magnetic flux lines, $\CM$, of a discrete gauge theory (both untwisted and twisted) with non-abelian gauge group, $G$, do not form a fusion subcategory. In particular, $\CM\not\simeq\Rep(G)$.

\medskip
\noindent
{\bf Proof:} Suppose the full set of flux lines form a subcategory. Then, we need $K$ to include at least one element of each conjugacy class in order to include all of $\CM$ in $\CS$. However, since $K$ is a normal subgroup, it must consist of full conjugacy classes. Therefore, $K=G$. Using theorem 4, we can label this putative subcategory as $\CS(G,H,B)$. Since $H$ has to commute with all elements in $G$, it has to be a subgroup of the center of the group $Z(G)$. Suppose the group has trivial center. This forces $B=1$, and $\CS(G,\mathbb{Z}_1,1)$ is the full discrete gauge theory, which means we also include objects with charge. This is a contradiction. 

Suppose $H$ is a non-trivial subgroup of $Z(G)$. We know that the function $B$, being a bicharacter, satisfies $B(e,h)=1 ~ \forall h \in H$. So the Wilson line $([e],\pi) \in \CS(G,H,B)$ if $\pi$ has $H$ in its kernel. Recall that the irreducible representations of $G/H$ are in one-to-one correspondence with irreducible representations of $G$ with $H$
in its kernel. Since $G$ is non-abelian, $Z(G)\neq G $. Hence, $G/H$ is a non-trivial group. It follows that there is at least one non-trivial irreducible representation $\pi'$ of $G$ such that $H$ is in its kernel. Hence, the Wilson line $([e],\pi')$ belongs to the subcategory $\CS(G,H,B)$ for any $B$. A contradiction. $\square$

\medskip
\noindent
The fact that $\CM\not\simeq\Rep(G)$ has consequences in section \ref{0formsec}. In particular, it explains why electric-magnetic self-dualities are non-trivial to engineer in theories with non-abelian gauge groups and trivial centers.\footnote{In any untwisted abelian gauge theory, this is not an issue as $\CM\simeq\Rep(G)$ and there is a canonical electric/magnetic duality.} If such a duality exists and involves magnetic flux lines, then they will necessarily be in a $\Rep(G)$-like subcategory with objects carrying electric charge (e.g., see the $S_3$ discrete gauge theory self-duality \cite{beigi2011quantum}, where the dimension-two flux line is in a $\Rep(S_3)$ subcategory with both dimension one Wilson lines).

Now, we turn to the question of primality. Here the following theorem of \cite{naidu2009fusion} is useful

\medskip
\noindent
{\bf Theorem 6 \cite{naidu2009fusion}:} A discrete gauge theory with gauge group, $G$, is a prime TQFT if and only if there is no triple $(K,H,B)$ with $K,H\triangleleft G$ normal subgroups centralizing each other, $HK=G$, $(G,\mathbb{Z}_1)\ne(K,H)\ne(\mathbb{Z}_1,G)$, and $B$ is a $G$-invariant bicharacter on $K\times H$ such that $BB^{\rm op}|_{(K\cap H)\times(K\cap H)}$ is non-degenerate. In the case of non-trivial twisting, $\omega$, the previous conditions still hold, but $B$ is also a $G$-invariant $\omega$-bicharacter.

\medskip
\noindent
{\bf Proof:} See proof of theorem 1.3 (though it is phrased using different, but equivalent, terminology) in \cite{naidu2009fusion}. $\square$

\medskip
\noindent
Note that in the statement of theorem 6, $B^{\rm op}(h,k):=B(k,h)$ for all $k\in K$ and $h\in H$.

Given this theorem, we may prove the following result that will be useful to us in section \ref{nonW}:

\medskip
\noindent
{\bf Theorem 7:} If $G$ is a non-direct product group with trivial center, then the corresponding (twisted or untwisted) gauge theory is a prime TQFT.

\medskip
\noindent
{\bf Proof:} We have a non-direct product group $G$ with trivial center. Let us assume that $\text{Rep}(D(G))$ has a modular subcategory. Then, there exists two normal subgroups, $K$ and $H$, commuting with each other and satisfying $KH=G$. So, every element of $G$ is a product of an element of $K$ with an element of $H$. Hence, any element in $K \cap H$ has to commute with all elements of $G$. Since the center of $G$ is trivial by choice, $K \cap H= \mathbb{Z}_1$. It follows that $G$ has to be a direct product of $K$ and $H$. A contradiction. Hence, for non-direct product groups $G$ with trivial center, $\text{Rep}(D(G))$ is prime. $\square$

A simple set of examples subject to this theorem include the $S_N$ discrete gauge theories analyzed above and the $\DZ_{15} \rtimes \DZ_4$ discrete gauge theory we will analyze further in section \ref{nonW}.

Finally, we conclude with a proposal for engineering an example of a theory of the type envisioned in question (2) in the introduction. In particular, consider a $G\times G$ discrete gauge theory, $\CZ({\rm Vec}_{G\times G}^{\omega})$. Clearly, for trivial twisting this is a non-prime theory since $\CZ({\rm Vec}_{G\times G})=\CZ({\rm Vec}_{G})\boxtimes \CZ({\rm Vec}_{G})$. Indeed, by theorem 6, we can take $K=G\times\mathbb{Z}_1$, $H=\mathbb{Z}_1\times G$, and $B=1$. However, if we turn on a twist, $\omega\in H^3(G\times G, U(1))$, we might be able to generate a prime theory. In particular, if we can find $G$ such that $\omega$ is non-trivial and does not factorize, then we would have an example of a prime theory with Wilson lines in $\Rep(G\times G)=\Rep(G)\boxtimes \Rep(G)$. Choosing one Wilson line in each $\Rep(G)$ factor and fusing would give a unique fusion outcome.\footnote{We thank D.~Aasen for suggesting the basis for this idea.} It would be interesting to see if this proposal can be realized. For example, we would like to see if there is an obstruction at the level of the existence of a $G$-invariant $\omega$-bicharacter (all other requirements of theorem 6 can be satisfied). A concrete example of a theory of the type discussed in question (2) is studied in section \ref{exZ3Q16}.

\subsection{Zero-form symmetries}\label{0formsec}
In sections \ref{unfaithful} and \ref{genlessons} we saw that zero-form symmetries played an important role in generating fusions rules of the form \eqref{abcfusion}. In this section we review some relevant results of \cite{nikshych2014categorical}  and prove a theorem that will be useful to us in section \ref{nonW}.

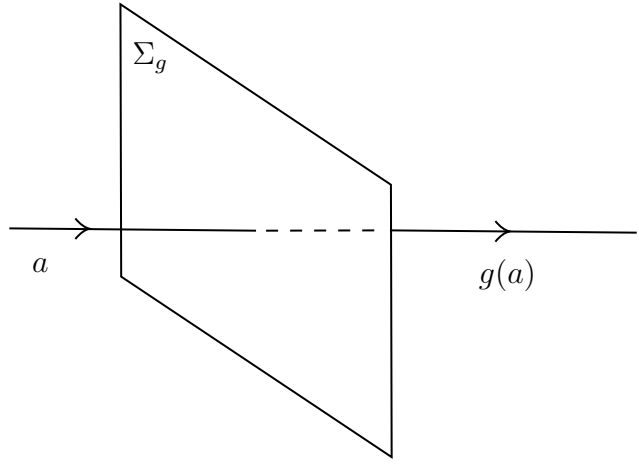
\begin{figure}[h!]
\centering

\tikzset{every picture/.style={line width=0.75pt}} 

\begin{tikzpicture}[x=0.75pt,y=0.75pt,yscale=-1,xscale=1]

\draw   (398.54,121.68) -- (398.9,258.69) -- (263.89,167.92) -- (263.53,30.91) -- cycle ;
\draw    (208.05,143.64) -- (331.22,144.8) ;
\draw    (398.05,144.64) -- (521.22,145.8) ;
\draw   (241,138.64) .. controls (243.35,141.52) and (245.7,143.24) .. (248.05,143.82) .. controls (245.7,144.39) and (243.35,146.12) .. (241,149) ;
\draw   (450.67,139.97) .. controls (453.02,142.85) and (455.36,144.58) .. (457.71,145.15) .. controls (455.36,145.73) and (453.02,147.45) .. (450.67,150.33) ;
\draw  [dash pattern={on 4.5pt off 4.5pt}]  (336.22,144.8) -- (396.05,144.64) ;

\draw (268,48.4) node [anchor=north west][inner sep=0.75pt]    {$\Sigma _{g}$};
\draw (218,157.4) node [anchor=north west][inner sep=0.75pt]    {$a$};
\draw (441,157.4) node [anchor=north west][inner sep=0.75pt]    {$g( a)$};

\end{tikzpicture}
\caption{The symmetry defect $\Sigma_g$, labelled by a zero-form symmetry group element $g$, acts on an anyon $a$.}
\label{symaction}
\end{figure}

In three spacetime dimensions, zero-form symmetries are implemented by dimension two topological defects (recall that one-form symmetries are generated by abelian lines). These defects act on lines that pierce them as in figure \ref{symaction}. 
We will say the corresponding symmetry group, $H$, is non-trivial iff it has a generator, $h\in H$, such that there is an anyon $a\in\CT$ satisfying $h(a)\ne a$.

Note that the automorphisms of the gauge group $G$, ${\rm Aut}(G)$, are a natural source of symmetries. Indeed, in the context of the $G$-SPT that we gauge to generate the discrete gauge theory, these automorphisms permute the symmetry defects. Therefore, we expect they will play a role in the discrete gauge theory. To be more precise, recall that we can distinguish between the inner automorphisms ${\rm Inn}(G)\trianglelefteq {\rm Aut}(G)$, generated by conjugations of the form $gxg^{-1}$ for $x,g\in G$, and outer automorphisms, ${\rm Out}(G):={\rm Aut}(G)/{\rm Inn}(G)$. Since the discrete gauge theory involves magnetic charges labeled by conjugacy classes and electric charges labeled by representations of centralizers, it is clear that inner automorphisms will act trivially on the discrete gauge theory (conjugacy classes are invariant under ${\rm Inn}(G)$ and the normalizers of different elements in a conjugacy class are isomorphic). Therefore, we can at best expect ${\rm Out}(G)$ to lift to a symmetry of the TQFT. Indeed, this is precisely what happens.

More formally, we have that, in a discrete gauge theory ${\rm Out}(G)$ lifts to a part of the group of braided autoequivalences of the discrete gauge theory, ${\rm Aut}^{\rm br}(\CZ({\rm Vec}_G))$:

\medskip
\noindent
{\bf Theorem 8 \cite{nikshych2014categorical}:} The subgroup of braided autoequivalences that fix the Wilson lines ${\rm Stab}({\rm Rep}(G))\le {\rm Aut}^{\rm br}(\CZ({\rm Vec}_G))$ takes the form
\begin{equation}
{\rm Stab}({\rm Rep}(G))\simeq H^2(G,U(1))\rtimes{\rm Out}(G)~.
\end{equation}

\medskip
\noindent
{\bf Proof:} See the proof of Corollary 6.9 (though it is phrased using different, but equivalent, terminology) in \cite{nikshych2014categorical}. $\square$

\medskip
\noindent
Note that ${\rm Out(G)}$ generally acts non-trivially on the conjugacy classes. Therefore, it will also generally act non-trivially on the Wilson lines. However, in certain more exotic cases, all of ${\rm Out (G)}$ preserves conjugacy classes.\footnote{The smallest group that has this feature has order $2^7$ \cite{brooksbank2013groups}. See \cite{Cui:2019lvb} for an application of groups that have at least some class-preserving outer automorphisms to quantum doubles.} In such cases, the Wilson lines are fixed. Note that elements $\zeta\in H^2(G,U(1))$ always leave the Wilson lines invariant since they act as follows \cite{nikshych2014categorical}
\begin{equation}\label{zetaAct}
\zeta(([a],\pi_a))=([a],\pi_g\rho_g)~, \ \ \ \rho_g(x):={\zeta(x,g)\over\zeta(g,x)}~,
\end{equation}
where $g\in[a]$ (in particular, $g=1$ for Wilson lines). Note that $\rho_g(x)$ depends only on the cohomology class of $\zeta$ (it is invariant under shifts by a 2-coboundary).

A second set of symmetries involves the exchange of electric and magnetic degrees of freedom. These are electric/manetic self-dualities and are inherently quantum mechanical in nature. These symmetries are closely related to the existence of Lagrangian subcategories. As we briefly mentioned at the beginning of section \ref{WilsonSec}, a Lagrangian subcategory, $\CL$, is a collection of bosons with trivial mutual braiding that is equal to its M\"uger center (e.g., like the subcategory of Wilson lines, $\CC_{\CW}\simeq\Rep(G)$) . This latter condition simply means that the only objects that braid non-trivially with {\it every} element of $\CL$ are elements of that subcategory.

To find the set of these symmetries, it turns out to be useful to construct the categorical Lagrangian Grassmannian, $\mathbb{L}(G)$. This is the collection of all Lagrangian subcategories. Each such subcategory, $\CL_{(N,\mu)}\simeq\Rep(G_{(N,\mu)})$ with $|G_{(N,\mu)}|=|G|$, is labeled by a normal abelian subgroup, $N\triangleleft G$, and a $G$-invariant $\mu\in H^2(N,U(1))$ (the Wilson line subcategory is $\CL_{1,1}$). For the purposes of understanding these symmetries, the important subcategory is \cite{nikshych2014categorical}
\begin{equation}
\mathbb{L}\supseteq\mathbb{L}_0:=\left\{\CL\in\mathbb{L}(G)|\CL\simeq\Rep(G)\right\}~.
\end{equation}
In particular, we have

\medskip
\noindent
{\bf Theorem 9 \cite{nikshych2014categorical}:} The action of ${\rm Aut}^{\rm br}(\CZ({\rm Vec}_G))$ on $\mathbb{L}_0(G)$ is transitive. Moreover,
\begin{equation}
|{\rm Aut}^{\rm br}(\CZ({\rm Vec}_G))|=|H^2(G,U(1))|\cdot |{\rm Out}(G)|\cdot |\mathbb{L}_0(G)|~.
\end{equation}

\medskip
\noindent
{\bf Proof:} See proposition 7.6 and corollary 7.7 of \cite{nikshych2014categorical}. $\square$

\medskip
\noindent
Examples of such dualities appear in the $S_3$ discrete gauge theory \cite{beigi2011quantum} and beyond \cite{hu2020electric}.

Let us now apply this theorem to prove a result that will be useful for us below

\medskip
\noindent
{\bf Theorem 10:} If $G\simeq N\rtimes K$, where $N$ is an abelian group, then the corresponding untwisted discrete gauge theory has an electric-magnetic self-duality.

\medskip
\noindent
{\bf Proof:} By theorem 9, in order to find a self-duality, we need to find a normal abelian subgroup $N\triangleleft G$ and a $G$-invariant 2-cocycle, $\mu\in H^2(N,U(1))$. Moreover, we need to find a corresponding $G_{(N,\mu)}\simeq G$. In particular, from remark 7.3 of \cite{nikshych2014categorical}, when $\mu$ is trivial, we have that $G_{(N,1)}\simeq\widehat{N}\rtimes G/N$, where $\widehat{N}$ is the character group of $N$. For an abelian group, $\widehat{N}\simeq N$. Therefore, we have that $G_{(\tilde N, 1)}\simeq N\rtimes K=G$ as desired. $\square$

\medskip
\noindent
This theorem will be useful in our symmetry searches in section \ref{nonW}. Note that one immediate consequence of the above discussion is that none of the examples discussed above have self-dualities. Indeed, theories with simple gauge groups have no non-trivial normal abelian subgroups. On the other hand, theories like $BOG$ and $GL(2,3)$ have $H^2(BOG,U(1))\simeq H^2(GL(2,3),U(1))\simeq\mathbb{Z}_1$ (and similarly for all normal abelian subgroups). Since these groups are not semi-direct products, we conclude they lack self-dualities.

\subsection{Quasi-zero-form symmetries}\label{quasi0}
In the previous subsections, we have seen that zero-form symmetries play an important role in generating fusion rules for non-abelian anyons with unique outcomes. However, since our interest is simply in the existence of such fusion rules, it is natural that we should generalize our notion of symmetry to include symmetries of the modular data (and hence, by Verlinde's formula, automorphisms of the fusion rules) that don't necessarily lift to symmetries of the TQFT.\footnote{In fact, most generally, we might expect automorphisms of the fusion rules that are not even symmetries of the modular data (e.g., as studied recently in \cite{Buican:2019evc}).} The basic reason such \lq\lq quasi zero-form symmetries" as we will call them exist is that the modular data does not define a TQFT (see \cite{mignard2017modular} for a consequence of this fact). In particular, the underlying $F$ and $R$ symbols may not be invariant (up to an allowed gauge transformation) under a quasi zero-form symmetry even if $S$ and $T$ are.

In fact, such \lq\lq quasi-zero-form symmetries" are common, with charge conjugation being a particular example \cite{davydov2016unphysical}. Indeed, even in the $A_N$ (with $N=k^2\ge9$) theories we discussed in section \ref{simpleW}, such quasi-charge conjugation symmetries exist. These symmetries are in addition to the genuine zero-form symmetries we described when analyzing these examples. In appendix \ref{A9app}, we study the particular case of $A_9$ discrete gauge theory in more detail and explicitly disentangle the quasi-symmetries from the genuine symmetries.

More generally, there are theories that have no genuine symmetries. One set of examples include discrete gauge theories based on the Mathieu groups. These are simple groups with trivial ${\rm Out(G)}$ and $H^2(G,U(1))$. Moreover, since these groups have no non-trivial normal abelian subgroups, $\mathbb{L}(G)=\mathbb{L}_0(G)\simeq\Rep(G)$, and so there are no non-trivial self-dualities.

The largest Mathieu groups, $M_{23}$ and $M_{24}$ are of particular interest to us since their discrete gauge theories have non-abelian Wilson lines that fuse together to produce a unique outcome.\footnote{By the results of \cite{Buican:2020oyy}, these theories cannot have such fusions involving lines that carry magnetic flux.}  Moreover, of the theories with fusions of type \eqref{abcfusion}, these are the only untwisted discrete gauge theories that have no modular symmetries that lift to symmetries of the full TQFTs.

For $M_{23}$ it is not hard to check that
\begin{equation}\label{WM23}
\CW_{22}\times\CW_{45_1}=\CW_{990_1}~, \ \ \ \CW_{22}\times\CW_{45_2}=\CW_{990_2}~,
\end{equation}
where $22$ is the real twenty-two dimensional representation, $45_{1,2}$ are two forty five dimensional complex representations, and $990_{1,2}$ are two nine hundred and ninety dimensional representations. Under charge conjugation
\begin{equation}\label{CM23}
\CW_{45_1}\leftrightarrow\CW_{45_2}~, \ \ \ \CW_{990_1}\leftrightarrow\CW_{990_2}~.
\end{equation}
For $M_{24}$, we have a particularly rich set of fusions\footnote{It would be interesting to know if our results here have any connection with moonshine phenomena observed involving $M_{24}$ as in \cite{Eguchi:2010ej,Cheng:2012tq,Gannon:2012ck}.}
\begin{eqnarray}\label{WM24}
\CW_{23}\times\CW_{45_1}&=&\CW_{1035_2}~, \ \CW_{23}\times\CW_{45_2}=\CW_{1035_3}~, \ \CW_{23}\times\CW_{231_1}=\CW_{5313}\cr\CW_{23}\times\CW_{231_2}&=&\CW_{5313}~, \ \CW_{45_1}\times\CW_{231_1}=\CW_{10395}~, \ \CW_{45_2}\times\CW_{231_1}=\CW_{10395}~,\cr\CW_{45_1}\times\CW_{231_2}&=&\CW_{10395}~,\ \CW_{45_2}\times\CW_{231_2}=\CW_{10395}~.
\end{eqnarray}
where $23$ is a real twenty-three dimensional representation, $45_{1,2}$ are complex forty-five dimensional representations, $231_{1,2}$ are two-hundred and thirty-one dimensional complex representations, and $1035_{2,3}$ are complex one-thousand and thirty-five dimensional representations, $5313$ is a real five-thousand three-hundred and thirteen dimensional representation, and $10395$ is a real ten-thousand three-hundred and ninety-five dimensional representation. Under charge conjugation, we have
\begin{equation}\label{CM23}
\CW_{45_1}\leftrightarrow\CW_{45_2}~, \ \ \ \CW_{231_1}\leftrightarrow\CW_{231_2}~,\ \ \ \CW_{1035_2}\leftrightarrow\CW_{1035_3}~.
\end{equation}
While we have seen similar actions in previous sections, but here the novelty is that charge conjugation is a quasi-symmetry. 

More generally, as we will discuss in greater detail below, all other examples of TQFTs that we have found with fusion rules involving non-abelian anyons with unique outcome have at least quasi zero-form symmetries.

Finally, let us conclude this section by discussing how twisting affects the quasi-zero-form symmetries. When the quasi-symmetry is charge conjugation and the group has complex representations, the quasi-symmetry lifts to an action on Wilson lines (see appendix \ref{A9app} for a discussion in a concrete example). In this case, the quasi-symmetry persists regardless of the twisting.

As a more complicated example, let us consider the case of $BOG$ first discussed in section \ref{unfaithful}. This theory only has real conjugacy classes and representations. However, there is still a non-trivial charge conjugation acting on certain dyons since elements in $BOG$ have centralizer groups $\mathbb{Z}_4$, $\mathbb{Z}_6$, and $\mathbb{Z}_8$. These latter groups admit complex representations. However, unlike the spectrum of Wilson lines, the spectrum of dyons generally changes as we change the twist. Therefore, we might imagine that the charge conjugation quasi symmetry can be twisted away.

In fact, this is not the case. The main point is that any twisting $\omega\in H^3(BOG,U(1))\simeq\mathbb{Z}_{48}$ of the $BOG$ discrete gauge theory is \lq\lq cohomologically trivial" in the following sense: the $\eta_g(h,k)\in H^2(N_g,U(1))$ phases defined in \eqref{H2f3} are all trivial. Indeed, this statement follows from the fact that $H^2(N_g,U(1))=\mathbb{Z}_1$ for all $g\in BOG$. Therefore, none of the anyons are lifted by the twisting, and the characters of $BOG$ change as follows
\be
\chi_{\pi_g^{\omega}}(h) \rightarrow \epsilon_g(h) \cdot \chi_{\pi_g^{\omega}}(h)
\ee
where $\epsilon_g$ is a 1-cochain that gives the 2-coboundary, $\eta_g$. It is not too hard to check that all choices of the twisting leave us with complex characters. Therefore, the charge conjugation quasi-symmetry persists (here it would be more accurate to term it a \lq\lq modular symmetry" since it is apriori possible---though we have not checked---that charge conjugation becomes a symmetry of the theory for certain choices of $\omega$).\footnote{One may also wonder about the fate of the genuine ${\rm Out}(BOG)\simeq\mathbb{Z}_2$ zero-form symmetry under twisting. First, consider $\omega$ corresponding to the order 2 element in $\DZ_{48}$. Since $\text{Out}(BOG)$ acts on $H^3(BOG,U(1))$ through $\text{Aut}(H^3(BOG,U(1)))$, $\omega$ should be fixed under it. Hence, it seems plausible that the twisted discrete gauge theory corresponding to this choice of $\omega$ has $\text{Out}(BOG)$ as a subgroup of its symmetries (while theorem 8 has nothing to say on this point since it assumes untwisted theories, we view the existence of a symmetry in this case as a plausible assumption). In fact, more generally, if the action of $\text{Out}(G)$ leaves $\omega \in H^3(G,U(1))$ invariant up to a 3-coboundary, then it can be shown that this is a symmetry of  the modular data of the twisted theory. It would be interesting to understand what happens for other twists as well.}

\subsection{Beyond Wilson lines}\label{nonW}
So far, we have only constructed fusion rules of the form \eqref{abcfusion} using Wilson lines. In the case of gauge theories with simple groups, this is all we can do \cite{Buican:2020oyy}. However, when we have non-simple gauge groups, the existence of self-dualities discussed in section \ref{0formsec} as well as the possibility of electric-magnetic dualities between theories with different gauge groups and Dijkgraaf-Witten twists \cite{naidu2007categorical,hu2020electric} suggests that we should also be able to involve non-abelian anyons carrying flux. Indeed, we will see this is the case.

To that end, let us study a fusion of the form
\be
\label{fusgen}
\CL_{([g], \pi^{\omega}_g)} \times \CL_{([h], \pi^{\omega}_h)}= \CL_{([k], \pi^{\omega}_k)}~,\ \ \ g,h\ne1~,
\ee
Carefully applying the machinery in section \ref{fusionrules} reveals the following contraints\footnote{We refer the interested reader to the derivation in section III of \cite{Buican:2020oyy} for further details.}
\begin{enumerate}
\item $[g] \cdot [h]=[k]=[h]\cdot [g]$
\item $\exists! ~ \pi^{\omega}_k$ such that $m(\pi^{\omega}_k |_{N_{g} \cap N_{h} \cap N_k},\pi^{\omega}_{g}|_{N_{g} \cap N_{h} \cap N_k} \otimes \pi^{\omega}_{h}|_{N_{g} \cap N_{h} \cap N_k} \otimes \pi^{\omega}_{(g,h,k)})=1$ 
\end{enumerate}
We will apply these constraints in what follows. 

For an untwisted discrete gauge theory based on a group $G$ with a non-trivial center $Z(G)$, the constraints above implies that if we have a fusion of Wilson lines giving a unique outcome
\be
\CW_{\pi} \times \CW_{\gamma}= \CW_{\pi \gamma}~,
\ee
then we have a fusion of dyons of the form
\be
\CL_{([g], \pi)} \times \CL_{([h], \gamma)}= \CL_{([gh], \pi \gamma)}~,
\ee
where for any $g,h \in Z(G)$. Hence, we can dress the Wilson lines with fluxes from the center of the group to obtain fusion rules involving dyons with unique outcomes. For example, we have already seen that the discrete gauge theories corresponding to $BOG$ and $GL(2,3)$ have Wilson lines fusing to give a unique outcome. Since these two groups have a non-trivial center (isomorphic to $\DZ_2$), the above discussion immediately implies the existence of dyonic fusions where the dyons are labelled by the non-trivial element of the centre. In fact, these two types of fusions exhaust all $a\times b=c$ type fusions in both $\CZ(\text{Vec}_{BOG})$ and $\CZ(\text{Vec}_{GL(2,3)})$.

In the case of the fusion of non-abelian Wilson lines with a unique outcome, we saw that we were not guaranteed to find fusion subcategories beyond the three universal subcategories present in any discrete gauge theory (the theory itself, the trivial TQFT, and the Wilson line sector, $\CC_{\CW}\simeq\Rep(G)$). On the other hand, when we have fusions of non-abelian anyons carrying flux with a unique outcome, we are guaranteed to have fusion subcategories. When the gauge group has a non-trivial center, $Z(G)$, this statement is trivial.\footnote{The discussion in section \ref{subcat} guarantees that $\CS(Z(G),\mathbb{Z}_1,1)$ and $\CS(\mathbb{Z}_1, Z(G),1)$ are non-trivial subcategories.} The following theorems guarantee this fact more generally:

\medskip
\noindent
\noindent {\bf Theorem 11:} Let $G$ be a non-simple finite non-abelian group. If we have a fusion rule involving two dyons or fluxes giving a unique outcome in the (twisted or untwisted) $G$ gauge theory, then $\CS(M_g,\mathbb{Z}_1,1)$ and $\CS(M_h,\mathbb{Z}_1, 1)$ (along with $\CS(\mathbb{Z}_1,M_g,1)$ and $\CS(\mathbb{Z}_1,M_h, 1)$) are proper fusion subcategories of the theory. Here, $g$ and $h$ are elements labelling the non-trivial conjugacy classes (of length $> 1$) involved in the fusion. $M_g$ is the normal subgroup generated by the elements in $[g]$. 

\medskip
\noindent
{\bf Proof:} We have an $a \times b=c$ type fusion rule involving the non-trivial conjugacy classes $[g]$ and $[h]$. Let $M_g$ be the normal subgroup generated by $[g]$. In fact, it has to be a proper normal subgroup. To see this, suppose $M_g=G$.  From Lemma 3.4 of \cite{naidu2009fusion}, we know that $[g]$ and $[h]$ commute element-wise. Hence, $[h]$ commutes with all elements in $M_g=G$. It follows that $[h]$ should be a subset of the elements in $Z(G)$. However, elements of $Z(G)$ form single element conjugacy classes.  A contradiction. Hence, $M_g$ has to be a proper normal subgroup of $G$. Since $g \neq e$, it is clear that $M_g$ is not the trivial subgroup either. We can use the same argument to show that $M_h$ is also a proper non-trivial normal subgroup of $G$. Therefore, by theorem 4, we have fusion subcategories corresponding to the choices $\CS(M_g, \mathbb{Z}_1, 1)$ and $\CS(M_h, \mathbb{Z}_1, 1)$ (and similarly $\CS(\mathbb{Z}_1,M_g,1)$ and $\CS(\mathbb{Z}_1,M_h, 1)$). $\square$

\medskip
\noindent
Note that we have, $\CL_{([g], \pi^{\omega}_g)} \in\CS(M_g,\mathbb{Z}_1, 1)$ and $\CL_{([h], \pi^{\omega}_h)} \in\CS(M_h,\mathbb{Z}_1, 1)$. Generically, we also expect $\CL_{([g], \pi^{\omega}_g)} \not\in\CS(M_h,\mathbb{Z}_1, 1)$ and $\CL_{([h], \pi^{\omega}_h)} \not\in\CS(M_g,\mathbb{Z}_1, 1)$. In such situations we have, in the spirit of section \ref{unfaithful}, an \lq\lq explanation" for the fusion rule.

In fact, the reasoning in the proof to theorem 11 immediately implies that if $[h]$ has at least one element $h'\in[h]$ such that $[h',h]\ne1$, then $\CL_{([g], \pi^{\omega}_g)}$ and $\CL_{([h], \pi^{\omega}_h)}$ lie in different subcategories

\medskip
\noindent
{\bf Corollary 12:} Given the conditions in theorem 11, if there exists $h'\in[h]$ such that $[h',h]\ne1$, $\mu_{[g]}\in\CS(M_g,\mathbb{Z}_1, 1)$, $\CL_{([g], \pi^{\omega}_g)} \not\in\CS(M_h,\mathbb{Z}_1, 1)$, and similarly for $h\leftrightarrow g$.

For $a \in M_g$ the fusion subcategory $\CS(M_g,\DZ_1,1)$ contains anyons $([a],\pi_a)$ where $\pi_a$ is any irrep of the centralizer $N_a$.  In an untwisted discrete gauge theory, for a fusion of fluxes labelled by conjugacy classes $[g]$ and $[h]$, we can define fusion subcategories $\CS(M_g,M_h,1)$ and $\CS(M_hM_g,1)$ which have a more restricted set of elements. For $a \in M_g$, the anyon $([a],\pi_a)$ is an element of $\CS(M_g,M_h,1)$ if and only if $M_h \subseteq \text{Ker}(\pi_a)$. Clearly, $([g],1_g) \in \CS(M_g,M_h,1)$ and $([h],1_h) \in \CS(M_h,M_g,1)$. However, in general, we don't expect $([g],1_g) \not\in \CS(M_h,M_g,1)$ and $([h],1_h) \not\in \CS(M_g,M_h,1)$. We will discuss an example of this below.

\medskip
If one of the operators involved in the fusion of non-abelian anyons with a unique outcome is a Wilson line, then we also have the following theorem:

\medskip
\noindent
{\bf Theorem 13:} Let $G$ be a non-simple group. If we have a fusion of a Wilson line and a  dyon giving a unique outcome, then $\CS(\text{Ker}(\chi_{\pi}),\mathbb{Z}_1,1)$ and $\CS(\mathbb{Z}_1,{\rm Ker}(\chi_{\pi}),1)$ are proper fusion subcategories of the (twisted or untwisted) discrete gauge theory. Here, $\pi$ is an irrep of $G$ labelling the Wilson line.

\medskip
\noindent
{\bf Proof:} Suppose $[b]$ is the non-trivial conjugacy labelling the flux line. Let $\chi_{\pi}$ be the character of an irreducible representation, $\pi$, of $G$ labelling the Wilson line. From note 3.5 of \cite{naidu2009fusion} we know that $\chi$ should be trivial on a subset of elements given by $[G,b]$. Since $b$ is not in the center, $[G,b]$ is guaranteed to have a non-trivial element. Hence, $\chi_{\pi}$ is not a faithful representation. $\text{Ker}(\chi_{\pi})$ is a non-trivial normal subgroup of $G$. Since $\chi_{\pi}$ is not the trivial representation, $\text{Ker}(\chi_{\pi}) \neq G$ is a non-trivial proper normal subgroup. Hence, by theorem 4, we have a fusion subcategory given by $\CS(\text{Ker}(\chi_{\pi}), \mathbb{Z}_1, 1)$ and $\CS(\mathbb{Z}_1,\text{Ker}(\chi_{\pi}), 1)$. $\square$

\medskip
\noindent
Note that in this case the Wilson line is an element of $\CS(\mathbb{Z}_1,\text{Ker}(\chi_{\pi}), 1)$ while the magnetic flux is not. In this sense, such fusions are \lq\lq natural." To illustrate the ideas above, let us consider the following examples.

\subsubsection{$\CZ(\text{Vec}_{\DZ_3 \rtimes Q_{16}})$}\label{exZ3Q16}

Let us consider the $\mathbb{Z}_{3}\rtimes Q_{16}$ discrete gauge theory. Even though this group has many non-trivial proper normal subgroups, we have $\mathbb{Z}_{3}\rtimes Q_{16}\neq HK$ for any proper normal subgroups $H,K$. Hence, using theorem 6, we have that $\CZ(\text{Vec}_{\DZ_3 \rtimes Q_{16}})$ is a prime theory.

This group has a length $2$ conjugacy class $[f_3]$ (here we are using the notation of GAP \cite{GAP}, where this group is entry $(48,18)$ in GAP's small group library) and a 2-dimensional representation $2_3$ (the third 2-dimensional representation in the character table of $\DZ_3 \rtimes Q_{16}$ on GAP). We have the following fusion of a Wilson line and a flux line giving a unique outcome. 
\be
\label{wfZ3Q16}
\CW_{2_3} \times \mu_{[f_3]}= \CL_{([f_3],2_3|_{N_{f_3}})}~,
\ee
where the restricted representation $2_3|_{N_{f_3}}$ is irreducible. 

Since we have a prime theory, the existence of this fusion rule is not due to a Deligne product. However, it can be explained using the subcategory structure of $\CZ(\text{Vec}_{\DZ_3 \rtimes Q_{16}})$. To that end, consider the fusion subcategory $\CS(\DZ_1, \text{Ker}(2_3)),1)$. This fusion subcategory contains only Wilson lines. A Wilson line $\CW_{\pi}$ belongs to this subcategory only if $\text{Ker}(2_3)$ is in $\text{Ker}(\pi)$. From the character table of $\DZ_3 \rtimes Q_{16}$, we find three representations satisfying this constraint: $1$, $1_3$ and $2_3$. Here $1$ is the trivial representation and $1_3$ is the third 1-dimensional representation in the character table. Hence, the anyons contained in the fusion subcategory  $\CS(\DZ_1, \text{Ker}(2_3), 1)$ are the Wilson lines $\CW_{1}$, $\CW_{1_3}$ as well as $\CW_{2_3}$.  Moreover, we can check the following

\be
1_3 \times 1_3 = 1; ~~ 1_3 \times 2_3 = 2_3; ~~ 
2_3 \times 2_3 = 1 + 1_3 + 2_3.
\ee

Now let us consider a fusion subcategory corresponding to the triple $\CS(M_{f_3},\text{Ker}(1_2),1)$ where $M_{f_3}$ is the normal subgroup generated by the elements of the conjugacy class $[f_3]$ and $1_2$ is the second 1 dimensional representation in the character table of $\DZ_3 \rtimes Q_{16}$. We have $M_{f_3}=\{e, f_3, f_4, f_3\cdot f_4\}$.  A Wilson line $\CW_{\pi}$ belongs to the set of generators of this subcategory only if $\text{Ker}(1_2)$ is in $\text{Ker}(\pi)$. Using the character table we can check that there are only two representations which satisfy this constraint: $1$ and $1_2$. Moreover, we have $1_2 \times 1_2=1$. Hence, the Wilson lines in  $\CS(M_{f_3}, \text{Ker}(1_2), 1)$ are $\CW_{1}$ and $\CW_{1_2}$. Note that the flux line $\mu_{[f_3]}$ belongs to this subcategory. 

Hence, we have two fusion subcategories $\CS(\DZ_1, \text{Ker}(2_3)),1)$ and $(M_{f_3}, \text{Ker}(1_2), 1)$  with the following structure
\bea
\CW_{2_3} \in  \CS(\DZ_1, \text{Ker}(2_3)),1); ~~ \mu_{[f_3]} \in  (M_{f_3}, \text{Ker}(1_2), 1);  \nonumber \\
\CS(\DZ_1, \text{Ker}(2_3)),1) \cap   \CS(M_{f_3}, \text{Ker}(1_2), 1)=\{\CW_{1}\}
\eea
Therefore, the fusions $\CW_{2_3} \times \overline{\CW}_{2_3}$ and $\mu_{[f_3]}\times \overline{\mu}_{[f_3]}$ have only $\CW_{1}$ in common. This trivial intersection explains the fusion \eqref{wfZ3Q16} and gives an example of the idea behind question (2) in the introduction.

\subsubsection{$\CZ(\text{Vec}_{\DZ_{15} \rtimes \DZ_4})$}\label{exZ15Z4}

Let us consider the $\mathbb{Z}_{15}\rtimes\mathbb{Z}_4$ discrete gauge theory. Since the center of the gauge group is trivial and the group involves a semi-direct product, we know from theorem 7 that this gauge theory is prime. 

This group has a length $5$ conjugacy class labelled by the element $f_2$ and a length $2$ conjugacy class labelled by the element $f_3$  (here we are using the notation of GAP, where this group is entry $(60,7)$ in GAP's small group library).  We also have a length $10$ conjugacy class labeled by $f_2f_3$. It is therefore clear that we have a fusion of flux lines giving a unique outcome corresponding to these conjugacy classes
\begin{equation}
\mu_{[f_2]}\times\mu_{[f_3]}=\mu_{[f_2f_3]}~.
\end{equation}

Based on our discussion above, let us consider the groups $M_{f_2}$ and $M_{f_3}$ generated by the elements in the corresponding conjugacy class. It is not too hard to show that
\bea 
M_{f_2}&=& [e] \cup [f_2] \\
M_{f_3}&=& [e] \cup [f_3] \cup [f_4]
\eea
Hence, the fusion subcategories $\CS(M_{f_2}, M_{f_3}, 1)$ and $\CS(M_{f_3},M_{f_2},1)$ can only have Wilson lines as common elements. The trivial Wilson line $\CW_1$ is of course a common element. As we saw in section \ref{subcat}, a Wilson line, $\CW_{\pi}$, is a member of the fusion subcategory, $\CS(M_{f_2}, M_{f_3}, 1)$, only if the condition
\be
\chi_{\pi}(h):= B(e,h) \text{ deg } \chi_{\pi}= \text{ deg } \chi_{\pi}~, ~ \ \ \forall ~ h \in M_{f_3} ~,
\ee 
is satisfied. Hence, $M_{f_3}$ should be in the kernel of $\chi_{\pi}$. Similarly, a Wilson line $\CW_{\pi'}$, is a member of $(M_{f_3}, M_{f_2}, 1)$ only if $M_{f_2}$ is in the kernel of $\chi_{\pi'}$. Therefore, the common elements of the two fusion subcategories are given by the Wilson lines $W_{\tilde\pi}$ for which $M_{f_2}$ and $M_{f3}$ are in the kernel of $\chi_{\tilde\pi}$. Using the character table of $\DZ_{15} \rtimes \DZ_{2}$, we find that there is only one representation $\pi_{1_2}$, which satisfies this constraint. 

Consider the fusions
\bea
\mu_{[f_2]} \times \mu_{[{f_2^{-1}}]}&=& \CW_1 + \cdots~, \\
\mu_{[f_3]} \times \mu_{[{f_3}^{-1}]}&=& \CW_1 + \cdots~. 
\eea
We know $\mu_{[f_2]}$ and $\mu_{[f_3]}$ belong to the fusion subcategories $(M_{f_2},M_{f_3},1)$ and $(M_{f_3},M_{f_2},1)$. Therefore, the only anyons common to both fusions above are $\CW_1$ and $\CW_{1_2}$. 
We would like to know whether the Wilson line, $\CW_{1_2}$, appears on the right hand side of these fusions. To that end, consider the fusion
\be
\CW_{1_2}\times \mu_{[f_3]}=\CL_{([f_3],1_2|_{N_{f_3}})}~.
\ee
It turns out that $1_2|_{N_{f_3}}$ is the trivial representation of $N_{f_3}$. Hence, $\mu_{[f_3]}$ is fixed under fusion with the one-form symmetry generator, $\CW_{1_2}$. So it is clear that $\CW_{1_2}$ should appear in the fusion $\mu_{[f_3]} \times \mu_{[{f_3}^{-1}]}$. Similarly, consider the fusion 
\be
 \CW_{1_2}\times \mu_{[f_2]}=\CL_{([f_2],1_2|_{N_{f_2}})}~.
\ee
It is easy to check that $1_2|_{N_{f_2}}$ is a non-trivial representation of $N_{f_2}$. Hence, $\mu_{[f_2]}$ is not fixed under the fusion with $\CW_{1_2}$. Since $\CW_{1_2}$ is an order two anyon, it cannot appear in the fusion  $\mu_{[f_2]} \times \mu_{[{f_2}^{-1}]}$ (because if $\CW_{1_2} \subset \mu_{[f_2]} \times\mu_{[{f_2}^{-1}]}$, then multiplying both sides on the left with $\CW_{1_2}$ implies that $\CL_{([f_2],1_2)}$ is the inverse of $\mu_{[f_2]}$ which is clearly false). 

We have that the fusions $\mu_{[f_2]} \times\mu_{[f_2^{-1}]}$ and $\mu_{[f_3]} \times\mu_{[f_3^{-1}]}$ only have the trivial anyon in common. Hence, the combination of subcategory structure and one-form symmetry explains the fusion rule
\be 
\mu_{[f_2]} \times \mu_{[f_3]}= \mu_{[f_2f_3]}~.
\ee
It is interesting to note that this discussion parallels the one for Wilson lines in section \ref{unfaithful}. 

This example is additionally illuminating because this theory also has a fusion involving a Wilson and  a flux line with unique outcome. Indeed, we have two 2-dimensional representations $2_1$ and $2_2$ of $\DZ_{15} \rtimes \DZ_{4}$ whose restriction to the centralizer $N_{f_2}= \DZ_{3} \rtimes \DZ_{4}$ are irreducible. Hence, we have the fusion rules
\be
\CW_{2_1} \times\mu_{[f_2]}=\CL_{([f_2],2_1|_{N_{f_2}})}~,\ \ \ \CW_{2_2}\times\mu_{[f_2]}=\CL_{([f_2],2_2|_{N_{f_2}})} ~.
\ee
Do we have trivial braiding between the anyons involved in this fusion? This question is equivalent to whether the dyons are bosons are not. For $\CL_{([f_2],2_i|_{N_{f_2}})}$ to be a boson, we want $f_2$ to be in the kernel of $2_i|_{f_2}$, which is equivalent to the condition that $f_2$ be in the kernel of $2_i$. Using this condition, we can easily check to see that the anyons $\CW_{2_1}$ and $\mu_{[f_2]}$ braid non-trivially with each other, while $\CW_{2_2}$ and $\mu_{[f_2]}$ braid trivially with each other.

Moreover, this theory has several fusions involving dyons which give a unique output. For example, consider the dyons $\CL_{([f_2], \tilde{1}_{f_2})}$ and $\CL_{([f_3], \tilde{1}_{f_3})}$, where $\tilde{1}_{f_2}$ and $\tilde{1}_{f_3}$ are the unique non-trivial real 1-dimensional representations of $N_{f_2}=\DZ_3 \rtimes \DZ_4$ and $N_{f_3}=\DZ_3 \times D_{10}$, respectively. We have the fusion
\be
\CL_{([f_2],\tilde{1}_{f_2})} \times \CL_{([f_3],\tilde{1}_{f_3})}= \CL_{([f_2f_3],\tilde{1}_{f_2f_3})}
\ee
where $\tilde{1}_{f_2f_3}$ is the unique non-trivial 1-dimensional representation of $N_{f_2f_3}=\DZ_6$.

Let us also explore the zero-form symmetry of this theory. We have $\text{Out}(\DZ_{15} \rtimes \DZ_4)=\DZ_2$ and $H^2(\DZ_{15} \rtimes \DZ_4)=\DZ_1$. From theorem 10, we know that this theory features non-trivial self-duality. In fact, the group $\DZ_{15} \rtimes \DZ_4$ has three non-trivial normal abelian subgroups $\DZ_{3}, \DZ_5, \DZ_{15}$ all of which have trivial $2^{\text{nd}}$ cohomology group. So we have the Lagrangian subcategories
\be
\{\CL_{(\DZ_1,1)},\CL_{(\DZ_3,1)},\CL_{(\DZ_5,1)},\CL_{(\DZ_{15},1)}\}
\ee
Using remark 7.3 in \cite{nikshych2014categorical}, we have
\be
\CL_{(N,1)}\simeq \text{Rep}((\DZ_{15} \rtimes \DZ_4)_{(N,1)}) \simeq \hat{N} \rtimes (\DZ_{15} \rtimes \DZ_4)/\hat{N}
\ee
where $\hat{N}$ is the group of representations of $N$ and $N=\DZ_3,\DZ_5,\DZ_{15}$. Also, we have the isomorphisms
\be
\DZ_{15} \rtimes \DZ_{4} \simeq \DZ_{3} \rtimes (\DZ_5 \rtimes \DZ_4) \simeq \DZ_5 \rtimes (\DZ_3 \rtimes \DZ_4)
\ee
Hence, all Lagrangian subcategories above are isomorphic to $\text{Rep}(\DZ_{15} \rtimes \DZ_{4})$. Hence, $|\mathbb{L}_0(\DZ_{15} \rtimes \DZ_4)|=4$. From theorem 9, we know that ${\rm Aut}^{\rm br}(\CZ({\rm Vec}_{\mathbb{Z}_{15}\rtimes\mathbb{Z}_4}))$ should act transitively on $|\mathbb{L}(\DZ_{15}\rtimes \DZ_4)|$. In fact, we can use proposition 7.11 of \cite{nikshych2014categorical} to show that $H^2(\DZ_{15} \rtimes \DZ_4,U(1)) \rtimes \text{Out}(\DZ_{15}\rtimes \DZ_{4})\simeq \DZ_{2}$ acts trivially on $|\mathbb{L}_0(\DZ_{15}\rtimes \DZ_4)|$. Using theorem 9, we have $|{\rm Aut}^{\rm br}(\CZ({\rm Vec}_{\mathbb{Z}_{15}\rtimes\mathbb{Z}_4}))|=8$. 

Finally, since $\DZ_{15} \rtimes \DZ_4$ has complex characters, $\CZ(\text{Vec}^{\omega}_{\DZ_{15} \rtimes \DZ_4})$ has a non-trivial quasi-zero-form symmetry given by charge conjugation.

\subsubsection{Symmetry and quasi-symmetry searches}

We have used the software GAP to search for groups for which the corresponding untwisted discrete gauge theories have fusions rules with unique outcomes. We present our results below. The relevant GAP code is given in Appendix \ref{GAPapp}.

\bigskip
\noindent {\bf Fusion of Wilson lines}
\bigskip

\noindent Irreducible representations of a direct product of groups is the product of representations of the individual groups. Hence, it is natural that the first example with two Wilson lines fusing to give a unique Wilson line is the quantum double of $S_3 \times S_3$ (however, this fusion arises because the discrete gauge theory factorizes; this follows from theorem 6). More interesting (non-direct-product) groups with this property only appear at order 48 (see Appendix \ref{appA}). For groups of order less than or equal to 639 (except orders 384, 512, 576)\footnote{We have not checked order 384, 512, 576 due to the huge number of groups (up to isomorphism) with these orders.} we have verified that whenever the corresponding untwisted discrete gauge theory has a fusion Wilson lines giving a unique outcome, $\text{Aut}^{\rm br}\CZ(\text{Vec}_G)$ is non-trivial. In this set of groups, there are two which have a trivial automorphism group. They are $S_3 \times (\DZ_5 \rtimes \DZ_4)$ and $(((\DZ_3 \times \DZ_3) \rtimes Q_8) \rtimes \DZ_3) \rtimes \DZ_2$. However, $H^2(S_3 \times (\DZ_5 \rtimes \DZ_4),U(1))=\DZ_2$ leading to non-trivial $\text{Aut}^{\rm br}(\CZ(\text{Vec}_{S_3 \times (\DZ_5 \rtimes \DZ_4)}))$. The group $(((\DZ_3 \times \DZ_3) \rtimes Q_8) \rtimes \DZ_3) \rtimes \DZ_2$ has trivial $H^2(G,U(1))$. So the theory $\CZ(\text{Vec}_{(((\DZ_3 \times \DZ_3) \rtimes Q_8) \rtimes \DZ_3) \rtimes \DZ_2})$ doesn't have classical symmetries. $(((\DZ_3 \times \DZ_3) \rtimes Q_8) \rtimes \DZ_3) \rtimes \DZ_2$ has only one abelian normal subgroup $N=\DZ_3 \times \DZ_3$. Moreover, we have $(((\DZ_3 \times \DZ_3) \rtimes Q_8) \rtimes \DZ_3) \rtimes \DZ_2 \simeq N \rtimes K$ where $K=GL(2,3)$. Therefore, using theorem 10, we know that this theory has non-trivial electric-magnetic duality. 
The groups $S_3 \times (\DZ_5 \rtimes \DZ_4)$ and $(((\DZ_3 \times \DZ_3) \rtimes Q_8) \rtimes \DZ_3) \rtimes \DZ_2$ have complex characters, hence the corresponding discrete gauge theories have quasi-zero-form symmetries.

\bigskip
\noindent {\bf Fusion of flux lines}
\bigskip

\noindent The simplest example of an untwisted discrete gauge theory with a fusion of two flux lines giving a single outcome is $\CZ ({\rm Vec}_{S_3 \times S_3})$. The conjugacy classes of a direct product is a product of conjugacy classes of the individual groups. Hence, it follows that quantum doubles of direct products naturally have such fusions. As mentioned above, it follows from theorem 6 that discrete gauge theories based on direct product groups are non-prime. Therefore, the fusion rules with unique outcome in this case are a consequence of the Deligne product. Since $\text{Out}(S_3 \times S_3)=\DZ_2$, $\CZ ({\rm Vec}_{S_3 \times S_3})$ has non-trivial zero-form symmetry.

\noindent After $S_3 \times S_3$, we have several groups of order 48 with flux fusions giving unique outcome. The examples discussed in Appendix \ref{appA} (except $BOG$ and $GL(2,3)$) exhaust all such groups of order 48. All of these groups have non-trivial automorphism group, and hence the corresponding discrete gauge theory has non-trivial symmetries. In fact, for groups of order less than or equal to 639 (except orders 384, 512, 576) we have verified that whenever the corresponding untwisted discrete gauge theory has a fusion of flux lines with a unique outcome, $\text{Aut}^{\rm br}(\CZ(\text{Vec}_G))$ is non-trivial. In fact, the only group with a trivial automorphism group in this set is $S_3 \times (\DZ_5 \rtimes \DZ_4)$. We already discussed above that this theory has non-trivial zero-form symmetries as well as non-trivial quasi-zero-form symmetries. 

\bigskip
\noindent {\bf Fusion of a Wilson line with a flux line}
\bigskip

\noindent The simplest example with a fusion of a Wilson line and a flux line giving a single outcome is $\CZ(\text{Vec}_{S_3 \times S_3})$. Then we have more examples in order $48$. The examples discussed in Appendix \ref{appA} (except $BOG$ and $GL(2,3)$) exhausts all such groups of order 48.
For groups of order less than or equal to 639 (except orders 384, 512, 576) we have verified that whenever the corresponding untwisted discrete gauge theory has a fusion of a Wilson line with a flux line giving a unique outcome, $\text{Aut}^{\rm br}\CZ(\text{Vec}_G)$ is non-trivial. In this set of groups, there are three which have a trivial automorphism group. They are $S_3 \times (\DZ_5 \rtimes \DZ_4)$, $(\DZ_3 \times \DZ_3) \rtimes QD_{16}$ (where $QD_{16}$ is the semi-dihedral group of order 16) and $(((\DZ_3 \times \DZ_3) \rtimes Q_8) \rtimes \DZ_3) \rtimes \DZ_2$. We discussed the groups $S_3 \times (\DZ_5 \rtimes \DZ_4)$ and $(((\DZ_3 \times \DZ_3) \rtimes Q_8) \rtimes \DZ_3) \rtimes \DZ_2$ above. The group $(\DZ_3 \times \DZ_3) \rtimes QD_{16}$ has trivial $H^2(G,U(1))$. So the theory $\CZ(\text{Vec}_{(\DZ_3 \times \DZ_3) \rtimes QD_{16}})$ doesn't have classical symmetries. However, $(\DZ_3 \times \DZ_3) \rtimes QD_{16}$ has one abelian normal subgroup $N=\DZ_3 \times \DZ_3$. Moreover, we have $(\DZ_3 \times \DZ_3) \rtimes QD_{16} \simeq N \rtimes K$ where $K=QD_{16}$. Therefore, using theorem 10, we know that the corresponding untwisted discrete gauge theory has non-trivial electric-magnetic self-duality. 

The group $(\DZ_3 \times \DZ_3) \rtimes QD_{16}$ has complex characters, hence the corresponding discrete gauge theory has quasi-zero-form symmetries.

\bigskip
\noindent {\bf Fusion of general dyons}
\bigskip

Being a Deligne product, $\CZ({\rm Vec}_{S_3 \times S_3})$ also has fusions involving dyons, and this is the smallest rank theory with such fusions. The next example is in order $48$. The examples discussed in Appendix \ref{appA} exhausts all such groups of order 48. For groups of order less than or equal to 100 we have verified that whenever the corresponding untwisted discrete gauge theory has a fusion of two dyons giving a unique outcome, $\text{Aut}^{\rm br}\CZ(\text{Vec}_G)$ is non-trivial. In fact, every group in this set has non-trivial automorphism group. Hence, they all have non-trivial classical 0-form symmetries. 

\section{$G_k$, cosets, and $a\times b=c$}\label{cosets}
In this section, we turn our attention to a (generally) very different set of theories: TQFTs based on $G_k$ Chern-Simons (CS) theories and cosets thereof (here $G$ is a compact simple Lie group). Unlike the theories discussed in section \ref{discrete}, the theories we discuss here are typically chiral (i.e., $c_{\rm top}\ne0\ ({\rm mod}\ 8)$).

In order to gain a sense of what such theories allow us to do in constructing TQFTs with fusion rules of the form \eqref{abcfusion} and \eqref{aabar},  it is useful to recall the basic representation theory of $SU(2)$. Somewhat surprisingly, this intuition will be quite useful for more general $SU(N)_k$ CS theories. To that end, consider the textbook matter of the fusion of $SU(2)$ spin $j_1$ and $j_2$ representations
\begin{equation}\label{spinadd}
j_1\otimes j_2=\sum_{j=|j_1-j_2|}^{j_1+j_2}j~.
\end{equation}
As in the case of the finite groups in the previous section, we would like to understand if we can have $j_1\otimes j_2=j_3$ for $j_1,j_2>0$ and fixed $j_3$ spin. Clearly this is impossible, since we would have $j_1+j_2>|j_1-j_2|$ and the sum \eqref{spinadd} will have at least two contributions.

While this result is rather trivial, it is useful to recast it using the group theory analog of the $F$-transformation described in the introduction (as well as in section \ref{discrete} for the case of discrete groups). To that end, we wish to consider
\begin{equation}\label{aabarSU2}
j_1\otimes j_1=\sum_{j=0}^{2j_1}j~, \ \ \ j_2\otimes j_2=\sum_{k=0}^{2j_2}k~, \ \ \ |j_{1,2}|>1~,
\end{equation}
where $|j_{1,2}|$ are the dimensions of the representations. In particular, we see that (since $j_{1,2}>0$) both products in \eqref{aabarSU2} must always contain the trivial representation and the adjoint representation. This observation also implies that $j_1\otimes j_2\ne j_3$ for fixed $j_3$ spin.

The discussion around \eqref{aabarSU2} easily generalizes to arbitrary compact simple Lie group, $G$. In particular, let us consider
\begin{equation}\label{suNFtr}
\alpha\otimes\bar\alpha=1+\sum_{\gamma\ne1}N_{\alpha\bar\alpha}^{\gamma}\ \gamma~, \ \ \ \beta\otimes \bar\beta=1+\sum_{\delta}N_{\beta\bar\beta}^{\delta}\ \delta~,\ \ \ |\alpha|,\ |\beta|>1~,
\end{equation}
where $\alpha, \beta$ and $\bar\alpha, \bar\beta$ are conjugate higher-dimensional irreducible representations of $G$, ${\rm Irr}(G)$. The number of times the adjoint appears in the product $\alpha\otimes\bar\alpha$ is \cite{BZ:1989}:
\begin{equation}
N_{\alpha\bar\alpha}^{{\rm adj}}=\left|\left\{\lambda^{(\alpha)}_j\ne0\right\}\right|\ge1~,
\end{equation}
where $\lambda_j^{(\alpha)}$ are the Dynkin labels of $\alpha$. Therefore, we learn that for all higher-dimensional representations of $G$ 
\begin{equation}
\alpha\otimes\beta\ne\gamma~,\ \forall\ |\alpha|,\ |\beta|>1~, \ \alpha,\beta,\gamma\in{\rm Irr}(G)~,
\end{equation}

Of course, our interest is in the fusion algebra of $G_k$. From this perspective, the above discussion is in the limit $k\to\infty$. As we will prove in the next section, taking $G_k=SU(N)_k$ and imposing finite level does not lead to fusions of the form \eqref{abcfusion} or \eqref{aabar}.

\subsection{$G_k$ CS theory}\label{SUNcs}
Let us now consider the finite-level deformation of the fusion rules discussed in the previous section. These are the fusion rules of Wilson lines in $G_k$ CS theory. We first consider $SU(2)_k$ as it is rather illustrative. We will then generalize to $SU(N)_k$ and comment on more general $G_k$.

In the case of $SU(2)_k$, \eqref{spinadd} becomes \cite{Gepner:1986wi,DiFrancesco:1997nk}
\begin{equation}
j_1\otimes j_2=\sum_{j=|j_1-j_2|}^{{\rm min}(j_1+j_2,k-j_1-j_2)}j~.
\end{equation}
In addition to truncating the spectrum to the spins $\left\{0,1/2,1,\cdots,k/2\right\}$, the above deformation abelianizes the spin $k/2$ representation (since $k/2\otimes k/2=0$). However, these changes do not alter the conclusion from the previous section: we cannot write $j_1\otimes j_2=j_3$ for $j_3$ non-abelian irreducible $j_{1,2,3}$. Indeed, consider
\begin{equation}\label{su2k}
j_1\otimes j_1=\sum_{j=0}^{{\rm min}(2j_1,k-2j_1)}j~, \ \ \ j_2\otimes j_2=\sum_{j=0}^{{\rm min}(2j_2,k-2j_2)}j~, \ \ \  j_{1,2}\ne0,\ {k\over2}~.
\end{equation}
The conditions $j_{1,2}\ne0,\ {k\over2}$ are to ensure that the representation is non-abelian. In particular, we again see that the adjoint representation appears in \eqref{su2k}.

While the fusion rules discussed in \cite{Gepner:1986wi,DiFrancesco:1997nk} apply to more general groups, they are rather difficult to implement. Instead, using proposals suggested in \cite{Kirillov:1992np,Kirillov:1992jc} and finally proven in \cite{Feingold:2007wz}, the authors of \cite{Urichuk:2016xau} show that for $\alpha$ an irreducible representation of $G_k$ (with $G$ a compact simple Lie group), we have
\begin{equation}
\label{fusadjconst}
{}^{(k)}N_{\alpha\bar\alpha}^{\rm adj}=\left|\left\{\hat\lambda^{(\alpha)}_j\ne0\right\}\right|-1~,
\end{equation}
where $\hat\lambda^{\alpha}_j$ are the associated affine Dynkin labels.

In particular, for $SU(N)_k$, if $|\alpha|>1$, then ${}^{(k)}N_{\alpha\bar\alpha}^{\rm adj}\ge1$.\footnote{Here we define $|\alpha|$ to be the quantum dimension.} Indeed, the abelian representations, $\gamma_i$, satisfy a $\mathbb{Z}_N$ fusion algebra and are characterized by $\hat{\lambda}^{(\gamma_i)}_j=k\delta_{ij}$, where $i\in\left\{0,1,...,N-1\right\}$. On the other hand, all non-abelian representations have at least two non-zero Dynkin labels. As a result, we learn that 
\begin{equation}
\alpha\otimes\beta\ne\gamma~,\ \ \ \forall \alpha,\beta,\gamma\in{\rm Irr}(SU(N)_k)~,\ \ \ |\alpha|,|\beta|>1~.
\end{equation}
Therefore, we see that we have the following fusions for non-abelian Wilson lines in $SU(N)_k$ CS TQFT
\begin{equation}\label{Gkfusion}
\CW_{\alpha}\times\CW_{\beta}=\CW_{\gamma}+\cdots~,\ \ \ |\alpha|,|\beta|>1~,
\end{equation}
where the ellipses necessarily include additional Wilson lines. This statement is more generally true in any $G_k$ CS theory (with $G$ a compact and simple Lie group) for which the lines in question correspond to affine representations with at least two non-zero Dynkin labels.

Note that for certain $G_k$, non-abelian representations can have a single non-vanishing Dynkin label. For example, consider the $(E_7)_2$ CS theory.\footnote{We thank a referee for pointing out this example.} It has Wilson lines $\CW_{\tau}$ and $\CW_{\sigma}$ with quantum dimensions $\frac{1+\sqrt{5}}{2}$ and $\sqrt{2}$, repectively, and they fuse to give a unique outcome. The existence of this fusion rule follows from the fact that $(E_7)_2$ is not a prime TQFT. In fact, it resolves into the product of prime theories $\text{Fib} \boxtimes \text{Ising}'$, where $\text{Fib}$ is the Fibonacci anyon theory and $\text{Ising}'$ is a TQFT with the same fusion rules as the the Ising model.    

We can apply the above arguments to learn about global properties of $G_k$ CS theory. For example, we can ask if $G_k$ CS theory is prime or not. The answer is no in general. Indeed, consider the case $G=SU(2)$. For $k\in\mathbb{N}_{\rm even}$, $SU(2)_k$ is prime. However, for $k\in\mathbb{N}_{\rm odd}$, the abelian anyon generating the $\mathbb{Z}_2$ one-form symmetry forms a modular subcategory. By M\"uger's theorem \cite{muger2003structure} (see also \cite{Intriligator:1989zw} for a discussion at the level of RCFT), it then decouples and the theory resolves into a product of two prime theories
\begin{equation}\label{su2fact}
SU(2)_k\simeq \begin{cases}
SU(2)_1\boxtimes SU(2)_k^{\rm int}~, &\text{if}\ k=1 \ {(\rm mod 4)} \\
\overline{SU(2)_1}\boxtimes SU(2)_k^{\rm int}~, &\text{if}\ k=3 \ {(\rm mod 4)}~.
\end{cases}
\end{equation}
where $SU(2)_k^{\rm int}$ is a TQFT built out of the integer spin $SU(2)_k$ representations. Here $\overline{SU(2)_1}$ is the TQFT conjugate to $SU(2)_1$ (these TQFTs are sometimes called the anti-semion and semion theories in the condensed matter literature).

While $G_k$ CS theory is not prime in general, our arguments above readily prove the following:

\medskip
\noindent
{\bf Claim 14:} Non-abelian Wilson lines in $SU(N)_k$ CS theory must all lie in the same prime TQFT factor. For more general $G_k$ CS theory (with $G$ compact and simple), all Wilson lines corresponding to affine representations with at least two non-zero Dynkin labels must be part of the same prime TQFT factor.

\medskip
\noindent
{\bf Proof:} Suppose this were not the case. Then, we would find fusion rules of the form \eqref{Gkfusion} with {\it no} Wilson lines in the ellipses. $\square$

\medskip
\noindent
Clearly, to produce fusion rules of the form \eqref{abcfusion} for non-abelian Wilson lines in the same prime TQFT, we will need to go beyond $SU(N)_k$ CS theory. One way to proceed is to consider coset theories and use some intuition from section \ref{discrete}. Indeed, since cosets can have fixed points (which we will describe below), it is natural to think they can lead to fusion rules of the form \eqref{abcfusion}.

\subsection{Virasoro minimal models and some cosets without fixed points}\label{vir}
We begin with a discussion of the Virasoro minimal models, as these are simple examples of theories that are related to cosets. While these cosets do not have fixed points, they turn out to produce factorized TQFTs that are nonetheless illustrative. In the next section, we will focus on cosets that have fixed points, and we will see how to engineer fusion rules of the form \eqref{abcfusion}.

One way to construct the Virasoro minimal models is to take a three-dimensional spacetime $\mathbb{R}\times\Sigma$ and place $SU(2)_{k-1}\times SU(2)_1$ CS theory on $I\times\Sigma$, where $I$ is an interval in $\mathbb{R}$. We can place $SU(2)_k$ CS theory outside this region. At the two $1+1$ dimensional interfaces between the CS theories (which form two copies of $\Sigma$, call them $\Sigma_{1,2}$), we obtain the left and right movers of the RCFT. Here the chiral (anti-chiral) primaries lie where endpoints of Wilson lines from the $SU(2)_k$ and $SU(2)_{k-1}\times SU(2)_1$ theories meet on $\Sigma_1$ ($\Sigma_2$).

Another way to think about the Wilson lines related to the Virasoro minimal models is to start with $SU(2)_{k-1}\times SU(2)_1$ CS theory and change variables to make an $SU(2)_k$ subsector manifest \cite{Isidro:1991fp}. Integrating this sector out leaves an effective coset TQFT.

The end result is that the TQFT we are interested in is\footnote{This is the TQFT analog of the classic result \cite{Goddard:1984vk} for the corresponding affine algebras: 
\begin{equation}
{\rm Vir}_p\simeq{\widehat{\mathfrak{su}}(2)_{p-2}\times\widehat{\mathfrak{su}}(2)_1\over\widehat{\mathfrak{su}}{(2)_{p-1}}}~.
\end{equation}
}
\begin{equation}\label{vircoset}
\CT_{p}={SU(2)_{p-2}\boxtimes SU(2)_1\over SU(2)_{p-1}}~,\ \ \ p\ge3~.
\end{equation}
Here, the natural number $p\ge3$ labels the corresponding Virasoro minimal model (so, for example, $p=3$ for the Ising model).\footnote{In writing \eqref{vircoset}, we have used the Deligne product to emphasize the fact that the $SU(2)_{p-2}\times SU(2)_1$ CS theory is a product TQFT.} We may construct the MTC data underlying the RCFT and the coset TQFT by taking products (e.g., see \cite{Ramadevi:1993np})
\begin{equation}\label{factFR}
F_{\CT_{p}}=F_{SU(2)_{p-2}}\cdot F_{SU(2)_1}\cdot \bar F_{SU(2)_{p-1}}~,\ \ \ R_{\CT_p}=R_{SU(2)_{p-2}}\cdot R_{SU(2)_1}\cdot \bar R_{SU(2)_{p-1}}~.
\end{equation}
In order to make \eqref{factFR} precise, we need to explain how the states in $\CT_p$ are related to those in the individual $SU(2)_k$ theories that make up the coset. Let us denote the $SU(2)_{p-2}$, $SU(2)_1$, and $SU(2)_{p-1}$ weights as $\lambda$, $\mu$, and $\nu$. Then, to build the coset we should identify Wilson lines as follows
\begin{equation}\label{abID}
\CW_{\{\lambda,\mu,\nu\}}:=\CW_{\lambda}\times\CW_{\mu}\times\CW_{\nu}\simeq\left(\CW_{p-2}\times\CW_{\lambda}\right)\times\left(\CW_1\times\CW_{\mu}\right)\times\left(\CW_{p-1}\times\CW_{\nu}\right)~,
\end{equation}
where $\CW_{p-2}$, $\CW_1$, and $\CW_{p-1}$ are abelian Wilson lines transforming in the weight $p-2$ (spin $(p-2)/2$), weight $1$ (spin $1/2$), and weight $p-1$ (spin $(p-1)/2$) representations of the different TQFT factors.\footnote{At the level of the corresponding affine algebras, this is the statement that \cite{DiFrancesco:1997nk}
\begin{equation}
\left\{\hat\lambda,\hat\mu,\hat\nu\right\}\simeq\left\{a\hat\lambda,a\hat\mu,a\hat\nu\right\}~,
\end{equation}
where the hat denotes affine weights, and $a$ is the generator of the (diagonal) $\CO(\widehat{\mathfrak{su}}(2))$ outer automorphism.} Moreover, in order to be a valid Wilson line in $\CT_p$, we should demand that our Wilson lines satisfy
\begin{equation}\label{rootconst}
\CW_{\left\{\lambda,\mu,\nu\right\}}\in\CT_p\ \Leftrightarrow\ \lambda+\mu-\nu\in Q\ \Leftrightarrow\ \lambda+\mu+\nu=0\ ({\rm mod}\ 2)~,
\end{equation}
where $Q$ is the $SU(2)$ root lattice. This relation guarantees that all lines that remain have trivial braiding with $\CW_{\left\{p-2,1,p-1\right\}}$ (which is a boson that is in turn identified with the vacuum). It is in terms of these degrees of freedom that \eqref{factFR} should be understood.

Before proceeding, let us stop and note that the fusion in \eqref{abID} has no fixed points. Indeed, this statement readily follows from the fact that $SU(2)_1$ is an abelian TQFT, and abelian theories cannot have fixed points since their fusion rules are those of a finite abelian group (in this case $\mathbb{Z}_2$).

Given this groundwork, we claim that $\CT_p$ factorizes as follows
\begin{equation}\label{VirFact}
\CT_p\simeq\begin{cases}
\left(SU(2)_{p-2}\boxtimes SU(2)_1\right)^{\rm int}\boxtimes SU(2)_{p-1}^{\rm int}~, &\text{if}\ p=0\ {(\rm mod\ 2)} \\
SU(2)_{p-2}^{\rm int}\boxtimes SU(2)_{p-1}^{\rm conj}~, &\text{if}\ p=1 \ {(\rm mod\ 2)}~.
\end{cases}
\end{equation}
The various TQFTs appearing in \eqref{VirFact} are 
\begin{eqnarray}\label{compVir}
\left(SU(2)_{p-2}\boxtimes SU(2)_1\right)^{\rm int}&:=&{\rm gen}\left(\left\{\CW_{\left\{\lambda,\mu\right\}}\in SU(2)_{p-2}\boxtimes SU(2)_1|\ \lambda+\mu=0\ {\rm(mod\ 2)}\right\}\right)~,\cr SU(2)_{p-1}^{\rm int}&:=&{\rm gen}\left(\left\{\CW_{\nu}\in SU(2)_{p-1}|\ \nu=0\ {\rm(mod\ 2)}\right\}\right)~, \cr SU(2)_{p-1}^{\rm conj}&:=&{\rm gen}\left(\left\{\CW_{\left\{\lambda,\mu,\nu\right\}}|\ \lambda+\mu+\nu=0\ {\rm(mod\ 2)~,\ \CW_{\lambda}, \CW_{\mu}\ {\rm abelian}}\right\}\right)~,\cr SU(2)_{p-2}^{\rm int}&:=&{\rm gen}\left(\left\{\CW_{\lambda}\in SU(2)_{p-2}|\ \lambda=0\ {\rm(mod\ 2)}\right\}\right)~,
\end{eqnarray}
where \lq\lq${\rm gen(\cdots)}$" means that the TQFT is generated by the Wilson lines enclosed. Notice that in the case that $p$ is even, $p-1$ is odd and $SU(2)_{p-1}^{\rm int}$ is precisely the decoupled TQFT factor required by M\"uger's theorem in \eqref{su2fact} containing integer spins (even Dynkin labels). Similar logic applies to $SU(2)_{p-2}^{\rm int}$ in the case that $p$ is odd. The TQFT $SU(2)_{p-1}^{\rm conj}$ has the same fusion rules as $SU(2)_{p-1}$, but it is a different TQFT. Finally, for the case that $p=3$ (i.e., the Ising model), we see that $\CT_3$ does not factorize.\footnote{Note also that Ising shares the same fusion rules as $SU(2)_2$, though they are not the same TQFTs. For example, the $\sigma$ fields have different twists.}

Our strategy to prove the factorization in \eqref{VirFact} is to construct the various factors and then argue that they are well-defined TQFTs by M\"uger's theorem \cite{muger2003structure}. Although we will not pursue it in this paper, this same approach leads to interesting generalizations for cosets built out of groups other than $SU(2)$.

To that end, let us first take the case of $p\ge3$ odd. Using the result in \eqref{factFR}, we have that the modular $S$ matrix also takes a product form
\begin{equation}\label{factST}
S_{\left\{\lambda,\mu,\nu\right\}\left\{\lambda',\mu',\nu'\right\}}=S^{(p-2)}_{\lambda\lambda'}\cdot S^{(1)}_{\mu\mu'}\cdot S^{(p-1)}_{\nu\nu'}~,\ \ \ \theta_{\left\{\lambda,\mu,\nu\right\}\left\{\lambda',\mu',\nu'\right\}}=\theta^{(p-2)}_{\lambda\lambda'}\cdot\theta^{(1)}_{\mu\mu'}\cdot\bar\theta^{(p-1)}_{\nu\nu'}~,
\end{equation}
where the superscripts on the righthand sides of the above equations refer to the corresponding factors in the coset \eqref{vircoset}. From the $S$ matrix, Verlinde's formula yields (see also the discussion in \cite{DiFrancesco:1997nk})
\begin{equation}\label{fusionVir}
N_{\left\{\lambda,\mu,\nu\right\}\left\{\lambda',\mu',\nu'\right\}}^{\left\{\lambda'',\mu'',\nu''\right\}}=N_{\lambda\lambda'}^{(p-2)\lambda''}\cdot N_{\mu\mu'}^{(1)\mu''}\cdot N_{\nu\nu'}^{(p-1)\nu''}~,
\end{equation}
where, again, the superscripts on the righthand side denote the different coset factors in \eqref{vircoset}. The factor $SU(2)_{p-2}^{\rm int}$ in the second line of \eqref{VirFact} is clearly closed under fusion. So too is $SU(2)_{p-1}^{\rm conj}$. To have factorization of the TQFT, we need only show that all Wilson lines can be written in this way and, by M\"uger's theorem, that one of these factors is modular. The second part is trivial: we have already seen that $SU(2)_{p-2}^{\rm int}$ is modular in the discussion surrounding \eqref{su2fact}. We can confirm this statement by looking at the modular $S$-matrix for $SU(2)_{p-2}$
\begin{equation}
S^{(p-2)}_{\lambda\lambda'}=\sqrt{2\over p}\sin\left({(\lambda+1)(\lambda'+1)\pi\over p}\right)~.
\end{equation}
and taking the submatrix involving the integer spins (even weights).

Therefore, we need only check that all states in the coset \eqref{vircoset} can be expressed in this way. To that end, we can see that
\begin{equation}
|SU(2)^{\rm int}_{p-2}|={p-1\over2}~,\ |SU(2)_{p-1}^{\rm conj}|=p~,
\end{equation}
where the norm denotes the number of simple elements within. Therefore, we see that we have $|\CT_p|=p(p-1)/2$, which is precisely the number of states in the coset \eqref{vircoset} (note that in these computations we have used \eqref{abID} and \eqref{rootconst}) and the corresponding $A$-type Virasoro minimal model.

To make contact with the fusion rules in \eqref{virabc}, we need to explain precisely how coset lines map onto the Virasoro primaries. The results above allow us to realize the, say, Virasoro left-movers as states on the boundary of the bulk TQFT, $\CT_p\simeq SU(2)_{p-2}^{\rm int}\boxtimes SU(2)_{p-1}^{\rm conj}$ with $p$ odd. Now, we need to see how we can map boundary endpoints of lines in this theory to Virasoro primaries, $\varphi_{(r,s)}$. To that end, by comparing the $S$-matrix for $\CT_p\simeq SU(2)_{p-2}^{\rm int}\boxtimes SU(2)_{p-1}^{\rm conj}$ with the corresponding expressions for those of the Virasoro minimal models, we have that the labels of the Virasoro primary, $\varphi_{(r,s)}$ map as follows (see also \cite{DiFrancesco:1997nk})
\begin{equation}\label{mappingtoVir}
r=\lambda+1~, \ \ \ s=\nu+1~.
\end{equation}
In particular, we see that the $\varphi_{(r,1)}$ primaries are endpoints of lines in $SU(2)_{p-2}^{\rm int}$ while the $\varphi_{(1,s)}$ are endpoints of lines in $SU(2)_{p-1}^{\rm conj}$. This reasoning explains the fact that non-abelian Virasoro primaries of these types have unique fusion outcomes\footnote{Though, again, we stress that this factorization is not a factorization of RCFT correlators.}
\begin{equation}\label{virabc2}
\varphi_{(r,1)}\times\varphi_{(1,s)}=\varphi_{(r,s)}~,
\end{equation}
discussed in the introduction (at least for $p$ odd). As an example, we have $\CT_3\simeq{\rm Ising}$ (i.e., the TQFT is the Ising MTC), which does not factorize. On the other hand, for $p=5$, we have
\begin{equation}\label{T5}
\CT_5=(G_2)_1\boxtimes SU(2)_4^{\rm conj}~,
\end{equation}
where $(G_2)_1$ is the so-called \lq\lq Fibnonacci" TQFT, and $SU(2)_4^{\rm conj}$ is a TQFT with the same fusion rules and $S$-matrix as $SU(2)_4$.

Let us now consider $p\ge4$ even. The modular data and fusion rules still take a product form as in \eqref{factST} and \eqref{fusionVir}. Now, however, we should examine the first line in \eqref{VirFact}. Using \eqref{fusionVir}, it is again easy to see that both $SU(2)_{p-1}^{\rm int}$ and $(SU(2)_{p-2}\boxtimes SU(2)_1)^{\rm int}$ are separately closed under fusion. Moreover, just as before, we can use the discussion around \eqref{su2fact} and M\"uger's theorem to conclude that $SU(2)_{p-1}^{\rm int}$ is indeed a decoupled TQFT as claimed in \eqref{VirFact}.

We should again check that all states in \eqref{vircoset} can be reproduced. To that end, we have
\begin{equation}
|SU(2)^{\rm int}_{p-1}|={p\over2}~,\ |(SU(2)_{p-2}\boxtimes SU(2)_1)^{\rm int}|=p-1~.
\end{equation}
As a result, we have $|\CT_p|=p(p-1)/2$, which is the correct number of states in the coset \eqref{vircoset} and the corresponding $A$-type Virasoro minimal model. 

Our mapping is again as in \eqref{mappingtoVir}, but now $\varphi_{(r,1)}$ primaries are endpoints of lines in $(SU(2)_{p-2}\boxtimes SU(2)_1)^{\rm int}$, and $\varphi_{1,s}$ are endpoints of lines in $SU(2)_{p-1}^{\rm int}$. This again explains the fusion outcomes in \eqref{virabc2} for the case of $p$ even as well. As an example, note that
\begin{equation}\label{T4}
\CT_4={\rm Ising'} \boxtimes (F_4)_1~,
\end{equation}
where the first factor is a rank three TQFT with the same fusion rules as Ising (and $SU(2)_2$), and the second factor is the time reversal of the Fibonacci theory in \eqref{T5}.

As a result, we conclude that, although the TQFTs discussed in this section do have non-abelian anyons fusing to give a unique outcome, this is due to the fact that the corresponding TQFTs factorize.

\subsection{Beyond Virasoro: cosets with fixed points}\label{fixedpoint}
In section \ref{discrete} we saw that fixed points of various kinds gave rise to fusion rules of the form \eqref{aabar} (in particular, see theorem 1 of section \ref{genlessons}). In the context of cosets, we can also naturally engineer fixed points under the action of fusion with abelian anyons generating identifications of fields. In the case of Virasoro, this didn't happen (see \eqref{abID}). Indeed, this statement followed from the fact that we had an abelian factor in the coset \eqref{vircoset}.

The simplest way to get around this obstacle and generate fixed points is to consider instead
\begin{equation}\label{supercoset}
\widehat{\CT}_p={SU(2)_{p-2}\boxtimes SU(2)_2\over SU(2)_{p}}~,
\end{equation}
where $p\ge3$ (we should take $p\ge4$ to avoid the problem of abelian factors). By further identifying some of these coset fields, we get theories related to the $\CN=1$ super-Virasoro minimal models \cite{Goddard:1986ee,Goddard:1984vk}. Note that the case of $p=3$ corresponds to the $\CT_4$ case discussed previously (i.e., to the TQFT related to the tri-critical Ising model). 

For the theories in \eqref{supercoset}, we find the following generalization of the identification condition in \eqref{abID}\footnote{We also require that $\lambda+\mu+\nu=0\ {\rm (mod\ 2)}$ so that the lines in the coset theory have trivial braiding with the bosonic line $\CW_{\left\{p-2,2,p\right\}}$. This line is in turn identified with the vacuum.}
\begin{eqnarray}\label{abIDN1}
\CW_{\{\lambda,\mu,\nu\}}&:=&\CW_{\lambda}\times\CW_{\mu}\times\CW_{\nu}\simeq\left(\CW_{p-2}\times\CW_{\lambda}\right)\times\left(\CW_2\times\CW_{\mu}\right)\times\left(\CW_{p}\times\CW_{\nu}\right)\cr&=&\CW_{p-2-\lambda}\times\CW_{2-\mu}\times\CW_{p-\nu}~,
\end{eqnarray}
In particular, if $\lambda=(p-2)/2$, $\mu=1$, and $\nu=p/2$, we can have a fixed point\footnote{Note that the fixed points discussed in section \ref{discrete} are fixed points under 1-form and 0-form symmetry action. In the coset examples studied here, fixed points refer to field identification fixed points.}. Of course, if $p$ is odd, we don't have a fixed point. In this case, we can again run logic similar to that used in the Virasoro case to argue that the TQFT factorizes.

However, if $p$ is even, then we need to properly define the coset. In particular, we should resolve the fixed point Wilson line as follows (see \cite{Schellekens:1989uf,Schellekens:1990xy} for the dual RCFT discussion)
\begin{equation}\label{Wfixed}
\CW_{\left\{(p-2)/2,1,p/2\right\}}\to\CW_{\left\{(p-2)/2,1,p/2\right\}}^{(1)}+\CW_{\left\{(p-2)/2,1,p/2\right\}}^{(2)}~.
\end{equation}
Let us consider what turns out to be the simplest interesting case, $p=6$
\begin{equation}
\widehat{\CT}_6={SU(2)_{4}\boxtimes SU(2)_2\over SU(2)_{6}}~.
\end{equation}
The fixed point resolution in \eqref{Wfixed} becomes $\CW_{\left\{2,1,3\right\}}\to\CW_{\left\{2,1,3\right\}}^{(1)}+\CW_{\left\{2,1,3\right\}}^{(2)}$. As in the cases of one-form gauging with fixed points discussed in section \ref{discrete}, it is natural that there should be a zero-form symmetry exchanging $\CW_{\left\{2,1,3\right\}}^{(1)}\leftrightarrow\CW_{\left\{2,1,3\right\}}^{(2)}$. 

As a first step to better understand the theory after resolving the fixed point, note that $\widehat{\CT}_6$ has the following number of lines
\begin{equation}
|\widehat{\CT}_6|=28~.
\end{equation}
Of these fields, twenty-six come from identifying full length-two orbits in \eqref{abIDN1} while two come from resolving the fixed point. In what follows, $\left\{\lambda,\mu,\nu\right\}$ will denote fields in full orbits, while labels of the form $\left\{2,1,3\right\}^{(i)}$ (with $i=1,2$) will denote the fixed point lines.

To understand the fusion rules and the question of primality after fixed point resolution, we can compute the $S$ matrix using the algorithm discussed in \cite{Schellekens:1989uf} (let us denote the result by $\tilde S$). It takes the form
\begin{eqnarray}\label{tildeS}
\tilde S_{\left\{\lambda,\mu,\nu\right\}\left\{\lambda',\mu',\nu'\right\}}&=&2S_{\left\{\lambda,\mu,\nu\right\}\left\{\lambda',\mu',\nu'\right\}}~,\ \ \ \tilde S_{\left\{2,1,3\right\}^{(i)}\left\{\lambda',\mu',\nu'\right\}}=S_{\left\{2,1,3\right\}\left\{\lambda',\mu',\nu'\right\}}~,\cr\tilde S_{\left\{2,1,3\right\}^{(i)}\left\{2,1,3\right\}^{(j)}}&=&{1\over2}\begin{pmatrix}
1 & -1 & \\
-1 &1 
\end{pmatrix}~,
\end{eqnarray}
where
\begin{equation}
S_{\left\{\lambda,\mu,\nu\right\}\left\{\lambda',\mu',\nu'\right\}}=S^{(p-2)}_{\lambda\lambda'}\cdot S^{(2)}_{\mu\mu'}\cdot S^{(p)}_{\nu\nu'}~,
\end{equation}
is the naive generalization of \eqref{factST} to the cosets at hand. Note that the fusion rules we obtain from $\tilde S$ for fields not involving $\left\{2,1,3\right\}^{(i)}$ are the naive ones we get from $S$ via the restrictions and identifications described above.

The above discussion is sufficient to prove that $\widehat{\CT}_6$ is prime. Indeed, we see from \eqref{tildeS} that the fields that come from identifying length-two orbits have the quantum dimensions they inherit from $S$. The fixed point resolution fields, on the other hand, have half the quantum dimension of the fixed point field. We therefore have the following four abelian anyons generating a $\mathbb{Z}_2\times\mathbb{Z}_2$ fusion algebra
\begin{equation}\label{abT6}
\CW_{\left\{0,0,0\right\}}\simeq\CW_{\left\{4,2,6\right\}}~,\ \CW_{\left\{4,0,0\right\}}\simeq\CW_{\left\{0,2,6\right\}}~,\ \CW_{\left\{0,2,0\right\}}\simeq\CW_{\left\{4,0,6\right\}}~,\ \CW_{\left\{0,0,6\right\}}\simeq\CW_{\left\{4,2,0\right\}}~.
\end{equation}
By \eqref{tildeS}, we see that the braiding amongst abelian anyons is not affected by taking $S\to\tilde S$. As a result, we see that the four abelian anyons all braid trivially. Therefore, they cannot form a decoupled TQFT.

\begin{table}
\begin{center}
\nonumber
     \begin{tabular}{| c| c |}
\hline  {\rm Wilson lines} & ${\rm Quantum\ dimensions}$\cr\hline \hline 
         $\CW_{\{0,0,0\}},\CW_{\{4,0,0\}},\CW_{\{0,2,0\}},\CW_{\{0,0,6\}}$ &$1$\cr \hline 
         $\CW_{\{0,0,2\}},\CW_{\{0,0,4\}},\CW_{\{4,0,2\}},\CW_{\{4,0,4\}}$ & $\cot \left(\frac{\pi }{8}\right)$\cr\hline
         $\CW_{\{1,0,1\}},\CW_{\{1,0,5\}},\CW_{\{3,0,1\}},\CW_{\{3,0,5\}}$ & $\sqrt{\frac{3}{2}} \csc \left(\frac{\pi }{8}\right)$\cr\hline
         $\CW_{\{0,1,3\}},\CW_{\{2,1,3\}^{(1)}},\CW_{\{2,1,3\}^{(2)}}$ & $\sqrt{2} \csc \left(\frac{\pi }{8}\right)$\cr\hline
         $\CW_{\{1,0,3\}},\CW_{\{3,0,3\}}$ & $\sqrt{3} \csc \left(\frac{\pi }{8}\right)$\cr\hline
         $\CW_{\{2,0,0\}},\CW_{\{2,0,6\}}$ & $2$\cr\hline
         $\CW_{\{0,1,1\}},\CW_{\{0,1,5\}}$ & $\csc \left(\frac{\pi }{8}\right)$\cr\hline
         $\CW_{\{1,1,0\}},\CW_{\{1,1,6\}}$ & $\sqrt{6}$\cr\hline
         $\CW_{\{2,0,2\}},\CW_{\{2,0,4\}}$ & $2 \cot \left(\frac{\pi }{8}\right)$\cr\hline
         $\CW_{\{1,1,2\}},\CW_{\{1,1,4\}}$ & $\sqrt{6} \cot \left(\frac{\pi }{8}\right)$\cr\hline
         $\CW_{\{2,1,1\}}$ & $2 \csc \left(\frac{\pi }{8}\right)$\cr\hline
      \end{tabular}
\caption{The twenty-eight Wilson lines and associated quantum dimensions in the $\widehat{\CT}_6$ TQFT.}\label{table1}
\end{center}
\end{table}

Given this discussion, what could a putative factorized theory look like? Since $\widehat{\CT}_6$ has order $28=7\cdot2^2$, we see that the only way to have a non-trivial factorization is to have a factorization of the form $\tilde\CT_{14}\boxtimes\tilde\CT_2$ into prime TQFTs with rank fourteen and rank two, or $\tilde\CT_7\boxtimes\tilde\CT_4$ with prime TQFTs of rank seven and four, or $\tilde\CT_7\boxtimes\tilde\CT_2\boxtimes\tilde\CT_2'$ with prime TQFTs of rank seven, two, and two.

Let us consider the first factorization first. Since the abelian anyons (and any subset thereof) cannot form a separate TQFT factor (this factor would be non-modular), the classification in \cite{rowell2009classification} implies that we have either $\tilde\CT_2\simeq(G_2)_1$ or $\tilde\CT_2\simeq(F_4)_1$. In any case, the non-trivial anyon in $\tilde\CT_2$ has quantum dimension $d_{\tau}=(1+\sqrt{5})/2$. It is easy to check that no such quantum dimension can be produced from products of quantum dimensions in the different coset factors (and so restrictions cannot produce them either). Moreover, one can check that the resolved fixed point fields cannot have this quantum dimension either. This same logic applies to the $\tilde\CT_7\boxtimes\tilde\CT_2\boxtimes\tilde\CT_2'$ factorization as well.

Therefore, it only remains to consider $\tilde\CT_{7}\boxtimes\tilde\CT_4$. The other factor, $\tilde\CT_4$, has four anyons. By \cite{rowell2009classification}, this theory is either $(G_2)_2$ or its time reversal. In either case, we cannot produce the requisite $d_{\alpha}=2\ {\rm cos}(\pi/9)$ quantum dimension from our coset. Therefore, we conclude that $\widehat{\CT}_6$ is indeed a prime TQFT.

Moreover, we find the following fusion rules of non-abelian Wilson lines with unique outcome
\begin{eqnarray}
\CW_{\left\{2,0,0\right\}}\times\CW_{\left\{0,0,2\right\}}&=&\CW_{\left\{2,0,2\right\}}~,\ \CW_{\left\{2,0,0\right\}}\times\CW_{\left\{0,0,4\right\}}=\CW_{\left\{2,0,4\right\}}~,\cr \CW_{\left\{1,1,0\right\}}\times\CW_{\left\{0,0,2\right\}}&=&\CW_{\left\{1,1,2\right\}}~, \ \CW_{\left\{1,1,0\right\}}\times\CW_{\left\{0,0,4\right\}}=\CW_{\left\{1,1,4\right\}}~,\cr \CW_{\left\{0,1,1\right\}}\times\CW_{\left\{2,0,0\right\}}&=&\CW_{\left\{2,1,1\right\}}~.
\end{eqnarray}
We can obtain additional such fusion rules by taking a product with some of the abelian lines in \eqref{abT6}.

Just as in the case of discrete gauge theories with fusion rules of the above type, our theory also has a non-trivial symmetry of the modular data. Indeed, from \eqref{tildeS}, it is clear that the $\tilde S$-matrix has a $\mathbb{Z}_2$ symmetry under the interchange
\begin{equation}
g\left(\CW_{\left\{2,1,3\right\}^{(1)}}\right)=\CW_{\left\{2,1,3\right\}^{(2)}}~, \ \ \ 1\ne g\in\mathbb{Z}_2~.
\end{equation}
Note that this symmetry is not charge conjugation since $\tilde S$ is manifestly real. Moreover, since we don't change the twists, this action lifts to a symmetry of the modular data (additionally, it should lift to a symmetry of the full TQFT).

If we wish to make contact with the $\CN=1$ minimal model, then we should note that the fermionic $\CW_{\left\{0,2,0\right\}}$ line corresponds to the supercurrent of the SCFT. We can then organize the Neveu-Schwarz (NS) sector into supermultiplets under fusion with this operator. Doing so (and paying careful attention to the fields in the resolution of the fixed point), we find nine NS sector fields and nine Ramond sector fields as required.

There are many ways to generalize the example we have given here. Indeed, when there are fixed points in the coset construction we expect to often be able to generate fusion rules of the form \eqref{abcfusion}. A deeper understanding of these theories and some more general methods to characterize whether the cosets are prime (along the lines of the general criteria we have in the case of discrete gauge theories) would be useful. In any case, we see that, as in the case of discrete gauge theories, symmetry fixed points and zero-form (quasi) symmetries are deeply connected with fusion rules of the form \eqref{abcfusion}.

\section{Summary and Conclusions}
In this paper, we have seen that the existence of fusions of non-abelian anyons having a unique outcome is intimately connected with the global structure of the corresponding TQFT. 

Let us summarize our results for continuous gauge groups (and continuous groups more generally):
\begin{itemize}
\item{Building on the well-known fact that $SU(2)$ spin addition / fusion of two non-abelian representations (i.e., higher-dimensional / spin non-singlet representations) is reducible (i.e., has multiple outcomes with different total spin), we argued that a similar result holds in all compact simple Lie groups.}
\item{We argued that the result in the previous bullet point on classical groups can be extended to a theorem constraining $SU(N)_k$ CS theory: fusions of non-abelian Wilson lines in these theories do not have unique outcomes. More generally, Wilson lines corresponding to affine representations with at least two non-vanishing Dynkin labels in any $G_k$ CS theory (for $G$ a compact simple Lie group) do not have unique outcomes. These results have implications for the global structure of these theories (claim 14): the Wilson lines discussed here must all lie in the same prime factor (although $G_k$ CS theories are not prime in general).}
\item{We showed that one way to produce $a\times b=c$ fusions involving non-abelian $a$ and $b$ is to consider cosets. In the case of TQFTs underlying Virasoro minimal models we argued that (as in the $(E_7)_2$ case) such rules arise from factorizations of the TQFTs into multiple prime factors. On the other hand, if we include cosets with fixed points, we can obtain prime theories with such fusion rules.}
\end{itemize}

Next, let us summarize our results for discrete gauge groups (and discrete groups more generally):
\begin{itemize}
\item{We argued that Zisser's construction of irreducible products of higher-dimensional irreducible $A_N$ representations \cite{zisser1993irreducible} can be lifted to fusions of non-abelian Wilson lines with unique outcomes in $A_N$ discrete gauge theory. From the perspective of the closely related $S_N$ group and corresponding discrete gauge theory, the $A_N$ result requires certain 1-form symmetry fixed points (where we define \lq\lq one-form symmetry" in the $S_N$ group to correspond to the $\mathbb{Z}_2\subset{\rm Rep(S_N)}$ generated by the sign representation). We then derived theorem 1 that generalizes this relation between the $A_N$ and $S_N$ discrete gauge theories to other TQFTs.}
\item{Going to the $S_N$ discrete gauge theory by gauging the $\mathbb{Z}_2$ $0$-form outer automorphism symmetry of the $A_N$ discrete gauge theory resolves the $a\times b=c$ non-abelian fusion rule into fusion rules not of this type. However, we saw that in the case of $O(5,3)$ discrete gauge theory such resolutions do not always occur via automorphism gauging. On the other hand, a symmetry fixed point again plays a role: in the resulting $O(5,3)\rtimes\mathbb{Z}_2$ discrete gauge theory, there is a 0-form symmetry fixed point. We then proved theorem 2, which explains why this phenomenon occurs in more general theories. In fact, the $O(5,3)\rtimes\mathbb{Z}_2$ discrete gauge theory relative of the $a\times b=c$ fusion equations in the $O(5,3)$ TQFT described in \eqref{O53WabcE} also has a 1-form symmetry fixed point for the anyon appearing on the right hand side. In the original $O(5,3)$ TQFT this latter anyon becomes a set of two anyons related by the 0-form symmetry. Our theorem 2A generalizes this observation to other TQFTs.}
\item{We showed that one can lift Gallagher's theorem to a statement on the fusion of non-abelian Wilson lines involving unfaithful representations with a unique outcome in TQFT. Moreover, we elucidated the roles that subcategory structure and symmetries play in this result for various specific TQFTs. We then proved theorem 3 that generalizes these observations to a broader set of theories. We also argued that this subcategory structure helps explain the large ratio of group orders in \eqref{groupgap}.}
\item{To gain a sense of how magnetic fluxes behave in general discrete gauge theories, we proved theorem 5. In particular, we showed that in discrete gauge theories with a non-abelian gauge group, $G$, the magnetic fluxes do not form a fusion subcategory. This result immediately places constraints on electric-magnetic self-dualities / quantum symmetries that constrain our symmetry searches later in section 2.}
\item{At a more constructive level, we also proved theorem 10. This result gives infinitely many generalizations of the well-known electric-magnetic self-duality of the $S_3$ discrete gauge theory. }
\item{In order to better understand which discrete gauge theories are prime, we proved theorem 7. This result allowed us to more easily analyze which prime discrete gauge theories have fusions of non-abelian anyons with unique outcomes.}
\item{In order to get a handle on the structure of discrete gauge theories with fusion rules of our desired type involving anyons carying non-trivial flux, we proved theorem 11 and corollary 12. These results give the subcategory structure that arises when such fusions occur. In turn, this structure gives an explanation of these fusion rules. Theorem 13 then partially extends these results to the case in which one of the non-abelian anyons involved is a Wilson line.}
\item{The software GAP was used to analyze the fusion rules of hundreds of untwisted discrete gauge theories. In all the cases we checked, we find that discrete gauge theories with $a\times b=c$ type fusion rules have quasi-zero-form symmetries. This suggests that symmetries of the modular data are a characteristic feature of such fusion rules.}
\end{itemize}

\noindent
The above discussion leads to various natural questions:
\begin{itemize}
\item{In the discussion around \eqref{groupgap} we explained the large hierarchy between the size of simple and non-simple groups whose corresponding discrete gauge theories have non-abelian Wilson lines satisfying \eqref{abcfusion} by using symmetries and subcategory structure. It would be interesting to explore whether other related hierarchies can be explained in a similar way.}
\item{We saw that in almost all the prime untwisted discrete gauge theories we studied, if there was a fusion rule of the form \eqref{abcfusion}, then the theory had non-trivial zero-form symmetries. The only exceptions where discrete gauge theories based on the $M_{23}$ and $M_{24}$ Mathieu groups discussed in section \ref{quasi0}. Here we argued that there were zero-form symmetries of the modular data that did not lift to symmetries of the full theory. It would be interesting to understand if gauge theories based on certain finite simple sporadic groups are the only prime theories with fusion rules of the form \eqref{abcfusion} that exhibit this phenomenon.}
\item{In section \ref{SUNcs}, we proved that the non-abelian lines of $SU(N)_k$ CS theory don't have fusion rules of the form \eqref{abcfusion}. While $(E_7)_2$ CS theory does have such fusion rules, we do not know of an example of such a fusion in a prime $G_k$ CS theory with $G$ a compact and simple Lie group. It would be interesting to either find an example of such a fusion or prove a more general theorem forbidding one. Given such fusions are common for discrete gauge theories, it would be interesting to understand how these two statements interact with each other.}
\item{As we saw in section \ref{fixedpoint}, it would be useful to develop new tools to understand primality in theories built on cosets. One promising direction is to study the role of Galois actions in such theories.}
\end{itemize}

\begin{acknowledgements}
We thank D.~Aasen, M.~Barkeshli, S.~Ramgoolam, and I.~Runkel for useful discussions and correspondence. M.~B. is funded by the Royal Society grant, \lq\lq New Constraints and Phenomena in Quantum Field Theory." M.~B. and R.~R. are funded by the Royal Society grant, \lq\lq New Aspects of Conformal and Topological Field Theories Across Dimensions." The STFC also partially supported our work under the grant, \lq\lq String Theory, Gauge Theory and Duality."
\end{acknowledgements}

\newpage

\appendix
\section{Wilson line $a\times b=c$ in gauge theories with order forty-eight discrete gauge group}\label{appA}

Let us study groups of order $48$ for which the corresponding discrete gauge theories have Wilson line $a \times b=c$ type fusions\footnote{We won't discuss the direct product groups $S_3 \times S_3, D_8 \times S_3$ and $Q_8 \times S_3$ which also have such fusions (the corresponding discrete gauge theories factorize). Since we have already discussed the case of $BOG$ and $GL(2,3)$, we won't be discussing them here}.

\bigskip\noindent
$(48,15)$ ($(\mathbb{Z}_3\times D_8)\rtimes\mathbb{Z}_2$); 

\begin{eqnarray}\label{4815WW}
\CW_{2_2}\times\CW_{2_4}&=&\CW_4~, \ \CW_{2_2}\times\CW_{2_5}=\CW_4~, \ \CW_{2_3}\times\CW_{2_4}=\CW_4~, \ \CW_{2_3}\times\CW_{2_5}=\CW_4\cr \CW_{2_4}\times\CW_{2_6}&=&\CW_4~,\ \CW_{2_4}\times\CW_{2_7}=\CW_4~, \ \CW_{2_5}\times\CW_{2_6}=\CW_4~,\ \CW_{2_5}\times\CW_{2_7}=\CW_4~.\ \ \ 
\end{eqnarray}

\noindent We have $\text{Out}((\DZ_3 \times D_8)\rtimes \DZ_2)=\DZ_2 \times \DZ_2$. Let $r_1$ and $r_2$ be the generators of this group. They act on the Wilson lines involved in the fusion above as follows
\bea
&&r_1: \CW_{2_2}\leftrightarrow \CW_{2_2}; ~\CW_{2_3}\leftrightarrow \CW_{2_3}; ~\CW_{2_4}\leftrightarrow \CW_{2_4}; ~\CW_{2_5}\leftrightarrow \CW_{2_5}; ~\CW_{2_6}\leftrightarrow \CW_{2_7}; ~\\
&&r_1: \CW_{2_2}\leftrightarrow \CW_{2_2}; ~\CW_{2_3}\leftrightarrow \CW_{2_3}; ~\CW_{2_4}\leftrightarrow \CW_{2_5}; ~\CW_{2_6}\leftrightarrow \CW_{2_6}; ~\CW_{2_7}\leftrightarrow \CW_{2_7}; ~
\eea
Since this group has complex characters we also have a non-trivial quasi-zero-form symmetry given by complex conjugation. 
$\CZ(\text{Vec}_{(\DZ_3 \times D_8)\rtimes \DZ_2})$ also has all other $a \times b=c$ type fusions (involving fluxes and dyons) discussed in this paper.

\bigskip\noindent
$(48,16)$ ($(\mathbb{Z}_3: Q_8)\rtimes\mathbb{Z}_2$); 
This has fusions identical to \eqref{4815WW}. The only difference is that now $\CW_{2_4}$ and $\CW_{2_5}$ are conjugates. The outer automorphism group and symmetry action is identical to $\CZ(\text{Vec}_{(\DZ_3 \times D_8)\rtimes \DZ_2})$. Since this group has complex characters we also have a non-trivial quasi-zero-form symmetry given by complex conjugation. We additionally have all other $a \times b=c$ type fusions (involving fluxes and dyons) discussed in this paper.

\bigskip\noindent
$(48,17)$ (($\mathbb{Z}_3\times Q_8)\rtimes\mathbb{Z}_2$); This has identical character table to $(48,16)$, so same fusion rules. The properties are identical to the two cases above.

\bigskip\noindent
$(48,18)$ ($\mathbb{Z}_3\rtimes Q_{16}$); Identical characters to $(48,15)$, so shares \eqref{4815WW}. The discussion is identical to the case above. 

\bigskip\noindent
$(48,39)$ ($(\mathbb{Z}_4\times S_3)\rtimes \mathbb{Z}_2$); 

\begin{eqnarray}\label{4839W}
\CW_{2_1}\times\CW_{2_5}&=&\CW_4~, \ \CW_{2_1}\times\CW_{2_6}=\CW_4~, \ \CW_{2_2}\times\CW_{2_5}=\CW_4~, \ \CW_{2_2}\times\CW_{2_6}=\CW_4\cr \CW_{2_3}\times\CW_{2_5}&=&\CW_4~,\ \CW_{2_3}\times\CW_{2_6}=\CW_4~, \ \CW_{2_4}\times\CW_{2_5}=\CW_4~,\ \CW_{2_4}\times\CW_{2_6}=\CW_4~.\ \ \ 
\end{eqnarray}

\noindent We have $\text{Out}((\DZ_4 \times S_3)\rtimes \DZ_2)=\DZ_2 \times \DZ_2$. Let $r_1$ and $r_2$ be the generators of this group. They act on the Wilson lines involved in the fusion above as follows
\bea
&&r_1: \CW_{2_1}\leftrightarrow \CW_{2_1}; ~\CW_{2_2}\leftrightarrow \CW_{2_2}; ~\CW_{2_3}\leftrightarrow \CW_{2_3}; ~\CW_{2_4}\leftrightarrow \CW_{2_4}; ~\CW_{2_5}\leftrightarrow \CW_{2_6}; ~\\
&&r_1: \CW_{2_1}\leftrightarrow \CW_{2_2}; ~\CW_{2_3}\leftrightarrow \CW_{2_3}; ~\CW_{2_4}\leftrightarrow \CW_{2_4}; ~\CW_{2_5}\leftrightarrow \CW_{2_5}; ~\CW_{2_6}\leftrightarrow \CW_{2_6}; ~
\eea
Since this group has complex characters we also have a non-trivial quasi-zero-form symmetry given by complex conjugation. 
$\CZ(\text{Vec}_{(\DZ_4 \times S_3)\rtimes \DZ_2})$ also have all other $a \times b=c$ type fusions (involving fluxes and dyons) discussed in this paper. 

\bigskip\noindent
$(48,41)$; ($(\mathbb{Z}_4\times S_3)\rtimes \mathbb{Z}_2$) 

Fusion of Wilson lines giving unique output is same as $\eqref{4839W}$. We have $\text{Out}((\DZ_4 \times S_3)\rtimes \DZ_2)=D_{12}$. 

Since this group has complex characters we also have a non-trivial quasi-zero-form symmetry given by complex conjugation. 
$\CZ(\text{Vec}_{(\DZ_4 \times S_3)\rtimes \DZ_2})$ also have all other $a \times b=c$ type fusions (involving fluxes and dyons) discussed in this paper.

\section{Genuine zero-form symmetries and quasi-zero-form symmetries in $A_9$ discrete gauge theory}\label{A9app}
Recall from section \ref{simpleW} that $A_9$ is the simplest example of an $A_N$ (with $N=k^2\ge9$) discrete gauge theory with fusion rules involving non-abelian Wilson lines having unique outcome. Here our goal is to disentangle the genuine zero form symmetries
\begin{equation}\label{symAnAPP}
{\rm Aut}^{\rm br}(\CZ({\rm Vec}_{A_9}))\simeq H^2(A_9,U(1))\rtimes{\rm Out}(A_9)\simeq \mathbb{Z}_2\times\mathbb{Z}_2~,
\end{equation} 
from a charge conjugation quasi zero-form symmetry \cite{davydov2016unphysical}.

Let us first discuss the outer automorphisms. To that end, recall that $A_9$ has an outer automorphism corresponding to conjugation by odd elements of $S_9\triangleright A_9$.  Acting with the outer automorphism generated by $(89)\in S_9$, we see that the following lines are exchanged
\begin{equation}\label{0form}
\CL_{([(123456789)],\pi_{p})}\leftrightarrow\CL_{([(123456798)],\pi_p)}~, \ \ \ \CL_{([(12345)(678)],\pi_n)}\leftrightarrow\CL_{([(12345)(679)],\pi_n)}~,
\end{equation}
where the relevant conjugacy classes are listed in table \ref{table1},  and $0\le p\le8$, $0\le n\le14$ label representations of the corresponding $\mathbb{Z}_9$ and $\mathbb{Z}_{14}$ centralizers (they are also listed in table \ref{table1}).

In fact, as described in the main text, the symmetry in \eqref{0form} generates an action on some of the Wilson lines involved in \eqref{Anabc}
\begin{equation}
\CW_{[3^3]_+}\leftrightarrow\CW_{[3^3]_-}~.
\end{equation}
This action can be read off from the character table of $A_9$ or, equivalently, from the braiding
\begin{eqnarray}
{S_{\CW_{[3^3]_+}\CL_{([(12345)(678)],\pi_n)}}\over S_{\CW_{1}\CL_{([(12345)(678)],\pi_n)}}}&=&\chi_{[3^3]_+}([(12345)(678)])^*=-{1\over2}(1-i\sqrt{15})~,\cr {S_{\CW_{[3^3]_-}\CL_{([(12345)(678)],\pi_n)}}\over S_{\CW_1\CL_{([(12345)(678)],\pi_n)}}}&=&\chi_{[3^3]_-}([(12345)(678)])^*=-{1\over2}(1+i\sqrt{15})~,\cr {S_{\CW_{[3^3]_+}\CL_{([(12345)(679)],\pi_n)}}\over S_{\CW_1\CL_{([(12345)(679)],\pi_n)}}}&=&\chi_{[3^3]_+}([(12345)(679)])^*=-{1\over2}(1+i\sqrt{15})~,\cr {S_{\CW_{[3^3]_-}\CL_{([(12345)(679)],\pi_n)}}\over S_{\CW_1\CL_{([(12345)(679)],\pi_n)}}}&=&\chi_{[3^3]_-}([(12345)(679)])^*=-{1\over2}(1-i\sqrt{15})~.
\end{eqnarray}

Note that, since the $[(12345)(678)]$ and $[12345)(679)]$ conjugacy classes are complex, we also have a non-trivial $\mathbb{Z}_2$ charge conjugation that acts on the modular data and swaps $\CW_{[3^3]_+}\leftrightarrow\CW_{[3^3]_-}$ and $\CL_{([(123456789)],\pi_{p})}\leftrightarrow\CL_{([(123456798)],\pi_p)}$. Recall from the discussion in \eqref{zetaAct} that elements of $H^2(A_9,U(1))\simeq\mathbb{Z}_2$ act trivially on the Wilson lines. Hence, we learn that charge conjugation cannot be a genuine symmetry of the TQFT (this statement is also confirmed by the analysis in \cite{davydov2016unphysical}).

However, this is not a contradiction with what we have written, because ${\rm Out}(A_9)$ also interchanges the real conjugacy classes $[(123456789)]$ and $[(123456798)]$ along with the corresponding lines in \eqref{0form}. Since charge conjugation leaves these degrees of freedom untouched, it is a distinct operation.

\begin{table}
\begin{center}
\nonumber
     \begin{tabular}{| c| c |c |}
\hline  {\rm Conjugacy class} & ${\rm Length}$ & ${\rm Centralizer}$ \cr\hline \hline 
         $1$ &$1$& {$A_9$} \cr \hline 
              $[(12)(34)]$ & $378$ & ${\rm SmallGroup}(480,951)$\cr\hline 
               $[(12)(34)(56)(78)]$ & $945$ & ${\rm SmallGroup}(192,1493)$  \cr\hline
               $[(123)]$ & $168$ & ${\rm SmallGroup}(1080,487)$\cr\hline
               $[(123)(45)(67)]$ & $7560$ & ${\rm SmallGroup}(24,10)~,\ (D_8\times\mathbb{Z}_3)$\cr\hline
               $[(123)(456)]$ & $3360$ & ${\rm SmallGroup}(54,13)$\cr\hline
               $[(123)(456)(789)]$ & $2240$ & ${\rm SmallGroup}(81,7)~,\ ((\mathbb{Z}_3\times\mathbb{Z}_3\times\mathbb{Z}_3)\rtimes\mathbb{Z}_3)$\cr\hline
               $[(1234)(56)]$ & $7560$ & ${\rm SmallGroup}(24,5)~,\ (S_3\times\mathbb{Z}_4)$\cr\hline
               $[(1234)(567)(89)]$ & $15120$ & ${\rm SmallGroup}(12,2)~,\ (\mathbb{Z}_{12})$\cr\hline
               $[(1234)(5678)]$ & $11340$ & ${\rm SmallGroup}(16,13)~,\ ({\rm central\ product}\ D_8,\ Z_4)$\cr\hline              
               $[(12345)]$ & $3024$ & ${\rm SmallGroup}(60,9)$\cr\hline
               $[(12345)(67)(89)]$ & $9072$ & ${\rm SmallGroup}(20,5)~,\ (\mathbb{Z}_{10}\times\mathbb{Z}_2)$\cr\hline              
               $[(12345)(678)]$ & $12096$ & ${\rm SmallGroup}(15,1)~,\ (\mathbb{Z}_{15})$\cr\hline
               $[(12345)(679)]$ & $12096$ & ${\rm SmallGroup}(15,1)~,\ (\mathbb{Z}_{15})$\cr\hline               
               $[(123456)(78)]$ & $30240$ & ${\rm SmallGroup}(6,2)~,\ (\mathbb{Z}_{6})$\cr\hline            
               $[(1234567)]$ & $25920$ & ${\rm SmallGroup}(7,1)~,\ (\mathbb{Z}_{7})$\cr\hline
               $[(123456789)]$ & $20160$ & ${\rm SmallGroup}(9,1)~,\ (\mathbb{Z}_{9})$\cr\hline
               $[(123456798)]$ & $20160$ & ${\rm SmallGroup}(9,1)~,\ (\mathbb{Z}_{9})$\cr\hline
      \end{tabular}
\caption{The eighteen conjugacy classes of $A_9$, their order, and their centralizers (recall that the centralizers of elements in the same conjugacy class are isomorphic). The centralizer is labeled by its GAP ID (for sufficiently small groups) as \lq\lq${\rm SmallGroup}(a,b)$" along with a more common name in certain cases.}\label{table1}
\end{center}
\end{table}

Note that in the $A_9$ discrete gauge theory we can also turn on a large variety of twists
\begin{equation}
\omega\in H^3(A_9,U(1))\simeq\mathbb{Z}_2\times\mathbb{Z}_3^2\times\mathbb{Z}_4\simeq\mathbb{Z}_6\times\mathbb{Z}_{12}~.
\end{equation}
Since the charge conjugation quasi-symmetry is a property of the Wilson line fusion rules, it remains regardless of the twist.

\section{GAP code}\label{GAPapp}

The following GAP code defines the function checkdyon() which takes in a group as an argument. It checks for $a \times b=c$ type fusions for non-abelian anyons $a,b,c \in \CZ(\text{Vec}_{G})$ and ouputs all such fusions. Moreover, if such fusions exist, it outputs $\text{Out}(G)$ as well as $H^2(G,U(1))$. Note that it requires the package HAP to function.  

In order to define checkdyon() we need to first define the functions comconj() and conjprof().
\begin{flalign*}
&>\text{conjcom:=function(a,b)}&\\
 &>\text{local com,i,j;}&\\
 &>\text{com:=[];}&\cr
 &>\text{for i in [1..Size(AsList(a))] do }&\cr
 &>\text{for j in [i..Size(AsList(b))] do}&\cr
 &>\text{Append(com, [AsList(a)[i]*AsList(b)[j]*Inverse(AsList(b)[j]*AsList(a)[i])]);} &\cr
& >\text{od; od;}&\cr
& >\text{return DuplicateFreeList(com)=[AsList(a)[1]*Inverse(AsList(a)[1])]; end;}&
\end{flalign*}

\noindent This function takes two conjugacy classes of a group $G$ as inputs and outputs true if they commute element-wise and false otherwise. Now, let us define the function conjprod()
\begin{flalign*}
&>\text{conjprod:=function(a,b,c)}&\cr
& >\text{local prod,i,j,k;}&\cr
& >\text{prod:=[];}&\cr
& >\text{for i in [1..Size(AsList(a))] do}&\cr
& >\text{for j in [i..Size(AsList(b))] do}&\cr
& >\text{for k in [1..Size(c)] do}&\cr
& >\text{if AsList(a)[i]*AsList(b)[j] in AsList(c[k]) then }\cr
& >\text{Append(prod, [k]); break; fi; od; od; od;}&\cr
& >\text{if Size(DuplicateFreeList(prod))=1 then}& \cr
& >\text{return DuplicateFreeList(prod)[1]; else return 0; fi; end;}&
\end{flalign*}

\noindent This function takes three arguments. The first two arguments $a,b$ are two conjugacy classes of a group $G$ and the third argument $c$ is the set of all conjugacy classes of $G$. The function outputs an integer $k>1$ if the product of two input conjugacy is a single conjugacy class (which is at position $k$ in the list of conjugacy classes $c$). The function outputs $0$ otherwise. 

\noindent Using these two functions, we finally define the checkdyon() function.
\begin{flalign*}
& \text{checkdyon:=function(G)}&\cr
& > \text{local cn,i,j,k,a,l,cen1,cen2,cen3,cenint,irrcenint,irrcen1,irrcen2,irrcen3,
} &\cr 
& \text{cen1res,cen2res,cen3res,x,y,z,w,a1,a2,A,I,F,R;
} &\cr
& > \text{cn:=ConjugacyClasses(G);}&\cr
& > \text{a:=0;}&\cr
& > \text{for i in [1..Size(cn)] do}&\cr
& > \text{for j in [i..Size(cn)] do}&\cr
& > \text{if conjcom(cn[i],cn[j]) then}&\cr
& > \text{k:=conjprod(cn[i],cn[j],cn); }&\cr
& > \text{if k$<>$0 then }&\cr
& > \text{cen1:=Centralizer(G,AsList(cn[i])[1]);}&\cr
& > \text{cen2:=Centralizer(G,AsList(cn[j])[1]);}&\cr
& > \text{cen3:=Centralizer(G,AsList(cn[k])[1]);}&\cr
& > \text{cenint:=Intersection(cen1,cen2,cen3);}&\cr
& > \text{irrcen1:=Irr(cen1);}&\cr
& > \text{irrcen2:=Irr(cen2);}&\cr
& > \text{irrcen3:=Irr(cen3);}&\cr
& > \text{cen1res:=RestrictedClassFunctions(irrcen1,cenint);}&\cr
& > \text{cen2res:=RestrictedClassFunctions(irrcen2,cenint);}&\cr
& > \text{cen3res:=RestrictedClassFunctions(irrcen3,cenint);}&\cr
& > \text{irrcenint:=Irr(cenint);}&\cr  
& > \text{for x in [1..Size(cen1res)] do}&\cr
& > \text{for y in [1..Size(cen2res)] do}&\cr
& > \text{if Size(AsList(cn[i]))*DegreeOfCharacter(cen1res[x])$>$1 and}&\cr
& \text{ Size(AsList(cn[j]))*DegreeOfCharacter(cen2res[y])$>$1 then }& \cr
& > \text{for z in [1..Size(cen3res)] do}&\cr
& > \text{a1:=[~]; a2:=[~];}&\cr
& > \text{for w in [1..Size(irrcenint)] do}&\cr
& > \text{Append(a1,[ScalarProduct(irrcenint[w],cen1res[x]*cen2res[y])]); }&\cr
& > \text{Append(a2,[ScalarProduct(irrcenint[w],cen3res[z])]);}&\cr
& > \text{od;}&\cr
& > \text{if a1*a2=1 and }&\cr
& \text{Size(AsList(cn[i]))*DegreeOfCharacter(cen1res[x])*} &\cr 
& \text{Size(AsList(cn[j]))*DegreeOfCharacter(cen2res[y])=}& \cr
&  \text{Size(AsList(cn[k]))*DegreeOfCharacter(cen3res[z]) then }&\cr
& > \text{a:=1;}&\cr
\end{flalign*}
\begin{flalign*}
& > \text{Print(IdSmallGroup(G), `` ", StructureDescription(G), ``\textbackslash n");}&\cr
& > \text{Print(``Anyon a: ", cn[i], `` , ", irrcen1[x], ``\textbackslash n");}&\cr
& > \text{Print(``Anyon b: ", cn[j], `` , ", irrcen2[y], ``\textbackslash n");}&\cr
& > \text{Print(``Anyon c: ", cn[k], `` , ", irrcen3[z], ``\textbackslash n","\textbackslash n");}&\cr
& > \text{fi; od; fi; od;od; fi; fi; od; od; }&\cr
& > \text{if a=1 then}&\cr
& > \text{A:=AutomorphismGroup(G);}&\cr
& > \text{I:=InnerAutomorphismsAutomorphismGroup(A);}&\cr
& > \text{F:=FactorGroup(A,I);}&\cr
& > \text{Print(``Out(G): ",StructureDescription(F), ``\textbackslash n");}&\cr
& > \text{R:=ResolutionFiniteGroup(G,3);}&\cr
& > \text{Print(``H2(G,U(1)): ",Homology(TensorWithIntegers(R),2),``\textbackslash n");}&\cr
& > \text{Print(``\textbackslash n",``\textbackslash n"); fi;}&\cr
& > \text{end;}&
\end{flalign*}

\newpage
\bibliographystyle{unsrtnat}
\bibliography{chetdocbib}

\end{document}